\documentclass[12pt,preprintnumbers,nofootinbib,letterpaper]{article}
\pdfoutput=1
\usepackage{amsmath,amssymb,amsfonts,cite,xcolor,epsf,epsfig,epstopdf,graphics,graphicx,mathtools,caption,subcaption,physics,verbatim}
\def\thefootnote{*\arabic{footnote}}
\definecolor{ultramarine}{rgb}{0.07, 0.04, 0.56}
\definecolor{cadmiumgreen}{rgb}{0.0, 0.42, 0.24}
\definecolor{indigo_dye}{rgb}{0.0, 0.25, 0.42} 

\usepackage[linktocpage=true,breaklinks]{hyperref}
\hypersetup{
	colorlinks=true,
	citecolor=ultramarine,
	linkcolor=cadmiumgreen,
	urlcolor=indigo_dye,
}

\usepackage{orcidlink}

\usepackage[utf8]{inputenc}

\usepackage{enumitem}

\usepackage{outlines}

\usepackage[margin=2cm]{geometry}
\usepackage{adjustbox}
\usepackage{multirow}
\usepackage[]{cleveref}

\usepackage{float}

\usepackage{empheq}

\usepackage[framemethod=TikZ]{mdframed}
\mdfsetup{roundcorner=3pt}
\newmdenv[
skipabove=2pt,
skipbelow=2pt,
rightline=false,
leftline=false,
topline=false,
bottomline=false,
backgroundcolor=gray!15,
innerleftmargin=10pt,
innerrightmargin=10pt,
innertopmargin=10pt,
innerbottommargin=10pt,
leftmargin=0cm,
rightmargin=0cm,
linewidth=4pt]{eBox}

\numberwithin{equation}{section}

\usepackage{array}
\newcolumntype{P}[1]{>{\centering\arraybackslash}p{#1}}

\usepackage{dcolumn}
\newcolumntype{M}[1]{>{\centering\arraybackslash}m{#1}}
\newcolumntype{N}{@{}m{0pt}@{}}

\newcommand{\Mpl}{M_{\rm Pl}}

\newcommand{\D}{{\rm d}}

\usepackage{tikz}

\usetikzlibrary{arrows.meta,calc}

\definecolor{hblue}{RGB}{0,114,178}
\definecolor{gorange}{RGB}{213,94,0}

\newcommand{\futurecone}[4]{
	\fill[#1,fill opacity=0.16] (0,0) -- (-#2,#3) -- (#2,#3) -- cycle;
	\fill[#1,fill opacity=0.28] (0,#3) ellipse ({#2} and {#4});
	\draw[#1,thick] (0,0) -- (-#2,#3);
	\draw[#1,thick] (0,0) -- (#2,#3);
	\draw[#1,thick] (0,#3) ellipse ({#2} and {#4});
}

\newcommand{\axes}{%
	\draw[->,gray!70,thin] (-2.45,0) -- (2.45,0) node[below right=-2pt] {\small $n_1$};
	\draw[->,gray!70,thin] (0,-2.45) -- (0,2.45) node[above left=-2pt]  {\small $U$};
}
\newcommand{\hcone}{%
	\fill[hblue,fill opacity=0.13] (0,0) -- (-2.2,2.2) -- (2.2,2.2) -- cycle;
	\fill[hblue,fill opacity=0.13] (0,0) -- (2.2,-2.2) -- (-2.2,-2.2) -- cycle;
	\draw[hblue,thick]  (-2.2,-2.2) -- (2.2,2.2);
	\draw[hblue,thick]  (-2.2,2.2)  -- (2.2,-2.2);
}
\newcommand{\tangentline}{%
	\draw[gorange,thick,dash pattern=on 5pt off 5pt] (-2.2,2.2) -- (2.2,-2.2);
}

\definecolor{hblue}{RGB}{0,114,178}
\definecolor{gorange}{RGB}{213,94,0}

\begin{document}
	
	\vspace*{.7cm}
	
	\begin{center}
		
		\def\thefootnote{\fnsymbol{footnote}}
		
		{\Large {\bf Disformal Maps: Classification and Singular Dynamics}}
		\\[1cm]
		
		{Mohammad Ali Gorji\,\orcidlink{0000-0001-8848-2283}$^{1}$, Pavel Jirou\v{s}ek\,\orcidlink{0000-0002-3047-2100}, Alexander Vikman\,\orcidlink{0000-0003-3957-2068}$^{2}$, Masahide Yamaguchi\,\orcidlink{0000-0002-2193-5273}$^{1,3,4}$}
		\\[.7cm]
		
		{\small \textit{$^1$Cosmology, Gravity, and Astroparticle Physics Group, Center for Theoretical Physics of the Universe, Institute for Basic Science (IBS), Daejeon, 34126, Korea
		}}\\[.1cm]
		
		{\small \textit{$^{2}$CEICO - Central European Institute for Cosmology and Fundamental Physics, \\ FZU - Institute of Physics of the Czech Academy of Sciences, \\Na Slovance 1999/2, 182 00 Prague 8, Czech Republic}}\\[.1cm]
		
		{\small \textit{$^3$Department of Physics, Institute of Science Tokyo, \\2-12-1 Ookayama, Meguro-ku, Tokyo 152-8551, Japan
		}}\\[.1cm]
		
		{\small \textit{$^{4}$Department of Physics \& Institute of Physics and Applied Physics (IPAP), Yonsei University, \\Seoul 03722, South Korea}}\\[.1cm]
		
	\end{center}
	
	\vspace{.8cm}
	
	\hrule \vspace{0.3cm}
	
\begin{abstract}
Being agnostic about the field content of a gravitational system, we consider a general disformal transformation of the metric, $g_{\mu\nu}=C h_{\mu\nu}+D t_{\mu\nu}$, on a four-dimensional Lorentzian manifold. Using the Cayley-Hamilton theorem, we derive an explicit formula for the inverse disformed metric. Implementing the Hawking-Ellis classification, we categorize disformal transformations into four types: Type I, II, III, and IV, based on possible Jordan block structures. We further identify the Segre subclasses from the eigenvalues. We derive explicit relations between the Cayley-Hamilton degree of $t_{\mu\nu}$, its Hawking-Ellis type, and its Segre subclass, restricting the possible types from the degree alone. In some cases, the type can be determined without even performing a full Jordan decomposition. For singular transformations, when new dynamical degrees of freedom emerge, we obtain the form of their corresponding mimetic energy-momentum tensor $T^\star_{\mu\nu}$. Within the polynomial sector, the Hawking-Ellis types of $t_{\mu\nu}$ and $T^\star_{\mu\nu}$ always coincide for Type I. For Types II and III it can differ, while Type IV is preserved generically but can reduce to Type I when the complex pair is mapped to a repeated real eigenvalue. This makes it possible to infer physical properties of $T^\star_{\mu\nu}$ directly from the Hawking-Ellis type of $t_{\mu\nu}$. We apply our setup to scalar and gauge fields. We find a new pure-disformal scalar-gauge family yielding irrotational perfect fluids, including barotropic fluids with an arbitrary constant equation-of-state parameter. Two scalar fields realize all four types and the reductions from Types II and IV to Type I. This framework applies to the kinematical and dynamical study of invertible and non-invertible disformal transformations with different field content.
\end{abstract}

	\vspace{0.5cm}
	
	\hrule
	\def\thefootnote{\arabic{footnote}}
	\setcounter{footnote}{0}
	
	\thispagestyle{empty}
	
	\newpage
	
	\vspace*{.7cm}
	
	\hrule
\begingroup
\small
\linespread{0.95}\selectfont
\tableofcontents
\endgroup
\addtocontents{toc}{\protect\setcounter{tocdepth}{2}}
\vspace{0.7cm}
\hrule
	
\newpage
	
	
\section{Introduction}\label{introduction}

General relativity is remarkably well tested at Solar-System scales and further probed by binary pulsars and gravitational-wave events. Nonetheless, the observed accelerated expansion of the Universe on the largest scales, together with the theoretical issues of quantum non-renormalizability at short distances, the cosmological constant problem and the singularity problem, motivates the search for deviations from general relativity. In particular, recent observations hint \cite{DESI:2024mwx,DESI:2025zgx,DESI:2025fii} that the dark energy driving late-time cosmic acceleration is dynamical, a feature that cannot be accommodated by general relativity extended with a cosmological constant. It is then widely believed that general relativity gets modified in the infrared and ultraviolet regimes. In this regard, many modified gravity theories like scalar-tensor theories	\cite{Horndeski:1974wa,Chiba:1999ka,Armendariz-Picon:1999hyi,Garriga:1999vw,Armendariz-Picon:2000nqq,Armendariz-Picon:2000ulo,Nicolis:2008in,Deffayet:2009mn,Deffayet:2009wt,Deffayet:2010qz,Kobayashi:2010cm,Kobayashi:2011nu,Deffayet:2011gz,Langlois:2018dxi,Kobayashi:2019hrl,Zumalacarregui:2013pma,Gleyzes:2014dya,Gleyzes:2014qga,Crisostomi:2016czh,BenAchour:2016cay,BenAchour:2016fzp,Langlois:2015cwa}, vector-tensor theories \cite{Tasinato:2014eka,Heisenberg:2014rta,Heisenberg:2016eld,Kimura:2016rzw,Heisenberg:2018vsk}, massive gravity \cite{deRham:2010kj,Hinterbichler:2011tt,deRham:2014zqa}, Ho\v{r}ava-Lifshitz gravity \cite{Horava:2009uw,Horava:2009if,Blas:2009qj,Blas:2010hb,Mukohyama:2010xz}, teleparallel gravity \cite{Hammond:2002rm,Maluf:2013gaa,Bahamonde:2015zma,BeltranJimenez:2017tkd,Krssak:2018ywd,Bahamonde:2021gfp}, and higher-dimensional theories \cite{Overduin:1997sri,Randall:1999ee,Randall:1999vf,Dvali:2000hr,Maartens:2010ar} have been investigated to address different issues at different scales, for reviews see, e.g. \cite{Clifton:2011jh,Joyce:2014kja}.
	
Most modified gravity theories introduce extra scalar, vector, or tensor degrees of freedom on top of the gravitons (see, however, \cite{Afshordi:2006ad,Lin:2017oow,Mukohyama:2019unx,Aoki:2020lig}). Even a single extra scalar degree of freedom gives rise to a rich variety of theories with applications ranging from cosmology to compact objects, including black holes. These are, for example, k-essence \cite{Armendariz-Picon:1999hyi,Garriga:1999vw,Chiba:1999ka,Armendariz-Picon:2000nqq,Armendariz-Picon:2000ulo}, kinetic gravity braiding \cite{Deffayet:2010qz,Kobayashi:2010cm}, Galileons\cite{Nicolis:2008in,Deffayet:2009wt,Deffayet:2009mn}, Horndeski theory \cite{Horndeski:1974wa,Deffayet:2011gz,Kobayashi:2011nu,Kobayashi:2019hrl}, beyond Horndeski \cite{Zumalacarregui:2013pma,Gleyzes:2014dya,Gleyzes:2014qga,Crisostomi:2016czh,BenAchour:2016cay,BenAchour:2016fzp,Takahashi:2021ttd}. It is worth reminding the reader that k-essence does not involve any higher derivatives or non-minimal couplings to curvature in the action, while Horndeski theory is the most general scalar-tensor theory with higher derivatives and direct couplings to curvature, yet yielding second-order equations of motion for both the metric and the scalar. It turns out that all of these theories can be understood as various subsets of a more general class, which is called Degenerate Higher-Order Scalar-Tensor (DHOST) theories \cite{Langlois:2015cwa,Langlois:2018dxi}. Disformal transformations \cite{Bekenstein:1992pj} played a crucial role in finding and classifying these theories (see \cite{BenAchour:2024hbg} and references therein). Like a conformal transformation, a disformal transformation is a map 
\begin{align}\label{dis-T-d}
g_{\mu\nu}= { C} \, h_{\mu\nu} + { D} \, t_{\mu\nu} \,,
\end{align}
from metric $h_{\mu\nu}$ to metric $g_{\mu\nu}$, where $t_{\mu\nu}$ is a symmetric rank-$2$ tensor which is not proportional to $h_{\mu\nu}$, while ${C}$ and ${D}$ are functions\footnote{One could absorb $D$ into $t_{\mu\nu}$, but for ease of comparison with the existing literature we keep $D$ explicit.} of the fields and the metric. The properties of the disformal part are encoded in $t_{\mu\nu}$, which is constructed out of the extra degrees of freedom under consideration. 
    
For example, in the case of a scalar field, $\phi$, one can take $t_{\mu\nu}=\partial_\mu\phi\,\partial_\nu\phi$, while the coefficients $C$ and $D$ depend on the contractions of $t_{\mu\nu}$ with $h_{\mu\nu}$ that is: ${ C}={ C}(\phi,h^{\alpha\beta}\partial_\alpha\phi\partial_\beta\phi)$ and ${ D}={ D}(\phi,h^{\alpha\beta}\partial_\alpha\phi\partial_\beta\phi)$. It is worth noting that, for some choices of the functions $C$ and $D$, the new metric obtained in this way is the acoustic metric which is the metric governing the propagation of small perturbations in k-essence, see e.g. \cite{Armendariz-Picon:2005oog,Babichev:2007dw,Sawicki:2024ryt,Barcelo:2026vlr}. 
    
On the other hand, applying disformal transformation \eqref{dis-T-d} with $t_{\mu\nu}=\partial_\mu\phi\partial_\nu\phi$ to a k-essence theory  one finds a subset of Horndeski theory \cite{Bettoni:2013diz,Ezquiaga:2017ner}. As long as such a transformation is non-singular, see e.g. \cite{Jirousek:2022rym,Jirousek:2022jhh}, the two theories are classically equivalent  in the absence of other matter fields \cite{Zumalacarregui:2013pma,Domenech:2015tca,Chiba:2020mte,Achour:2021pla,Takahashi:2022mew,Takahashi:2022ctx,Ikeda:2023ntu}. Furthermore, applying disformal transformations to Horndeski theories generates the beyond Horndeski theories \cite{Zumalacarregui:2013pma}. Later, the same strategy, implemented in the context of vector-tensor theories with $t_{\mu\nu}=A_\mu{A}_\nu$, was used to construct the generalized Proca theories \cite{Kimura:2016rzw}, which go beyond the standard Proca theory without introducing extra degrees of freedom. In a similar fashion, new vector-tensor theories can be generated \cite{Gorji:2018okn,Gumrukcuoglu:2019ebp,DeFelice:2019hxb} using gauge fields in the transformation\footnote{Gauge fields can also appear in the transformation in a different way, provided one also uses a scalar field \cite{Vikman:2017gxs}.}, so that $t_{\mu\nu}=h^{\alpha\beta}F_{\alpha\mu}F_{\beta\nu}$. In this respect, disformal transformations are very useful tools to clarify relations and equivalences between apparently different modified gravity theories. On the other hand, performing a singular disformal transformation on a theory substantially changes its dynamics, even when no new dynamical degrees of freedom are generated and -- more surprisingly -- even when the transformation is invertible \cite{Jirousek:2022rym,Jirousek:2022jhh}. 
    
For instance, an irrotational dust-like fluid, similar to \cite{Lim:2010yk} and reproducing dark matter on large scales, is realized in mimetic theories \cite{Chamseddine:2013kea,Chamseddine:2014vna} via a singular disformal transformation involving a scalar field \cite{Golovnev:2013jxa,Barvinsky:2013mea,Deruelle:2014zza,Jirousek:2022rym}. As the next step, this dark-matter-like fluid can be extended by higher derivatives to an imperfect dark matter \cite{Mirzagholi:2014ifa,Capela:2014xta,Ramazanov:2015pha,Ramazanov:2016xhp,Gorji:2025ajb}, which corresponds \cite{Blas:2010hb,Ramazanov:2016xhp} to the IR limit  of the projectable version of the Ho\v{r}ava-Lifshitz gravity. For further, most recent extension of the imperfect dark matter, see \cite{Gorji:2025ajb}. 
    
Moreover, the mimetic construction can be extended to vector fields  to find a Weyl-invariant, higher derivative formulation  \cite{Jirousek:2018ago,Hammer:2020dqp,Jirousek:2022rym} of unimodular gravity \cite{Einstein:1919gv,Henneaux:1989zc}, which is arguably the simplest dynamical model of dark energy. For further extensions and related constructions see also \cite{Gorji:2017cai,Takahashi:2017pje,Vikman:2017gxs,Langlois:2018jdg,Firouzjahi:2018xob,Gorji:2020ten,Domenech:2023ryc,Domenech:2025gao,Gorji:2025fbv,Allahyari:2026rnm,Braga:2026fkn}. Thus, singular disformal transformations with different field contents can be used to construct new modified-gravity theories, or to find novel formulations of old ones. Such reformulations are useful for a better understanding of the theories' origins and their potential extensions.

Most of the studies in the literature focus on a particular disformal transformation within the context of a theory with known field content. In this paper, we remain agnostic about the particular field content and study disformal transformations in a generic fashion. Our general results can be applied to many modified gravity theories to classify them through non-singular disformal transformations, as well as to find new theories by performing singular disformal transformations.

Throughout the paper we use the Lorentzian signature convention $(+,-,-,-)$.


\section{Disformal maps}\label{sec: invertibility}
Indeed, not all expressions of the type \eqref{dis-T-d} produce a viable metric tensor. For generic forms of the coefficients $C$ and $D$, there is no guarantee that $g_{\mu\nu}$ is non-degenerate or that it possesses a proper Lorentzian signature, both of which are fundamental requirements for any physical spacetime metric. Hence it is crucial to understand when these conditions are satisfied. Furthermore, in order to apply a disformal transformation in practice, it is advantageous to have an explicit expression for the inverse metric $g^{\mu\nu}$. Here we present a systematic way to find such an inverse.
	
We consider a straightforward generalization of disformal transformation \eqref{dis-T-d} on a four-dimensional Lorentzian manifold, where we include higher powers of the tensor $t_{\mu\nu}$ as
\begin{equation}
g_{\mu\nu}=Ch_{\mu\nu}+\sum_{I=1}^{3}D_{I} t^{I}_{\mu\nu}\ ,
\qquad
\mbox{standard form} \,,
\label{sec inv: general disformal transformation}
\end{equation}
which we refer to as the {\it standard form}. In the above expression, the powers of $t_{\mu\nu}$ are defined as
\begin{equation}\label{t-HO}
t^{1}_{\mu\nu}=t_{\mu\nu}\ ,\qquad t^{2}_{\mu\nu}=t_{\mu\alpha}h^{\alpha\beta}t_{\beta\nu}\ ,\qquad t^{3}_{\mu\nu}=t_{\mu\sigma}h^{\sigma\rho}t_{\rho\alpha}h^{\alpha\beta}t_{\beta\nu}\ ,
\end{equation}
and so on. Note that transformation \eqref{sec inv: general disformal transformation} represents the most general transformation of $g_{\mu\nu}$ in terms of $h_{\mu\nu}$ and $t_{\mu\nu}$, written in purely tensorial fashion, which does not involve derivatives of these tensors. This can be seen from the fact that we can always use the \emph{Cayley-Hamilton theorem} to express the fourth (or higher) power of $t_{\mu\nu}$ in terms of its lower powers. For a $4\times 4$ matrix the Cayley-Hamilton theorem equips us with an expression of the following type
\begin{align}
&t^{4}_{\mu\nu}=\delta_{0}h_{\mu\nu}+\sum_{I=1}^{3}\delta_{I}t^{I}_{\mu\nu} \,, 
&{\boldsymbol{\rm CH}}_4\,\, \mbox{condition} \,,
\label{sec inv: CH4 theorem}
\end{align}
where ${\boldsymbol{\rm CH}}_4$ indicates that the {\it Cayley-Hamilton degree} is 4, for reasons that will become clear shortly. In the above formula the coefficients can be found to be
\begin{align}
\begin{split}
\delta_{3} &\equiv [t]
\,, \\
\delta_{2} &\equiv \frac{1}{2} \left( [t^2] - [t]^2 \right)
\,, \\
\delta_{1} &\equiv \frac{1}{6} \left(2 [t^3] - 3[t] [t^2] + [t]^3  \right)
\,, \\
\delta_{0} &\equiv \frac{1}{24} \left( 6 [t^4] - 8 [t] [t^3] - 3 [t^2]^2 + 6 [t]^2 [t^2] - [t]^4\right)
= -\det(t^\mu{}_{\nu})
\,,
\end{split}
\label{sec inv: CH4 deltas}
\end{align}
where the square bracket denotes the trace
\begin{equation}\label{t-Trace-def}
[t^{I}]=t^{I}_{\alpha\beta}h^{\alpha\beta}\ .
\end{equation}
We stress that the above relations for $\delta_{I}$ are valid exclusively in the context of \eqref{sec inv: CH4 theorem}. In the rest of the paper the coefficient $\delta_{I}$ can and will take different forms. By contracting \eqref{sec inv: CH4 theorem} with $t^{\mu\nu}$, with indices raised using $h^{\mu\nu}$, we obtain expressions which allow us to reduce $[t^{I}]$ for $I>4$ in terms of its lower-order counterparts. In particular, this yields $[t^{5}]=\delta_{0}[t]+\sum_{I=1}^{3}\delta_{I}[t^{I+1}]$. Hence $[t^{I}]$ is dependent for $I>4$. Under the trace-dependence assumption, the metric-dependent arguments of $C$ and $D_I$ can therefore be restricted to $[t^I]$ with $1\leq I\leq4$. Additional invariants of the underlying fields are not removed by the Cayley-Hamilton theorem and will be included only where stated explicitly.
	
The above reasoning also tells us that we can always look for an inverse metric $g^{\mu\nu}$ in the form
\begin{equation}\label{DT-inverse}
g^{\mu\nu}=\tilde{C}h^{\mu\nu}+\sum_{I=1}^{3}\tilde{D}_{I}t^{I\,\mu\nu}\ ,
\end{equation}
where the indices on the right-hand side have been raised using $h^{\mu\nu}$ and the coefficients $\tilde{C}$ and $\tilde{D_{I}}$ depend on the traces. To find the explicit forms of $\tilde{C}$ and $\tilde{D_{I}}$, we impose the following condition
\begin{equation}
g_{\mu\alpha}g^{\alpha\nu}=\delta^{\nu}_{\mu}\ ,
\label{def-inverse-metric}
\end{equation}
whose left-hand side contains up to a $6$th power of $t_{\mu\nu}$. Using \eqref{sec inv: CH4 theorem} we can reduce these higher powers until there are only up to the $3$rd powers of $t_{\mu\nu}$. By comparing the coefficients of $t^{I}_{\mu\nu}$ on left-hand and right-hand sides, we obtain four linear equations for the four coefficients $\tilde{C}$ and $\tilde{D_{I}}$, with a non-trivial right-hand side. These equations can be solved straightforwardly as long as the associated Jacobian, which depends on $C$ and $D_{I}$, is non-vanishing. This gives us a sufficient and necessary condition for the existence of a unique inverse $g^{\mu\nu}$. The explicit form of the solution is a lengthy expression, so we have moved it to Eq.~\eqref{sol-inverse}. The details of the derivation are presented in \Cref{app-inverse}.
	
\subsection{Simplified Cayley-Hamilton theorem}
The Cayley-Hamilton formula \eqref{sec inv: CH4 theorem} always holds for a $4\times 4$ matrix, however, it is sometimes possible to find a simpler condition which allows us to reduce the power of $t_{\mu\nu}$ even further. Such a condition can then be used instead of \eqref{sec inv: CH4 theorem} in a completely analogous fashion, leading to considerable simplifications in the above procedure for finding an inverse $g^{\mu\nu}$. A prime example of this is the original disformal transformation \cite{Bekenstein:1992pj}, which is given as
\begin{equation}
g_{\mu\nu}=C\,h_{\mu\nu}+D\,t_{\mu\nu}\ ;
\qquad
t_{\mu\nu} = \partial_{\mu}\phi\partial_{\nu}\phi\ .\label{sec simple CH: scalar field ordinary disformal transformation}
\end{equation}
While $\partial_{\mu}\phi\partial_{\nu}\phi$ is indeed a $4\times 4$ matrix, it is straightforward to see that we can find a simple condition to reduce its higher powers. Indeed, it suffices to take a square of $t_{\mu\nu}$ to see that
\begin{align}
t^{2}_{\mu\nu}= \delta_{1}t_{\mu\nu}\,;
\qquad \delta_{1}=h^{\alpha\beta}\partial_{\alpha}\phi\partial_{\beta}\phi \,.
\end{align}
Consequently, it is sufficient to include only the first power of $t_{\mu\nu}$ in the general form of the transformation \eqref{sec inv: general disformal transformation} as well as in the expression for the inverse $g^{\mu\nu}$. A similar simplification can be found for many disformal transformations considered in the literature.
	
In order to understand the above simplification systematically, it is useful to take a moment to recap the Cayley-Hamilton theorem. This is a statement about the characteristic polynomial of a matrix, which, for  $t^{\mu}_{\ \nu}=h^{\mu\alpha}t_{\alpha\nu}$, is given as
\begin{equation}
P(\lambda)\equiv\mathrm{det}\left (t^{\mu}_{\ \nu}-\lambda \delta^{\mu}_{\ \nu}\right)=\prod_{i=1}^{q}(\lambda_{i}-\lambda)^{n_{i}}\ ,\label{sec simple CH: characteristic polynomial}
\end{equation}
where $q\leq4$ and $n_i$, with $\sum_i n_i=4$, is the algebraic multiplicity of the distinct eigenvalue $\lambda_i$. The Cayley-Hamilton theorem states that the characteristic polynomial evaluated at the matrix itself always vanishes
\begin{equation}
P\left(t^{\mu}_{\ \nu}\right ) = 0\ .\label{sec simple CH: CH theorem}
\end{equation}
We can see this directly by bringing the matrix $t^{\mu}_{\ \nu}$ into Jordan normal form (see \eqref{Jordan-decomposition-app})
\begin{equation}
t^{\mu}_{\ \nu}=\begin{pmatrix}
\boldsymbol{J}_{1}&0&\cdots&0\\
0&\boldsymbol{J}_{2}&\ddots&\vdots\\
\vdots&\ddots&\ddots&0\\
0&\cdots&0&\boldsymbol{J}_{p}
\end{pmatrix} \,.
\end{equation}
Let $\boldsymbol J_k$ have eigenvalue $\lambda_{i(k)}$ and size $s_k$. Evaluating the characteristic polynomial blockwise gives
\begin{equation}\label{eq:polynomial-matrix}
P(t^{\mu}_{\ \nu})=\begin{pmatrix}
P(\boldsymbol{J}_{1})&0&\cdots&0\\
0&P(\boldsymbol{J}_{2})&\ddots&\vdots\\
\vdots&\ddots&\ddots&0\\
0&\cdots&0&P(\boldsymbol{J}_{p})
\end{pmatrix} \,.
\end{equation}
For the block $\boldsymbol J_k$,
\begin{equation}
\left(\boldsymbol{J}_k-\lambda_{i(k)}\mathbf{1}\right)^{s_k}=0 \,.
\end{equation}
The corresponding polynomial factor can be isolated exactly:
\begin{equation}
\begin{split}
P(\boldsymbol J_k)
&=(\lambda_{i(k)}\mathbf 1-\boldsymbol J_k)^{n_{i(k)}}
\prod_{j\ne i(k)}(\lambda_j\mathbf 1-\boldsymbol J_k)^{n_j}\\
&=0 \, ,
\end{split}
\end{equation}
because the algebraic multiplicity satisfies $n_{i(k)}\geq s_k$. This proves \eqref{sec simple CH: CH theorem} block by block. If $m_i$ is the size of the largest Jordan block associated with the distinct eigenvalue $\lambda_i$, the minimal polynomial is, up to an irrelevant overall sign,
\begin{equation}
Q(\lambda)\equiv\prod_{i}(\lambda_{i}-\lambda)^{m_{i}}\ .\label{sec simple CH: simple polynomial}
\end{equation}
The same blockwise argument gives
\begin{equation}
Q\left(t^{\mu}_{\ \nu}\right)=0\ .
\end{equation}

In four dimensions, the degree of the polynomial $Q$ can range from $1$ to $4$. We will refer to this degree as the Cayley-Hamilton degree of $t_{\mu\nu}$ (or of the transformation) and denote it as ${\boldsymbol{\rm CH}}_1$ to ${\boldsymbol{\rm CH}}_4$. The ${\boldsymbol{\rm CH}}_4$ transformations satisfy the full formula \eqref{sec inv: CH4 theorem} and no simpler one can be found, while the other degrees satisfy the following respectively:
\begin{align}\label{sec simple CH: CH3 condition}
&t^{3}_{\mu\nu}=\delta_{0}h_{\mu\nu}+\delta_{1}t_{\mu\nu}+\delta_{2}t^{2}_{\mu\nu}\ ,
&{\boldsymbol{\rm CH}}_3\,\, \mbox{condition} \,,
\\
\label{sec simple CH: CH2 condition}
&t^{2}_{\mu\nu}=\delta_{0}h_{\mu\nu}+\delta_{1}t_{\mu\nu}\ ,
&{\boldsymbol{\rm CH}}_2\,\, \mbox{condition} \,,
\\
\label{sec simple CH: CH1 condition}
&t_{\mu\nu}=\delta_{0}h_{\mu\nu}\ ,
&{\boldsymbol{\rm CH}}_1\,\, \mbox{condition} \,.
\end{align}
Clearly, for ${\boldsymbol{\rm CH}}_1$ the disformal transformation automatically reduces to a purely conformal one. Note also that depending on the level of the simplification we can cut off the unnecessary powers of $t_{\mu\nu}$ in the transformation law \eqref{sec inv: general disformal transformation} as well as in the corresponding expression for the inverse metric. Comparing \eqref{sec simple CH: scalar field ordinary disformal transformation} with \eqref{sec simple CH: CH2 condition}, we see that the scalar field example is ${\boldsymbol{\rm CH}}_2$. We further explore the scalar field case in section \ref{sec-examples}.
	
\subsection{Normal form}
	
Although the disformal transformation \eqref{sec inv: general disformal transformation} includes all possible combinations constructed out of $h_{\mu\nu}$ and $t_{\mu\nu}$, in practice, this is not the form that one usually deals with. For example, let us consider the disformal transformation of the form
\begin{align}\label{DT-SV}
g_{\mu\nu}=Ch_{\mu\nu}+D_\phi \partial_\mu\phi\partial_\nu\phi+D_A A_\mu A_\nu+2 D_{\phi{A}} \partial_{(\mu}\phi {A}_{\nu)} \,,
\end{align}
where we temporarily allow the coefficients to depend on three scalar quantities of the underlying fields, rather than only on the traces of a single tensor
\begin{align}\label{invariants-SV}
Y_\phi \equiv h^{\alpha\beta}\partial_{\alpha}\phi\partial_{\beta}\phi \,,
\qquad
Y_A \equiv h^{\alpha\beta}A_{\alpha}{A}_{\beta} \,,
\qquad
Y_{\phi A} \equiv h^{\alpha\beta}\partial_{\alpha}\phi{A}_\beta\,.
\end{align}
Clearly \eqref{DT-SV} is not written in the desired standard form \eqref{sec inv: general disformal transformation}. So, how can we find the Cayley-Hamilton degree of \eqref{DT-SV}? Assuming $D_\phi\neq0$, a straightforward way is to rewrite the above transformation as follows
\begin{align}\label{DT-simplified-SV}
g_{\mu\nu} = C h_{\mu\nu} + D_{\phi} t_{\mu\nu} \,,
\end{align}
with the following identifications
\begin{align}
D_1=D_{\phi} \,,
\qquad
D_2=0 \,,
\qquad
D_3 = 0 \,,
\end{align}
and
\begin{align}\label{t-SV}
t_{\mu\nu} = \partial_\mu\phi\partial_\nu\phi + d_A A_\mu A_\nu + 2 d_{\phi{A}} \partial_{(\mu}\phi {A}_{\nu)} \,;
\qquad
d_A \equiv \frac{D_A}{D_\phi} \,,
\quad
d_{\phi{A}} \equiv \frac{D_{\phi{A}}}{D_\phi} \,.
\end{align}
We emphasize that although \eqref{DT-simplified-SV} might look to have the ${\boldsymbol{\rm CH}}_2$ form shown in \eqref{sec simple CH: CH2 condition}, this is not the case. This can be easily seen by computing the square of \eqref{t-SV}
\begin{align}
\begin{split}
t^2_{\mu\nu} &= \left(
Y_\phi+2 Y_{\phi A} d_{\phi A}+Y_A d_{\phi A}^2
\right) \partial_\mu\phi \partial_\nu\phi
+ \left(
Y_\phi d_{\phi A}^2+2 Y_{\phi A} d_A d_{\phi A}+Y_A d_A^2
\right)
A_{\mu} A_{\nu}
\\
&\quad + 2
\left[
Y_\phi d_{\phi A}+Y_{\phi A}\left(d_A+d_{\phi A}^2\right)+Y_A d_A d_{\phi A}
\right]
\partial_{(\mu}\phi {A}_{\nu)} \,,
\end{split}
\end{align}
which clearly shows that $t^2_{\mu\nu}$ is not proportional to $t_{\mu\nu}$. On the other hand, \eqref{DT-SV} always satisfies the degree-at-most-three identity \eqref{sec simple CH: CH3 condition} with
\begin{align}
\delta_0 = 0 \,,
\qquad
\delta_1 = -\left(d_A-d_{\phi {A}}^2\right) \left(Y_\phi Y_A-Y_{\phi A}^2\right) \,,
\qquad
\delta_2 = Y_\phi + d_A Y_A +2 Y_{\phi A} d_{\phi{A}} \,.
\end{align}
The Cayley-Hamilton degree is three ${\boldsymbol{\rm CH}}_3$ on generic patches where this cubic is the minimal polynomial. At exceptional values, the minimal polynomial can have lower degree.

Note that coefficients $C$ and $D_\phi$ in \eqref{DT-simplified-SV} are functions of $Y_\phi, Y_A$, and $Y_{\phi A}$, while the corresponding traces read
\begin{align}
\begin{split}
[t] &= Y_\phi + d_A Y_A +2 Y_{\phi A} d_{\phi{A}} \,,
\\
[t^2] &= Y_\phi ^2 +4 Y_\phi  Y_{\phi A} d_{\phi{A}} +2 Y_\phi  Y_A d_{\phi{A}}^2
+2 Y_{\phi A}^2 \left(d_A+d_{\phi{A}}^2\right)
+4 Y_A Y_{\phi A} d_A d_{\phi{A}}+Y_A^2 d_A^2 \,,
\\
[t^3] &= [t] \left(
Y_\phi ^2 +4 Y_\phi Y_{\phi A} d_{\phi {A}}+\left(3 d_{\phi {A}}^2-d_A\right) Y_\phi Y_A
+\left(3 d_A+d_{\phi {A}}^2\right) Y_{\phi A}^2
+4 Y_A Y_{\phi A} d_A d_{\phi {A}}+Y_A^2 d_A^2
\right) \,,
\end{split}
\end{align}
Here $t^{\mu}{}_{\nu}$ has rank at most two, so that its traces are not all independent: they identically satisfy $[t^{3}]=\frac{1}{2}[t]\big(3[t^{2}]-[t]^{2}\big)$, and at most $[t]$ and $[t^{2}]$ are independent. One can work with $Y_\phi, Y_A$, and $Y_{\phi A}$, or replace them by $[t]$, $[t^{2}]$, and one of these three quantities on patches where the corresponding change of scalar variables is locally invertible. In either case the reduction to the normal form is unaffected: the coefficients $\delta_I$ of the Cayley-Hamilton relation \eqref{sec simple CH: CH3 condition}, which are all that the inversion procedure requires, are already fixed by the two independent traces as $\delta_0=0$, $\delta_1=\frac{1}{2}\left([t^{2}]-[t]^{2}\right)$, and $\delta_2=[t]$.
	
A similar reduction is possible for the general form \eqref{sec inv: general disformal transformation}. If at least one $D_I$ is nonzero, choose such a coefficient as an overall disformal factor. For example, on a patch with $D_1\neq0$, defining
\begin{align}
d_{\mu\nu}=t_{\mu\nu}+\frac{D_2}{D_1}t^2_{\mu\nu}+\frac{D_3}{D_1}t^3_{\mu\nu},
\qquad D=D_1 \, ,
\end{align}
we find the simple form $g_{\mu\nu}=C h_{\mu\nu}+D d_{\mu\nu}$. The coefficients $C$ and $D$ still depend on the original scalar invariants $[t^I]$, which can be expressed in terms of $[d^I]$ only when the corresponding change of variables is locally invertible. Relabeling $d_{\mu\nu}$ as $t_{\mu\nu}$ gives the form obtained by setting $D_1=D$ and $D_2=0=D_3$ in \eqref{sec inv: general disformal transformation},
\begin{align}\label{DT-simplified}
g_{\mu\nu} = C h_{\mu\nu} + D t_{\mu\nu} \,,
\qquad
\mbox{normal form} \,,
\end{align}
which has the same form as \eqref{dis-T-d}. We call the above form the {\it normal form} in comparison with the standard form defined in Eq.~\eqref{sec inv: general disformal transformation}. We emphasize that the fact that we can always bring a disformal transformation into the normal form \eqref{DT-simplified} does not mean that the effect of the higher powers is completely redundant: even when the transformation contains only $t_{\mu\nu}$, the higher powers \eqref{t-HO} generally appear in the corresponding inverse contravariant metric. In the general analysis below we retain the trace-dependence assumption unless additional scalar arguments are explicitly introduced.
	
For practical purposes it is always easier to work with the normal form \eqref{DT-simplified} rather than the standard form \eqref{sec inv: general disformal transformation}.

	
\section{Classification}\label{sec-classification}
	
\subsection{Hawking-Ellis types}
In this section, we will show that the various Cayley-Hamilton degrees (${\boldsymbol{\rm CH}}_1$ to ${\boldsymbol{\rm CH}}_4$) are closely connected to the Hawking-Ellis classification (or Segre-Pleba\'{n}ski classification) of the tensor $t_{\mu\nu}$. While the Hawking-Ellis classification is usually utilized in the context of the energy-momentum tensor, it is in fact suitable for studying properties of any symmetric rank-$2$ tensor in a given geometry. Hence, it is not surprising that we can determine many properties of the transformation \eqref{sec inv: general disformal transformation} based on this classification.
	
According to the Hawking-Ellis classification, there exist only four canonical types of symmetric rank-$2$ tensors in a given four-dimensional Lorentzian geometry. More specifically, for any $t_{\mu\nu}$ there exists an orthonormal basis in which the tensor $t_{\mu\nu}$ takes one of the following forms\footnote{We have included an explanation of how one can arrive at this conclusion in appendices \ref{app:SC} and \ref{app:HEC}.}
\begin{equation}\label{sec class: HE canonical forms}
	\setlength{\arraycolsep}{4pt}
	\begin{gathered}
		\begin{array}{c}
			\left(\begin{array}{cccc}
				\lambda_1&0&0&0\\
				0&-\lambda_2&0&0\\
				0&0&-\lambda_3&0\\
				0&0&0&-\lambda_4
			\end{array}\right)\\[2mm]
			\mbox{Type I}
		\end{array}
		\quad
		\begin{array}{c}
			\left(\begin{array}{cccc}
				\lambda_1\pm f&\pm f&0&0\\
				\pm f&-\lambda_1\pm f&0&0\\
				0&0&-\lambda_2&0\\
				0&0&0&-\lambda_3
			\end{array}\right)\\[2mm]
			\mbox{Type II}
		\end{array}
		\quad
		\begin{array}{c}
			\left(\begin{array}{cccc}
				\lambda_1&f&0&0\\
				f&-\lambda_1&f&0\\
				0&f&-\lambda_1&0\\
				0&0&0&-\lambda_2
			\end{array}\right)\\[2mm]
			\mbox{Type III}
		\end{array}
		\\[5mm]
		\begin{array}{c}
			\left(\begin{array}{cccc}
				\mathrm{Re}[z]&-\mathrm{Im}[z]&0&0\\
				-\mathrm{Im}[z]&-\mathrm{Re}[z]&0&0\\
				0&0&-\lambda_1&0\\
				0&0&0&-\lambda_2
			\end{array}\right)\\[2mm]
			\mbox{Type IV}
		\end{array}
	\end{gathered}
\end{equation}
Note that the eigenvalues of the associated mixed tensor $t^a{}_{b}$ are $\lambda_i$, which are real, while $z$ is non-real.\footnote{We take non-real numbers to be complex numbers with a non-vanishing imaginary part.} $\lambda_{i}$ and $z$ are invariants of the tensor $t_{\mu\nu}$, while the parameter $f$ reflects a residual Lorentz freedom. Some authors fix this freedom by setting $f=1$ \cite{Hawking:1973uf}, while others keep it explicit \cite{Martin-Moruno:2017iqw, Martin-Moruno:2018eil, Martin-Moruno:2018coa, Martin-Moruno:2019kzc}. It is important to stress that $f>0$ and the plus and minus signs in Type II represent two distinct possibilities, which cannot be related by a Lorentz transformation. It is a less-known fact that these canonical types directly correspond to the possible Jordan normal forms of the associated matrix $t^{\mu}_{\ \nu}$. As we have shown in appendices \ref{app:SC} and \ref{app:HEC}, the Hawking-Ellis types \eqref{sec class: HE canonical forms} correspond to the following Jordan normal forms
\begin{equation}\label{sec class: Jordan forms}
\begin{gathered}
		\begin{array}{c}
			\left(\begin{array}{cccc}
				\lambda_1&0&0&0\\
				0&\lambda_2&0&0\\
				0&0&\lambda_3&0\\
				0&0&0&\lambda_4
			\end{array}\right)\\[2mm]
			\mbox{Type I;}\ [1,111]
		\end{array}
		\quad
		\begin{array}{c}
			\left(\begin{array}{cccc}
				\lambda_1&1&0&0\\
				0&\lambda_1&0&0\\
				0&0&\lambda_2&0\\
				0&0&0&\lambda_3
			\end{array}\right)\\[2mm]
			\mbox{Type II;}\ [211]
		\end{array}
		\\[5mm]
		\begin{array}{c}
			\left(\begin{array}{cccc}
				\lambda_1&1&0&0\\
				0&\lambda_1&1&0\\
				0&0&\lambda_1&0\\
				0&0&0&\lambda_2
			\end{array}\right)\\[2mm]
			\mbox{Type III;}\ [31]
		\end{array}
		\quad
		\begin{array}{c}
			\left(\begin{array}{cccc}
				z&0&0&0\\
				0&\bar z&0&0\\
				0&0&\lambda_1&0\\
				0&0&0&\lambda_2
			\end{array}\right)\\[2mm]
			\mbox{Type IV;}\ [z\bar z11]
		\end{array}
\end{gathered}
\end{equation}
It is important to note that this classification scheme does not capture the properties of $t_{\mu\nu}$ alone. Rather, it characterizes the properties of $t_{\mu\nu}$ with respect to a given metric, in this case, $h_{\mu\nu}$, as $t^\mu{}_\nu = h^{\mu\alpha} t_{\alpha\nu}$. For example, the existence of Types II-IV is only possible because $h_{\mu\nu}$ has a Lorentzian signature. With a Euclidean signature metric, these types are excluded due to the spectral theorem.
	
The Hawking-Ellis types can be identified by the causal character and Jordan structure of their eigenvectors (see appendix~\ref{app:HEC}). Type I has a complete real eigenbasis with one timelike and three spacelike directions. Type II has a size-two real Jordan block with a null eigenvector and two additional spacelike eigenvectors. Type III has a size-three real Jordan block with a null eigenvector and one additional spacelike eigenvector. Type IV has one complex-conjugate eigenvalue pair with complex null eigenvectors and two real spacelike eigenvectors. These tensors can be further classified based on the degeneracy of their eigenvalues. This more detailed classification is usually referred to as the Segre-Pleba\'{n}ski classification. Here we also adopt the Segre bracket notation, which allows us to characterize both the Hawking-Ellis types and the degeneracy in their eigenvalues. The notation is formed as follows: To a matrix we can associate a set of numbers describing the sizes of its associated Jordan blocks. Along with this notation we can include a round bracket enclosing some of these numbers. This denotes that the associated Jordan blocks share the same eigenvalue. Finally, for non-real eigenvalues, we write $z$ instead of $1$ and the complex conjugate $\bar{z}$ as in Type IV in \eqref{sec class: Jordan forms}. This does not produce any conflicts since the non-real eigenvalues cannot be degenerate, nor can they be associated with a non-trivial Jordan block. For example
\begin{equation}
\begin{pmatrix}
\lambda_1 & 1 & 0 & 0 \\
0 & \lambda_1 & 0 & 0 \\
0 & 0 & \lambda_1 & 0 \\
0 & 0 & 0 & \lambda_2 \\
\end{pmatrix}\ \qquad\mathrm{would\ be\ denoted\ as}\qquad [(21)1]\ .
\end{equation}
It should be fairly clear that the above notation contains all the information needed to determine the Cayley-Hamilton degree of $t_{\mu\nu}$. Since there are only a limited number of possibilities, we have summarized them in \Cref{sec class: table of CH degrees}. We can see that Type II and Type III tensors cannot be ${\boldsymbol{\rm CH}}_1$ or, in the latter case, ${\boldsymbol{\rm CH}}_2$ due to their non-trivial Jordan blocks. Similarly, Type IV has two non-real eigenvalues $z\neq \bar{z}$ which cannot be degenerate. Hence the degree is at least ${\boldsymbol{\rm CH}}_3$.
	
\begin{table}[!htb]
\centering
\renewcommand{\arraystretch}{1.5}
\caption*{\bf Hawking-Ellis classification of disformal transformations}
\begin{tabular}{|c|c|c|c|c|c|c|}
			\hline
			\multicolumn{1}{|c|}{Hawking-Ellis}
			&\multirow{2}{*}{Block structure}
			&\multirow{2}{*}{Segre bracket}
			&\multicolumn{4}{|c|}{Cayley-Hamilton degree}
			\\
			\cline{4-7}
			type/class & &
			& ${\boldsymbol{\rm CH}}_1$
			& ${\boldsymbol{\rm CH}}_2$
			& ${\boldsymbol{\rm CH}}_3$
			& ${\boldsymbol{\rm CH}}_4$
			\\ \hline
			\multirow{7}{*}{I}
			& \multirow{7}{*}{$[1,111]$}
			& $[1,111]$
			& & & & $\checkmark$
			\\ \cline{3-7}
			& & $[(1,1)11]$ & & & $\checkmark$ &
			\\ \cline{3-7}
			& & $[1,1(11)]$ & & & $\checkmark$ &
			\\ \cline{3-7}
			& & $[(1,1)(11)]$ & & $\checkmark$ & &
			\\ \cline{3-7}
			& & $[(1,11)1]$ & & $\checkmark$ & &
			\\ \cline{3-7}
			& & $[1,(111)]$ & & $\checkmark$ & &
			\\ \cline{3-7}
			& & $[(1,111)]$ & $\checkmark$ & & & \\ \hline
			\multirow{4}{*}{II}
			& \multirow{4}{*}{$[211]$}
			& $[211]$
			& & & & $\checkmark$
			\\ \cline{3-7}
			& & $[(21)1]$ & & & $\checkmark$ &
			\\ \cline{3-7}
			& & $[2(11)]$ & & & $\checkmark$ &
			\\ \cline{3-7}
			& & $[(211)]$ & & $\checkmark$ & & \\ \hline
			\multirow{2}{*}{III}
			& \multirow{2}{*}{$[31]$}
			& $[31]$
			& & & & $\checkmark$
			\\ \cline{3-7}
			& & $[(31)]$ & & & $\checkmark$ &
			\\
			\hline
			\multirow{2}{*}{IV}
			& \multirow{2}{*}{$[z{\bar z}11]$}
			& $[z{\bar z}11]$
			& & & & $\checkmark$
			\\ \cline{3-7}
			& & $[z{\bar z}(11)]$ & & & $\checkmark$ &
			\\
			\hline
\end{tabular}
\vspace{.5cm}
\caption{Hawking-Ellis classification and possible Cayley-Hamilton degrees.}
\label{sec class: table of CH degrees}
\end{table}
	
\subsection{Cayley-Hamilton degree}

At this point it is worth briefly revisiting the formulas \eqref{sec simple CH: CH3 condition}, \eqref{sec simple CH: CH2 condition} and \eqref{sec simple CH: CH1 condition}. As we have mentioned, the coefficients $\delta_{I}$ featured in these equations cannot be determined universally; however, it turns out they are uniquely fixed for each Segre-Pleba\'{n}ski class. Let us illustrate this on the ${\boldsymbol{\rm CH}}_2$ tensors. In this class there are at most two distinct eigenvalues and the polynomial \eqref{sec simple CH: simple polynomial} takes the form
\begin{equation}
Q(\lambda)=(\lambda_{1}-\lambda)(\lambda_{2}-\lambda)\ .
\end{equation}
Hence $\delta_{0,1}$ can be written in terms of eigenvalues as
\begin{equation}\label{notes: delta_0 delta_1}
\delta_{0}=-\lambda_{1}\lambda_{2} \,,
\qquad
\delta_{1}=\lambda_{1}+\lambda_{2}\ .
\end{equation}
As it can be seen from \Cref{sec class: table of CH degrees}, there are only four possible tensors that are ${\boldsymbol{\rm CH}}_2$: $[(1,1)(11)]$, $[(211)]$, $[(1,11)1]$, and $[1,(111)]$. Note that for our current purpose the last two are virtually identical. We will show on a case by case basis the expressions for $\delta_{0,1}$ in terms of $[t]$ and $[t^{2}]$. For $[(1,1)(11)]$ we find the following traces
\begin{align}\label{CH2-t-t2}
\begin{split}
[t]=2\left(\lambda_{1}+\lambda_{2}\right)\ ,
\qquad
[t^{2}]=2\left(\lambda_{1}^{2}+\lambda_{2}^{2}\right)\ .
\end{split}
\end{align}
These relations can be inverted to find the eigenvalues in terms of $[t]$ and $[t^{2}]$, which can be then plugged into the relations for $\delta_{I}$ to obtain
\begin{align}
\begin{split}
\delta_{0}=\frac{2[t^{2}]-[t]^{2}}{8},
\qquad
\delta_{1}=\frac{1}{2}[t]\ .
\end{split}
\end{align}
The calculation for the other classes is exactly the same; however, it yields different results. For $[(211)]$ there is only a single eigenvalue as $\lambda_{1}=\lambda_{2}$, hence the expressions simplify even more
\begin{align}
\begin{split}
\delta_{0}=-\frac{[t]^{2}}{16}\ ,
\qquad
\delta_{1}=\frac{1}{2}[t]\ .
\end{split}
\end{align}
Finally, for $[(1,11)1]$ and $[1,(111)]$ we find a more involved answer
\begin{align}
\begin{split}
\delta_{1}=\frac{[t]}{2}\mp\frac{1}{6}\sqrt{12[t^{2}]-3[t]^{2}}\ ,
\qquad
\delta_{0}=\frac{[t^{2}]}{4}-\frac{[t]^{2}}{8}\pm\frac{[t]}{24}\sqrt{12[t^{2}]-3[t]^{2}}\ .
\end{split}
\end{align}
An analogous strategy can be used for ${\boldsymbol{\rm CH}}_3$ tensors, where there are at most three different eigenvalues. Consequently, this requires the dependence of $\delta_{I}$ on $[t^{3}]$. Unfortunately, the calculations, in some cases, get quite complicated, hence we will not present the results here. For ${\boldsymbol{\rm CH}}_1$, the result is trivial with $\delta_{0}=[t]/4$.
	
\subsection{Composition of disformal maps}\label{sec:inverse-map}
Before closing this section, let us study the closure of the disformal transformations \eqref{sec inv: general disformal transformation} under the functional composition. Indeed, it is not clear whether this is always the case. This is because the tensor $t_{\mu\nu}$ could depend on $h_{\mu\nu}$. In this subsection we retain the trace-dependence assumption and take $t_{\mu\nu}$ to be completely independent of $h_{\mu\nu}$ such that
\begin{equation}\label{condition t-h-derivative}
\frac{\delta t_{\mu\nu}}{\delta h_{\sigma\rho}}=0\ .
\end{equation}
It is then straightforward to see that a sequence of disformal transformations results in another disformal transformation. We will demonstrate this for ${\boldsymbol{\rm CH}}_2$ transformations as this case already captures the novel features without introducing needless complexity. For the cases of ${\boldsymbol{\rm CH}}_3$ and ${\boldsymbol{\rm CH}}_4$ transformations, the steps are completely analogous.

For a disformal transformation with ${\boldsymbol{\rm CH}}_2$ Cayley-Hamilton degree, we consider transformation
\begin{equation}
g_{\mu\nu}=C\,f_{\mu\nu}+D\,t_{\mu\nu}\ ,\label{sec composition: disformal transformation 2}
\end{equation}
as well as
\begin{equation}
f_{\mu\nu}=c\,h_{\mu\nu}+d\,t_{\mu\nu}\ ,\label{sec composition: disformal transformation 1}
\end{equation}
where $C,\,D$ depend on the traces of $t_{\mu\nu}$ and $t^{2}_{\mu\nu}$ evaluated with respect to the metric $f_{\mu\nu}$ while for $c,\,d$ the traces are evaluated with respect to metric $h_{\mu\nu}$. Plugging \eqref{sec composition: disformal transformation 1} into \eqref{sec composition: disformal transformation 2}, we get the composed transformation
\begin{equation}
g_{\mu\nu}=C\,c\,h_{\mu\nu}+\left (C\,d+D\right )t_{\mu\nu}\ ,\label{sec composition: disformal transformation 1+2}
\end{equation}
which has the correct form but coefficients $C$ and $D$ still depend on $f_{\mu\nu}$ and not $h_{\mu\nu}$. These can, however, be easily re-expressed in terms of $[t]$ and $[t^{2}]$. To do so, we use the results of subsection \ref{subsec inv CH2} to find the inverse $f^{\mu\nu}$ as
\begin{equation}
f^{\mu\nu}=\tilde{c}\,h^{\mu\nu}+\tilde{d}\,t^{\mu\nu} \,,
\end{equation}
where
\begin{align}
\tilde{c} = \frac{c+d \delta _1}{c^2+c d \delta _1-d^2 \delta _0} \,,
\qquad
\tilde{d} = -\frac{d}{c^2+c d \delta _1-d^2 \delta _0} \,.
\end{align}
Note that $\tilde{c}$ and $\tilde{d}$ can be completely expressed in terms of $[t]$ and $[t^2]$. Contracting this with $t_{\mu\nu}$ we find
\begin{equation}
[t]_{f}\equiv t_{\alpha\beta}f^{\alpha\beta}=\tilde{c}\,[t]+\tilde{d}\,[t^{2}]\ .\label{sec composition: trace linear}
\end{equation}
In the same manner we can relate the traces of $t^{2}_{\mu\nu}$
\begin{equation}
[t^{2}]_{f}\equiv t_{\alpha\beta}t_{\sigma\rho}f^{\alpha\sigma}f^{\beta\rho}=\tilde{c}^{2}\,[t^{2}]+2\tilde{c}\tilde{d}\,[t^{3}]+\tilde{d}^{2}\,[t^{4}]\ .
\end{equation}
The last two terms can be simplified using \eqref{sec simple CH: CH2 condition}
\begin{equation}
[t^{2}]_{f}=\left(\tilde{d}^{2}\delta_{0}\delta_{1}+2\tilde{c}\tilde{d}\delta_{0}\right)[t]+\left(\tilde{d}^{2}(\delta_{0}+\delta_{1}^{2})+2\tilde{c}\tilde{d}\delta_{1}+\tilde{c}^{2}\right)[t^{2}]\ .\label{sec composition: trace quadratic}
\end{equation}
Plugging these into \eqref{sec composition: disformal transformation 1+2} yields a ${\boldsymbol{\rm CH}}_2$
disformal transformation of the form of \eqref{sec inv: general disformal transformation}. Hence the composition of two disformal maps results in another disformal map. Interestingly, since the resulting transformation again features the pair $h_{\mu\nu}$ and $t_{\mu\nu}$, the Hawking-Ellis/Segre-Pleba\'{n}ski classification of the disformal transformations remains intact under the action of composition.

We can use the result \eqref{sec composition: disformal transformation 1+2} to see under which conditions the disformal transformation is invertible. Indeed, the transformation \eqref{sec composition: disformal transformation 2} is the inverse of \eqref{sec composition: disformal transformation 1} when the composition \eqref{sec composition: disformal transformation 1+2} gives $g_{\mu\nu}=h_{\mu\nu}$. This clearly occurs when
\begin{align}
Cc=1 \, ,
\qquad
Cd+D=0 \,.
\end{align}
For $c\neq 0$, the above non-homogeneous linear equations have a unique solution for $C$ and $D$
\begin{align}
C=\frac{1}{c} \,,
\qquad
D=-\frac{d}{c} \,.
\label{sec composition: functional inverse coef solution}
\end{align}
The coefficient solution \eqref{sec composition: functional inverse coef solution} is only one part of functional invertibility. Three conditions are required on the patch under consideration:
\begin{enumerate}
\item $c^2+cd\delta_1-d^2\delta_0\neq0$, so that $f_{\mu\nu}$ is pointwise non-degenerate and the trace relations are defined;
\item $c\neq0$, so that the algebraic composition equations can be solved for the inverse coefficients in the same normal form;
\item the map between the independent trace variables is locally invertible, $\det\frac{\partial([t]_f,[t^2]_f)}{\partial([t],[t^2])}\neq0$.
If the traces are not independent, the invertibility condition must be imposed on the independent invariants of the corresponding subclass.
\end{enumerate}
At fixed $c$ and $d$, the linear part of the trace map has determinant
$\left(\tilde c+\delta_1\tilde d\right) \left(\tilde c^2+\delta_1\tilde c\tilde d-\delta_0\tilde d^2\right) =c/\left(c^2+cd\delta_1-d^2\delta_0\right)^2$. When $c$ and $d$ depend on the traces, the full Jacobian also contains their derivatives. Failure of this full trace-map condition is the usual source of functional non-invertibility in mimetic examples. For example the mimetic transformations studied in \cite{Chamseddine:2013kea, Chamseddine:2014vna, Zumalacarregui:2013pma, Deruelle:2014zza} fail invertibility precisely through this trace map.

The above strategy can be employed to find the criterion on invertibility of ${\boldsymbol{\rm CH}}_3$ and ${\boldsymbol{\rm CH}}_4$ transformations. However, while we have laid out very specific steps to find this criterion, the necessary calculations become tedious very quickly so we do not present them here.
	
\subsection{Preservation of Lorentzian signature}\label{sec:signature}
Contrary to conformal transformations, which preserve the Lorentzian signature, disformal transformations do not, in general, preserve it. To have a consistent map, it is therefore important to determine under which conditions the Lorentzian signature is preserved under disformal transformations.
In order to do so, we work with the normal form of the disformal transformation \eqref{DT-simplified}. Working with the tetrad basis
\begin{align}\label{h_ab-def}
h_{\mu\nu}=\eta_{ab}e^{a}{}_{\mu}e^{b}{}_{\nu} \,,
\end{align}
the Lorentzian signature for $h_{\mu\nu}$ is automatically imposed through the local Minkowski metric $\eta_{ab}$ as long as the tetrad basis is regular. This is the elegance of tetrad formalism. Now, instead of directly solving the eigenvalue problem for $g_{\mu\nu}$ defined in Eq.~\eqref{DT-simplified} to impose the conditions to have a Lorentzian signature, we can look for conditions under which,
\begin{align}\label{g_ab-def}
g_{ab} = C \eta_{ab} + D t_{ab} \,,
\qquad
g_{\mu\nu}&=g_{ab}e^{a}{}_{\mu}e^{b}{}_{\nu} \,,
\end{align}
with $t_{ab}=t_{\mu\nu}e^\mu{}_a e^\nu{}_b$, preserves the Lorentzian signature. Note that, unlike \eqref{h_ab-def}, where $\eta_{ab}$ guarantees a Lorentzian signature for $h_{\mu\nu}$, the combination $C\eta_{ab} + D t_{ab}$ in \eqref{g_ab-def} does not necessarily guarantee a Lorentzian signature for $g_{\mu\nu}$: whether the Lorentzian signature is preserved or not completely depends on the signs of $C$ and $D$, and also on the eigenvalue structure of $t_{ab}$. As mentioned before, the eigenvalue structure of $t_{ab}$ determines the Hawking-Ellis type and, therefore, we consider different types of disformal transformations case by case.
A powerful invariant is obtained directly from the mixed matrix form,
\begin{align}\label{det-normal-form}
\frac{\det g}{\det h}
= \det\left(C\delta^\mu{}_{\nu}+D t^\mu{}_{\nu}\right)
= \prod_i \left(C+D\lambda_i\right)^{n_i} \, ,
\end{align}
where the product is over the distinct eigenvalues of $t^\mu{}_{\nu}$ with algebraic multiplicities $n_i$. For Type IV the complex pair contributes the real positive factor
\begin{align}\label{det-normal-form-typeIV}
\left(C+D z\right)\left(C+D\bar z\right)
= \left(C+D\mathrm{Re}[z]\right)^2+D^2\mathrm{Im}[z]^2 \, .
\end{align}
In four dimensions, the sign of \eqref{det-normal-form} completely settles whether $g_{\mu\nu}$ is a Lorentzian metric: since $\det h<0$, the condition
\begin{align}\label{det-condition}
\frac{\det g}{\det h}>0
\end{align}
is equivalent to $\det g<0$, which holds if and only if $g_{\mu\nu}$ has one positive and three negative directions, i.e.\ the signature $(+,-,-,-)$, or the reversed one, $(-,+,+,+)$, whereas the non-Lorentzian signatures $(+,+,+,+)$, $(-,-,-,-)$ and $(+,+,-,-)$ all yield $\det g>0$.\footnote{This equivalence between the sign of the determinant and the Lorentzian character of the metric is special to four dimensions. For instance, in six dimensions the non-Lorentzian signature $(+,+,+,-,-,-)$ also yields a negative determinant.} Condition \eqref{det-condition} is therefore necessary and sufficient
for $g_{\mu\nu}$ to be Lorentzian, up to an overall sign, and hence to
possess a metric null cone. Note that \eqref{det-condition} is insensitive in two ways: to the overall orientation, since $g_{\mu\nu}$ and $-g_{\mu\nu}$ define the same cones, and to the causal structure of $h_{\mu\nu}$, which enters \eqref{det-normal-form} only through the Jordan normal form. Both insensitivities are appropriate as long as the map is invertible and non-singular. A worked-out example of this situation is the acoustic metric $Z^{\mu\nu}$ governing the propagation of scalar perturbations in scalar-tensor theories \cite{Sawicki:2024ryt}: there the analogue of \eqref{det-condition}, $\det Z^{\mu}{}_{\nu}>0$, is the full hyperbolicity requirement, while the reversed orientation is not excluded but rather provides the invariant definition of a ghost, and the relative configurations of the acoustic and light cones are classified rather than constrained. The same field-redefinition logic applies to the transformation \eqref{g_ab-def} provided that the complete action, including the matter couplings and conventions, is transformed consistently. An active metric reversal $g_{\mu\nu}\to-g_{\mu\nu}$ does not, in general, multiply the total action by a common overall sign.

It is worth emphasizing the logic behind the determinant criterion \eqref{det-condition}. In our setup there are not really two independent metrics: the disformal map \eqref{g_ab-def} is an invertible field redefinition relating two descriptions of one and the same physical system. Whenever the map is invertible and non-singular, physical predictions do not depend on whether they are computed in the $h$-frame or in the $g$-frame, and the only intrinsic requirement on $g_{\mu\nu}$ is that it be a Lorentzian metric, that is, that \eqref{det-condition} hold. A degree of freedom that is healthy with respect to $h_{\mu\nu}$ therefore remains healthy after the map, even when its description in the $g$-frame looks unusual, for instance when the $g$-cone is strongly tilted with respect to the $h$-cone, or when a direction that is timelike for one metric is spacelike for the other. Such features are frame-dependent and carry no physical content, so that \eqref{det-condition} is all that our framework requires.

One may nonetheless ask a finer, frame-dependent question: under what conditions do $h_{\mu\nu}$ and $g_{\mu\nu}$ share a common time direction and a common family of spacelike hypersurfaces, so that the two frames admit compatible local $3+1$ decompositions? This refinement is not needed for the consistency of the map, and imposing it goes beyond the field-redefinition logic used above; it becomes relevant only if one chooses to treat both frames as simultaneously physical. For completeness, we carry out this analysis, type by type, in appendix~\ref{app:causal}, where we find that the requirement of a common causal structure sharpens \eqref{det-condition} into the single family of inequalities $C+D\lambda_i>0$, one for each real eigenvalue of $t^{\mu}{}_{\nu}$.


\section{Dynamics of singular maps: Mimetic scenarios}\label{sec: singular transformations}

Performing {\it singular disformal transformations} on known gravity theories may lead to new modified gravity theories \cite{Chamseddine:2013kea,Deruelle:2014zza,Gorji:2018okn,Firouzjahi:2018xob,Jirousek:2018ago,Gorji:2020ten,Jirousek:2022jhh,Jirousek:2022rym,Gorji:2025ajb}. In particular, they can provide new dynamical degrees of freedom. The effective energy-momentum tensor $T^\star_{\mu\nu}$ that characterizes these new degrees of freedom is what we refer to as the mimetic energy-momentum tensor. In this section, we look for the relation between the Hawking-Ellis classification of $t_{\mu\nu}$, presented in section \ref{sec-classification}, and the Hawking-Ellis classification of the mimetic energy-momentum tensor $T^\star_{\mu\nu}$.

We consider a gravitational metric theory coupled to matter fields which are collectively represented by $\Psi_{\rm M}$. The action of the theory is given by $S_{\rm seed}[g,\Psi_{\rm M}]$. Performing the disformal transformation \eqref{sec inv: general disformal transformation} on the seed action $S_{\rm seed}[g,\Psi_{\rm M}]$ is equivalent to considering the disformed action
\begin{align}\label{action}
S_{\rm dis}[h,t,\Psi_{\rm M}] = S_{\rm seed}[g,\Psi_{\rm M}]
+ \frac{1}{2}\int \D^4x \sqrt{-g}\, \lambda^{\alpha\beta}
\bigg[
g_{\alpha\beta}
- \bigg( C \, h_{\alpha\beta}
+ \sum_{I=1}^3D_I \, t^I_{\alpha\beta} \bigg)
\bigg] \,,
\end{align}
where $t$ schematically shows dependency on $t_{\mu\nu}$. For example, as we will explicitly show, $t$ may include a scalar field, a vector field, gauge fields and their derivatives. In the above action, $\lambda^{\alpha\beta}$ are ten Lagrange multipliers and their equations of motion enforce the disformal transformation \eqref{sec inv: general disformal transformation}. We assume that the transformation depends algebraically on $h_{\mu\nu}$, with no derivatives of this metric. Varying at fixed underlying fields, we find
\begin{align}\label{EoM-h}
\frac{\delta{S_{\rm dis}}}{\delta{h_{\mu\nu}}}
= - \frac{1}{2} \sqrt{-g} \lambda^{\alpha\beta}
J^{\mu\nu}_{\alpha\beta} = 0 \,,
\end{align}
where
\begin{align}
\label{J-def}
J^{\mu\nu}_{\alpha\beta}
&\equiv \frac{\partial{g_{\alpha\beta}}}{\partial{h}_{\mu\nu}} \,,
\end{align}
is the Jacobian of the transformation between $g_{\mu\nu}$ and $h_{\mu\nu}$.

Equation \eqref{EoM-h} always admits the branch $\lambda^{\alpha\beta}=0$. When the Jacobian is non-singular, this is the only branch and the transformation is locally invertible, which leads to $S_{\rm dis}[h,t,\Psi_{\rm M}] = S_{\rm seed}[g,\Psi_{\rm M}]$. This is the trivial case when the disformal transformation \eqref{sec inv: general disformal transformation} is a mere field redefinition. When $J^{\mu\nu}_{\alpha\beta}$ is singular, its kernel is non-trivial and branches with nonzero $\lambda^{\alpha\beta}$ in this kernel become possible. Let us elaborate on this non-trivial possibility. The eigentensor equation for the Jacobian is
\begin{align}\label{eigen-Eq}
J^{\alpha\beta}_{\mu\nu} \xi^{(a)}_{\alpha\beta} = \lambda^{(a)} \xi^{(a)}_{\mu\nu} \,,
\qquad
J^{\mu\nu}_{\alpha\beta} \zeta_{(a)}^{\alpha\beta} = \lambda^{(a)} \zeta_{(a)}^{\mu\nu} \,,
\end{align}
where $\xi^{(a)}_{\mu\nu}$ and $\zeta_{(a)}^{\mu\nu}$ are the eigentensors and dual eigentensors corresponding to the eigenvalues $\lambda^{(a)}$. Consider the situation when the disformal transformation \eqref{sec inv: general disformal transformation} is not invertible so that one of the eigenvalues in Eq. \eqref{eigen-Eq} vanishes
\begin{align}\label{lambda-star}
\lambda_{\star}={0}\,,
\qquad \mbox{s.t.} \qquad
J^{\mu\nu}_{\alpha\beta} \zeta_{\star}^{\alpha\beta} = 0 \,.
\end{align}
	
For a one-dimensional dual kernel, comparing Eqs. \eqref{EoM-h} and \eqref{lambda-star} gives
\begin{align}\label{lambda-munu-def}
\lambda^{\mu\nu}\equiv {\tilde\rho}\, \zeta_{\star}^{\mu\nu} \,,
\end{align}
where ${\tilde \rho}$ is an arbitrary function at the moment. For several independent dual zero modes, the right-hand side is replaced by $\sum_a\tilde\rho_a\zeta_{(a)}^{\mu\nu}$. The corresponding contributions are summed in $T^\star_{\mu\nu}$ below, and the conservation law refers to this total source. Taking variation with respect to $g^{\mu\nu}$ gives, after imposing the constraint enforced by $\lambda^{\alpha\beta}$,
\begin{align}\label{singularity-condition}
\frac{2}{\sqrt{-g}}\frac{\delta{S_{\rm seed}}}{\delta{g^{\mu\nu}}}
= g_{\mu\alpha} g_{\nu\beta} \lambda^{\alpha\beta} \,.
\end{align}
Note that all quantities are expressed in terms of the metric $g_{\mu\nu}$.
	
We specify the seed action as the Einstein-Hilbert action which is minimally coupled to the matter
\begin{align}
S_{\rm seed}[g,\Psi_{\rm M}] = \int\D^4x\sqrt{-g}\left(\frac{\Mpl^2}{2}R+L_{\rm M}[g,\Psi_{\rm M}] \right) \,.
\end{align}
Using the above seed action in \eqref{action} and then taking the variation w.r.t. metric, we find
\begin{align}
\Mpl^2G_{\mu\nu} = T_{\mu\nu}^{\rm M} + T^\star_{\mu\nu} \,,
\end{align}
where $T^{\rm M}_{\mu\nu}=-\tfrac{2}{\sqrt{-g}}\,\delta\!\left(\sqrt{-g}\,L_{\rm M}\right)\!/\delta g^{\mu\nu}$ is the usual energy-momentum tensor while
\begin{align}\label{EMT-star}
T^\star_{\mu\nu} \equiv g_{\mu\alpha} g_{\nu\beta} \lambda^{\alpha\beta}
=
{\tilde \rho}\, g_{\mu\alpha} g_{\nu\beta} \zeta_{\star}^{\alpha\beta} \,,
\end{align}
is the mimetic energy-momentum tensor induced by the singular disformal transformation through the variation of the second term in the right-hand side of Eq. \eqref{action}. Therefore, to find the energy-momentum tensor of the new degrees of freedom \eqref{EMT-star}, we first need to solve the eigentensor Eq. \eqref{eigen-Eq} to find the solution for $\zeta_\star^{\mu\nu}$ in terms of $h_{\mu\nu}$ which is subject to the condition $\lambda_\star=0$. Since the disformal transformation is singular, we cannot express $h_{\mu\nu}$ in terms of $g_{\mu\nu}$. However, solving $\lambda_\star=0$, we find a relation between $C$ and $D_I$ which after substituting in \eqref{action} allows us to rewrite the disformed action completely in terms of the metric $g_{\mu\nu}$ even if we do not have an explicit expression of $h_{\mu\nu}$ in terms of $g_{\mu\nu}$. In the next section, we will do this explicitly for some particular examples.
	
Note also that the diffeomorphism invariance of the action \eqref{action} implies
\begin{align}\label{EMT-star-cons}
\nabla_\alpha T^{\star\alpha}{}_{\mu}
= \nabla_\alpha \left( {\tilde \rho}\, g_{\mu\beta} \zeta_{\star}^{\alpha\beta} \right)
= 0 \,,
\end{align}
where we have assumed that the usual matter energy-momentum tensor is separately conserved $\nabla_\alpha T^{{\rm M}\alpha}{}_{\mu} = 0$.
	
Based on the above discussion, to find the mimetic energy-momentum tensor, we need to solve the eigenvalue system \eqref{lambda-star} to find the explicit solution for $\zeta_{\star}^{\mu\nu}$. Using the normal form \eqref{DT-simplified} in \eqref{J-def}, we find
\begin{align}\label{J}
J^{\rho\sigma}_{\mu\nu}
=\, C\delta^{(\rho}_\mu\delta^{\sigma)}_\nu
+D\frac{\partial t_{\mu\nu}}{\partial h_{\rho\sigma}} +\frac{\partial C}{\partial h_{\rho\sigma}} h_{\mu\nu}
+\frac{\partial D}{\partial h_{\rho\sigma}} t_{\mu\nu} \, .
\end{align}
Here all metric derivatives are taken at fixed underlying fields, including the dependence of $C$ and $D$ on any metric-dependent scalar invariants. A nonzero dual zero mode satisfying \eqref{lambda-star} exists at a given configuration if and only if $\det J=0$, where $J$ is regarded as the $10\times10$ Jacobian relating the independent components of $h_{\mu\nu}$ and $g_{\mu\nu}$. Substituting \eqref{J} in \eqref{lambda-star}, we find
\begin{align}\label{eigenvalue-system}
C\zeta_\star^{\rho\sigma}
+D\frac{\partial t_{\alpha\beta}}{\partial h_{\rho\sigma}}
\zeta_\star^{\alpha\beta}
+\left(h_{\alpha\beta}\zeta_\star^{\alpha\beta}\right)
\frac{\partial C}{\partial h_{\rho\sigma}}
+\left(t_{\alpha\beta}\zeta_\star^{\alpha\beta}\right)
\frac{\partial D}{\partial h_{\rho\sigma}}=0\, .
\end{align}
We now restrict the analysis to dual zero modes that are linear combinations of $h^{\mu\nu}$ and powers of $t_{\mu\nu}$. This polynomial form does not follow from the Cayley-Hamilton degree alone. As shown in \Cref{app:polynomial}, it is guaranteed for metric-independent $t_{\mu\nu}$ when $C\neq 0$ and $C,D$ depend only on the traces $[t^I]$. For more general metric dependence, including additional scalar invariants, the appendix provides sufficient conditions based on closure of the full metric variations and invertibility of the tensor map \eqref{poly-B-app}. Within this polynomial sector, we write
\begin{align}\label{zeta-CH4}
&\zeta_\star^{\mu\nu}= a h^{\mu\nu} + \sum_{I=1}^3 b_I t^{I\,\mu\nu} \,,
&{\boldsymbol{\rm CH}}_4\,\, \mbox{condition} \,,
\\ \label{zeta-CH3}
&\zeta_\star^{\mu\nu}= a h^{\mu\nu} + \sum_{I=1}^2 b_I t^{I\,\mu\nu} \, ,
&{\boldsymbol{\rm CH}}_3\,\, \mbox{condition} \,,
\\
\label{zeta-CH2}
&\zeta_\star^{\mu\nu}= a h^{\mu\nu} + b\, t^{\mu\nu} \, ,
&{\boldsymbol{\rm CH}}_2\,\, \mbox{condition} \,,
\end{align}
where $a$ and $b_I$ are functions of the independent traces and of any other scalar arguments retained in the transformation. Each independent zero mode has an arbitrary overall normalization, which is not fixed by the homogeneous equation \eqref{eigenvalue-system}. Here we focus on Cayley-Hamilton degrees $n=2,3,4$. Non-trivial zero modes can occur also for $D=0$, as illustrated in \Cref{sec-examples}. Substituting these polynomial zero modes into \eqref{EMT-star} and using the Cayley-Hamilton relations gives
\begin{align}\label{EMT-CH4}
&T^{\star\mu}{}_{\nu} = \tilde{a}\, \delta^{\mu}_{\nu} + \sum_{I=1}^3 \tilde{b}_I t^{I\,\mu}{}_{\nu} \,,
&{\boldsymbol{\rm CH}}_4\,\, \mbox{condition} \,,
\\
\label{EMT-CH3}
&T^{\star\mu}{}_{\nu} = \tilde{a}\, \delta^{\mu}_{\nu} + \sum_{I=1}^2 \tilde{b}_I t^{I\,\mu}{}_{\nu}  \, ,
&{\boldsymbol{\rm CH}}_3\,\, \mbox{condition} \,,
\\
\label{EMT-CH2}
&T^{\star\mu}{}_{\nu} = \tilde{a}\, \delta^{\mu}_{\nu} + \tilde{b}\, t^{\mu}{}_{\nu}  \, ,
&{\boldsymbol{\rm CH}}_2\,\, \mbox{condition} \,.
\end{align}
The explicit forms of the coefficient functions $\tilde{a}$ and $\tilde{b}_I$ can be easily found by substituting \eqref{zeta-CH4}, \eqref{zeta-CH3}, and \eqref{zeta-CH2} in Eq. \eqref{EMT-star} and then using the Cayley-Hamilton relations \eqref{sec inv: CH4 theorem}, \eqref{sec simple CH: CH3 condition}, and \eqref{sec simple CH: CH2 condition}, respectively.

\subsection{Hawking-Ellis type of mimetic energy-momentum tensor}

In this subsection, we retain the polynomial restriction \eqref{EMT-CH4}-\eqref{EMT-CH2} and determine the allowed Hawking-Ellis types of $T^{\star\mu}{}_{\nu}$ from that of $t^{\mu}{}_{\nu}$. The starting point is the observation that Eqs.~\eqref{EMT-CH4}, \eqref{EMT-CH3}, and \eqref{EMT-CH2} all share the same structure: in each case, the mixed tensor $T^{\star\mu}{}_{\nu}$ is built out of the identity and powers of the single matrix $t^{\mu}{}_{\nu}$. In other words, $T^{\star\mu}{}_{\nu}$ is a polynomial in the matrix $t^{\mu}{}_{\nu}$,
\begin{align}\label{poly-map-Tstar}
T^{\star\mu}{}_{\nu}=p_n\!\left(t\right)^{\mu}{}_{\nu} \,,
\qquad
p_n(x)\equiv \tilde a+\sum_{I=1}^{n-1}\tilde b_I\, x^I \,,
\end{align}
where $n$ is the Cayley-Hamilton degree of the transformation. The index on $T^{\star\mu}{}_{\nu}$ is raised with the physical metric $g^{\mu\nu}$, whereas the powers of $t^\mu{}_{\nu}$ use $h^{\mu\nu}$ as in \eqref{t-HO}. Here a polynomial of a matrix is understood in the obvious way: the constant term multiplies the identity matrix and $x^I$ is replaced by the $I$-th matrix power of $t^{\mu}{}_{\nu}$. The advantage of this rewriting is that the action of a polynomial on the Jordan decomposition of a matrix is completely understood, so the entire relation between the Hawking-Ellis types of $t^{\mu}{}_{\nu}$ and $T^{\star\mu}{}_{\nu}$ follows from two elementary facts, which we now recall.

First, a polynomial respects similarity transformations. Writing the Jordan decomposition of the disformal tensor as $t^{\mu}{}_{\nu}=\left({\boldsymbol P}\cdot{\boldsymbol J}\cdot{\boldsymbol P}^{-1}\right)^{\mu}{}_{\nu}$, as in \eqref{Jordan-decomposition-app}, every matrix power satisfies $\boldsymbol{t}^I={\boldsymbol P}\cdot{\boldsymbol J}^I\cdot{\boldsymbol P}^{-1}$ and we have
\begin{align}
p_n\!\left(\boldsymbol{t}\right)={\boldsymbol P}\cdot {p}_n\!\left({\boldsymbol J}\right)\cdot{\boldsymbol P}^{-1} \,,
\nonumber
\end{align}
so $T^{\star\mu}{}_{\nu}$ is brought to (block) canonical form by the {\it same} transformation ${\boldsymbol P}$ that brings $t^{\mu}{}_{\nu}$ to its Jordan normal form. Moreover, since $\boldsymbol{J}$ is block diagonal, so is any power of it, and therefore
\begin{equation}\nonumber
p_n\!\left({\boldsymbol J}\right)
=\begin{pmatrix}
	p_n\!(\boldsymbol{J}_{1})&0&\cdots&0\\
	0&p_n\!(\boldsymbol{J}_{2})&\ddots&\vdots\\
	\vdots&\ddots&\ddots&0\\
	0&\cdots&0&p_n\!(\boldsymbol{J}_q)
\end{pmatrix} \,,
\end{equation}
i.e. the polynomial acts on each Jordan block separately. The problem is thus reduced to the action of $p_n$ on a single Jordan block.

Second, the action of a polynomial on a single Jordan block is governed by the derivatives of the polynomial. Decomposing a block of size $s$ with eigenvalue $\lambda$ as 
\begin{align}\nonumber
\boldsymbol{J}_s(\lambda)=\lambda\boldsymbol{1}+\boldsymbol{N}_s \,,
\end{align}
where $\boldsymbol{N}_s$ has unit entries on the first superdiagonal and zeros elsewhere, the matrix $\boldsymbol{N}_s$ is nilpotent: each multiplication by $\boldsymbol{N}_s$ shifts its entries one step further from the diagonal, so that $\boldsymbol{N}_s^s=0$. Expanding each power $\left(\lambda\boldsymbol{1}+\boldsymbol{N}_s\right)^I$ with the binomial theorem, which is legitimate because $\boldsymbol{1}$ and $\boldsymbol{N}_s$ commute, and collecting equal powers of $\boldsymbol{N}_s$, one finds the exact, finite expansion
\begin{align}\label{poly-map-Jordan}
p_n\!\left(\boldsymbol{J}_s(\lambda)\right)
= p_n(\lambda)\,\boldsymbol{1}+p_n'(\lambda)\,\boldsymbol{N}_s+\frac{1}{2}\,p_n''(\lambda)\,\boldsymbol{N}_s^2+\cdots
+\frac{1}{(s-1)!}\,p_n^{(s-1)}(\lambda)\,\boldsymbol{N}_s^{s-1} \,,
\end{align}
where the primes and the superscript $(s-1)$ denote derivatives with respect to the argument. Although \eqref{poly-map-Jordan} looks like a Taylor expansion, it is exact: all higher terms vanish identically by nilpotency, $\boldsymbol{N}_s^{s}=0$. For the block sizes relevant in four dimensions, $s=2$ and $s=3$, Eq.~\eqref{poly-map-Jordan} reads explicitly
\begin{align}\label{poly-map-blocks-2}
p_n\!\left(\begin{array}{cc}
\lambda&1\\
0&\lambda
\end{array}\right)
&=\left(\begin{array}{cc}
p_n(\lambda)&p_n'(\lambda)\\
0&p_n(\lambda)
\end{array}\right),
\\ \label{poly-map-blocks-3}
p_n\!\left(\begin{array}{ccc}
\lambda&1&0\\
0&\lambda&1\\
0&0&\lambda
\end{array}\right)
&=\left(\begin{array}{ccc}
p_n(\lambda)&p_n'(\lambda)&\tfrac12p_n''(\lambda)\\
0&p_n(\lambda)&p_n'(\lambda)\\
0&0&p_n(\lambda)
\end{array}\right).
\end{align}

Equations \eqref{poly-map-Jordan}, \eqref{poly-map-blocks-2}, and \eqref{poly-map-blocks-3} immediately yield: Every eigenvalue is mapped as $\lambda\to p_n(\lambda)$. Since the coefficients $\tilde a$ and $\tilde b_I$ are real, real eigenvalues remain real and the complex pair of Type IV is mapped into the complex pair $\big(p_n(z),\,p_n(\bar z)\big)$ with $p_n(\bar z)=\overline{p_n(z)}$.

\begin{itemize}
\item {\bf Type I:} A diagonalizable real operator remains diagonalizable with the same eigenvectors, although distinct eigenvalues can be mapped to the same value. Hence Type I always maps to Type I.

\item {\bf Type II:} By \eqref{poly-map-blocks-2}, the off-diagonal entry of the mapped $2\times2$ block is $p_n'(\lambda_1)$, where $\lambda_1$ is the eigenvalue of the block. If $p_n'(\lambda_1)\neq0$, the entry can be normalized back to unity by the residual similarity transformation $\mathrm{diag}\big[1,\,p_n'(\lambda_1)^{-1}\big]$, which simply rescales the vectors of the Jordan chain; the block structure, and hence Type II, is preserved.\footnote{Note that the two Lorentz-inequivalent branches $\pm$ of Type II in \eqref{sec class: HE canonical forms} are distinguished by the sign of the scalar product of the two vectors of the Jordan chain, see appendix~\ref{app:HEC}. Since the Hawking-Ellis classification of $T^{\star\mu}{}_{\nu}$ is defined with respect to the physical metric, the relevant sign is multiplied by both $p_n'(\lambda_1)$ and the metric factor $C+D\lambda_1$. Thus the two branches are exchanged when $(C+D\lambda_1)p_n'(\lambda_1)<0$.} If instead $p_n'(\lambda_1)=0$, the mapped block is diagonal and the tensor reduces to Type I.

\item {\bf Type III:} The nilpotent part of the mapped $3\times3$ block in \eqref{poly-map-blocks-3} has $p_n'(\lambda_1)$ on the first superdiagonal and $\frac{1}{2}p_n''(\lambda_1)$ on the second. If $p_n'(\lambda_1)\neq0$, the mapped matrix still has a single Jordan chain of length three and Type III is preserved. If $p_n'(\lambda_1)=0$ but $p_n''(\lambda_1)\neq0$, the mapped block is $\frac{1}{2}p_n''(\lambda_1)\,N_3^2$ shifted by a multiple of the identity, whose Jordan normal form consists of one size-2 block and one size-1 block: the tensor reduces to Type II, of subclass $[(21)1]$ when $p_n(\lambda_2)\neq p_n(\lambda_1)$ and $[(211)]$ otherwise, the latter being automatic for the class $[(31)]$. If both derivatives vanish, the block becomes proportional to the identity and the tensor reduces to Type I.

\item {\bf Type IV:} In this case, the relevant question is whether the mapped pair $\big(p_n(z),\,\overline{p_n(z)}\big)$ is still genuinely complex. It is convenient to introduce the divided difference
\begin{align}\label{divided-difference}
\Delta_n\equiv\frac{p_n(z)-p_n(\bar z)}{z-\bar z} \,,
\end{align}
which is a real number, as can be checked directly from the explicit expressions in \Cref{table-dictionary} below, and which plays the role of a ``derivative'' evaluated on the complex pair. Since $\mathrm{Im}[z]\neq0$, we have $p_n(z)\neq p_n(\bar z)$ if and only if $\Delta_n\neq0$: in this case Type IV is preserved. If $\Delta_n=0$, the pair collapses to a repeated \emph{real} eigenvalue, $p_n(z)=p_n(\bar z)\in\mathbb{R}$, carried by two trivial Jordan blocks, and the tensor reduces to Type I. 
\end{itemize}
Note that the converse degradations never occur: a polynomial with real coefficients can neither create nilpotent blocks nor turn real eigenvalues complex, so the Hawking-Ellis type can only move ``downwards'', III $\to$ II $\to$ I and IV $\to$ I, and Type I is always mapped to Type I.

Applying the above results to the Jordan normal forms \eqref{sec class: Jordan forms} of the four Hawking-Ellis types, the Jordan structures of the mimetic energy-momentum tensor $T^{\star\mu}{}_{\nu}$ read, for every Cayley-Hamilton degree at once,
\begin{equation}\label{mapped-Jordan-forms}
	\begin{gathered}
		\begin{array}{c}
			\left(\begin{array}{cccc}
				p_n(\lambda_1)&0&0&0\\
				0&p_n(\lambda_2)&0&0\\
				0&0&p_n(\lambda_3)&0\\
				0&0&0&p_n(\lambda_4)
			\end{array}\right)\\[2mm]
			\mbox{Type I}
		\end{array}
		\qquad
		\begin{array}{c}
			\left(\begin{array}{cccc}
				p_n(\lambda_1)&p_n'(\lambda_1)&0&0\\
				0&p_n(\lambda_1)&0&0\\
				0&0&p_n(\lambda_2)&0\\
				0&0&0&p_n(\lambda_3)
			\end{array}\right)\\[2mm]
			\mbox{Type II}
		\end{array}
		\\[5mm]
		\begin{array}{c}
			\left(\begin{array}{cccc}
				p_n(\lambda_1)&p_n'(\lambda_1)&\tfrac12p_n''(\lambda_1)&0\\
				0&p_n(\lambda_1)&p_n'(\lambda_1)&0\\
				0&0&p_n(\lambda_1)&0\\
				0&0&0&p_n(\lambda_2)
			\end{array}\right)\\[2mm]
			\mbox{Type III}
		\end{array}
		\qquad
		\begin{array}{c}
			\left(\begin{array}{cccc}
				p_n(z)&0&0&0\\
				0&p_n(\bar z)&0&0\\
				0&0&p_n(\lambda_1)&0\\
				0&0&0&p_n(\lambda_2)
			\end{array}\right)\\[2mm]
			\mbox{Type IV}
		\end{array}
	\end{gathered}
\end{equation}
where in each case the eigenvalues refer to those of $t^{\mu}{}_{\nu}$. These four matrices replace the case-by-case substitutions for each Cayley-Hamilton degree: the entire dependence on the degree $n$ is carried by the explicit forms of $p_n(\lambda)$, $p_n'(\lambda)$, $\frac{1}{2}p_n''(\lambda)$, and $\Delta_n$ in terms of the coefficients $\tilde a$ and $\tilde b_I$. This ``dictionary'' is easy to compute. For instance, $\Delta_4$ follows from $(z^2-\bar z^2)/(z-\bar z)=2\,\mathrm{Re}[z]$ and $(z^3-\bar z^3)/(z-\bar z)=3\,\mathrm{Re}[z]^2-\mathrm{Im}[z]^2$, and is collected in \Cref{table-dictionary}.

\begin{table}[!htb]
\centering
\renewcommand{\arraystretch}{1.5}
\setlength{\tabcolsep}{3.5pt}
\begin{tabular}{|c|c|c|c|c|}
		\hline
		& $p_n(\lambda)$
		& $p_n'(\lambda)$
		& $\frac{1}{2}p_n''(\lambda)$
		& $\Delta_n$
		\\ \hline
		${\boldsymbol{\rm CH}}_2$
		& $\tilde a+\tilde b\,\lambda$
		& $\tilde b$
		& $0$
		& not needed
		\\ \hline
		${\boldsymbol{\rm CH}}_3$
		& $\tilde a+\tilde b_1\lambda+\tilde b_2\lambda^2$
		& $\tilde b_1+2\tilde b_2\lambda$
		& $\tilde b_2$
		& $\tilde b_1+2\,\mathrm{Re}[z]\,\tilde b_2$
		\\ \hline
		${\boldsymbol{\rm CH}}_4$
		& $\tilde a+\tilde b_1\lambda+\tilde b_2\lambda^2+\tilde b_3\lambda^3$
		& $\tilde b_1+2\tilde b_2\lambda+3\tilde b_3\lambda^2$
		& $\tilde b_2+3\tilde b_3\lambda$
		& $\tilde b_1+2\,\mathrm{Re}[z]\,\tilde b_2+\left(3\,\mathrm{Re}[z]^2-\mathrm{Im}[z]^2\right)\tilde b_3$
\\ \hline
\end{tabular}
\vspace{.5cm}
\caption{Dictionary between the coefficients of the polynomial map \eqref{poly-map-Tstar} and the quantities controlling the Jordan structure of $T^{\star\mu}{}_{\nu}$ in \eqref{mapped-Jordan-forms}: the mapped eigenvalues $p_n(\lambda)$, the derivatives $p_n'(\lambda)$ and $\frac{1}{2}p_n''(\lambda)$ governing the nilpotent parts, and the divided difference $\Delta_n$ defined in \eqref{divided-difference} governing the complex pair of Type IV. For ${\boldsymbol{\rm CH}}_2$ the entries $\frac{1}{2}p_2''$ and $\Delta_2$ are not needed, since Type III and Type IV tensors cannot be ${\boldsymbol{\rm CH}}_2$, cf.~\Cref{sec class: table of CH degrees}.}\label{table-dictionary}
\end{table}

The correspondence within the polynomial sector is summarized, for all Cayley-Hamilton degrees simultaneously, in \Cref{table-EMT}. In particular, the type of the mimetic energy-momentum tensor coincides with that of the disformal tensor unless the corresponding derivative or divided difference of $p_n$ vanishes.

\begin{table}[!htb]
\centering
\renewcommand{\arraystretch}{1.5}
\begin{tabular}{|c|c|c|}
\hline
$t^\mu{}_\nu$ & Condition & $T^{\star\mu}{}_\nu$ \\
\hline
		Type I & not needed & Type I \\
		\hline
		\multirow{2}{*}{Type II}
		& $p_n'(\lambda_1)\neq0$ & Type II \\
		\cline{2-3}
		& $p_n'(\lambda_1)=0$ & Type I \\
		\hline
		\multirow{3}{*}{Type III}
		& $p_n'(\lambda_1)\neq0$ & Type III \\
		\cline{2-3}
		& $p_n'(\lambda_1)=0$\,, $p_n''(\lambda_1)\neq0$ & Type II \\
		\cline{2-3}
		& $p_n'(\lambda_1)=0$\,, $p_n''(\lambda_1)=0$ & Type I \\
		\hline
		\multirow{2}{*}{Type IV}
		& $\Delta_n\neq0$ & Type IV \\
		\cline{2-3}
		& $\Delta_n=0$ & Type I \\
\hline
\end{tabular}
\vspace{.5cm}
\caption{Relation between the Hawking-Ellis types of the disformal tensor $t^\mu{}_\nu$ and the mimetic energy-momentum tensor $T^{\star\mu}{}_\nu=p_n(t)^\mu{}_\nu$, valid for every Cayley-Hamilton degree $n$. Here $\lambda_1$ denotes the eigenvalue of the non-trivial Jordan block and $z$ the complex eigenvalue of Type IV; the explicit expressions of $p_n'$, $p_n''$, and $\Delta_n$ in terms of the coefficients $\tilde a$ and $\tilde b_I$ are given in \Cref{table-dictionary}. Which types can occur for a given degree is dictated by \Cref{sec class: table of CH degrees}; for instance, for ${\boldsymbol{\rm CH}}_2$ only the first two rows are relevant.} \label{table-EMT}
\end{table}

Two remarks are in order. First, in \eqref{mapped-Jordan-forms} the eigenvalues $\lambda_i$ are kept general: the map $\lambda_i\to p_n(\lambda_i)$ can produce accidental coincidences among the mapped eigenvalues, $p_n(\lambda_i)=p_n(\lambda_j)$ for $\lambda_i\neq\lambda_j$, which refine the Segre subclass of $T^{\star\mu}{}_{\nu}$ but do not affect the Hawking-Ellis type statements of \Cref{table-EMT}. Second, we stress that \Cref{table-EMT} presupposes only the polynomial structure \eqref{poly-map-Tstar}; the values of the coefficients $\tilde a$ and $\tilde b_I$, and hence whether the degenerate cases are actually realized, must be determined by solving the eigentensor equation \eqref{eigenvalue-system} for the specific transformation at hand, as we do in \Cref{sec-examples}.
	
\subsection{\texorpdfstring{Eigenvalue problem for ${\boldsymbol{\rm CH}}_2$}{Eigenvalue problem for CH2}}
	
To determine the relationship between the Hawking-Ellis types of $t^\mu{}_\nu$ and $T^{\star\mu}{}_\nu$, we must solve the eigenvalue system \eqref{eigenvalue-system}, which is generally non-trivial. In this subsection, we work within the trace-dependent polynomial sector and outline the coefficient calculation for the ${\boldsymbol{\rm CH}}_2$ case. The same strategy applies to the more involved ${\boldsymbol{\rm CH}}_3$ and ${\boldsymbol{\rm CH}}_4$ cases.
	
For the ${\boldsymbol{\rm CH}}_2$ case, where $t^\mu{}_\nu$ satisfies \eqref{sec simple CH: CH2 condition}, all powers $t^n_{\mu\nu}$ with $n > 1$ reduce to linear combinations of $h_{\mu\nu}$ and $t_{\mu\nu}$. On a fixed subclass, at most $[t]$ and $[t^2]$ are independent, while higher traces can be reduced accordingly. Using the trace dependence assumed in this subsection and applying
the chain rule in \eqref{eigenvalue-system}, we obtain
\begin{align}
\label{eigenvalue-system-CH2}
\begin{split}
C \zeta_{\star}^{\rho\sigma} + D \frac{\partial{t_{\alpha\beta}}}{\partial{h_{\rho\sigma}}} \zeta_{\star}^{\alpha\beta}
+ \sum_{I=1}^{2} \left[ C_{[t^I]} \left( h_{\alpha\beta} \zeta_{\star}^{\alpha\beta} \right) + D_{[t^I]} \left( t_{\alpha\beta} \zeta_{\star}^{\alpha\beta} \right) \right] \frac{\partial{[t^I]}}{\partial{h_{\rho\sigma}}} = 0 \,,
\end{split}
\end{align}
where $C_{[t^I]}=\partial{C}/\partial{[t^I]}$ and so on. Using the definition of $[t]$ and $[t^2]$ in \eqref{t-Trace-def} and \eqref{t-HO}, we find
\begin{align}\label{Dth-CH2}
\frac{\partial{[t]}}{\partial{h_{\rho\sigma}}}
= - t^{\rho\sigma} + h^{\alpha\beta} \frac{\partial{t_{\alpha\beta}}}{\partial{h_{\rho\sigma}}} \,,
\qquad
\frac{\partial{[t^2]}}{\partial{h_{\rho\sigma}}}
= - 2 t^{2\,\rho\sigma} + 2 t^{\alpha\beta} \frac{\partial{t_{\alpha\beta}}}{\partial{h_{\rho\sigma}}}
\,.
\end{align}
For the following calculation, we restrict to the subclass for which
\begin{align}\label{Dth-contract-CH2}
h^{\alpha\beta} \frac{\partial{t_{\alpha\beta}}}{\partial{h_{\rho\sigma}}}
= c_2 t^{\rho\sigma}
\,,
\qquad
t^{\alpha\beta} \frac{\partial{t_{\alpha\beta}}}{\partial{h_{\rho\sigma}}}
= c_0 h^{\rho\sigma} + c_1 t^{\rho\sigma}
\,.
\end{align}
This is an additional restriction on the metric dependence of
$t_{\mu\nu}$ and does not follow from the ${\boldsymbol{\rm CH}}_2$
relation alone. For a given $t_{\mu\nu}$ in this subclass, one can then
find explicit forms of $c_{0,1,2}$ and $\delta_{0,1}$.

Within the polynomial invariant subspace, the ${\boldsymbol{\rm CH}}_2$ ansatz is \eqref{zeta-CH2}
\begin{align}\label{dual-tensor-CH2}
\zeta_{\star}^{\mu\nu} = a\, h^{\mu\nu} + b\, t^{\mu\nu} \,,
\end{align}
where $a$ and $b$ are some functions of $[t]$ and $[t^2]$. For a one-dimensional polynomial kernel, Eq.~\eqref{eigenvalue-system-CH2} fixes the relative coefficients, leaving their overall normalization arbitrary. Substituting \eqref{Dth-CH2} and \eqref{dual-tensor-CH2} in \eqref{eigenvalue-system-CH2} and using \eqref{sec simple CH: CH2 condition} we find
\begin{align}\label{eigenvalue-system-CH2-final}
\left( a \, {\cal C}_h + b\, {\cal D}_h \right) h^{\mu\nu} + \left(
a \, {\cal C}_t + b\, {\cal D}_t
\right) t^{\mu\nu} = 0 \,,
\end{align}
where
\begin{align}
\begin{split}
{\cal C}_h &\equiv
C + 2 \left(c_0-\delta _0\right) \left(4 C_{[t^2]} + D_{[t^2]} [t] \right) \,,
\\
{\cal D}_h &\equiv
c_0 D + 2 (c_0-\delta_0) \left(C_{[t^2]}
[t] + D_{[t^2]} [t^2] \right)
\,, 
\\
{\cal C}_t &\equiv
c_2 D
+ \left(c_2-1\right) \big( 4 C_{[t]}+ [t] D_{[t]} \big)
+2 \left(c_1-\delta _1\right) \left(4 C_{[t^2]} + D_{[t^2]} [t] \right) \,,
\\
{\cal D}_t &\equiv
C + c_1 D + \left(c_2-1\right) \left([t] C_{[t]}+[t^2] D_{[t]} \right)
+ 2 (c_1-\delta_1) \left(C_{[t^2]}
[t]+D_{[t^2]} [t^2] \right)
\,,
\end{split}
\end{align}

Since we assume that $t^{\mu\nu}$ is not proportional to $h^{\mu\nu}$, Eq.~\eqref{eigenvalue-system-CH2-final} implies that the coefficients of both $h^{\mu\nu}$ and $t^{\mu\nu}$ must vanish. Solving these two conditions gives the allowed relative coefficients and the singularity condition for a nonzero polynomial mode. Under the sufficient conditions established in \Cref{app:polynomial}, the polynomial ansatz captures all dual zero modes.


\section{Examples}\label{sec-examples}

In this section, we first revisit the well-studied scalar and gauge field transformations, and then consider scalar-gauge and two-scalar examples. The scalar-gauge transformation gives a new singular pure-disformal family yielding irrotational perfect fluids, including barotropic cases with an arbitrary constant equation-of-state parameter. Although its coefficients depend on an additional invariant, its dual zero mode remains polynomial. The two-scalar disformal tensor realizes all four Hawking-Ellis types, but its general dual zero mode need not be polynomial. We therefore restrict to trace-dependent coefficients when applying \Cref{table-EMT}, and explicitly realize the Type II $\to$ Type I and Type IV $\to$ Type I reductions.

\subsection{Scalar field}\label{sec-example-scalar}

One of the most well-known examples of a disformal transformation is the scalar-field transformation in Eq.~\eqref{sec simple CH: scalar field ordinary disformal transformation}, originally introduced in~\cite{Bekenstein:1992pj}
\begin{equation}\label{dis-T-S}
g_{\mu\nu} = C(\phi,Y_\phi)\,h_{\mu\nu}+D(\phi,Y_\phi)\, \partial_\mu \phi \partial_\nu \phi\,.
\end{equation}
Here $Y_\phi$ is the scalar contraction defined in \eqref{invariants-SV}.
Comparing it with the normal form of the disformal transformation \eqref{DT-simplified}, we find
\begin{align}\label{t-phi}
t_{\mu\nu} = u_\mu u_\nu \, ,\qquad
u_\mu\equiv\partial_\mu\phi\,,
\end{align}
which gives
\begin{align}
t^2_{\mu\nu}=t_{\mu}{}^{\alpha}t_{\alpha\nu}=Y_\phi \partial_\mu \phi \partial_\nu \phi \,.
\end{align}
The above result clearly shows that the disformal transformation \eqref{dis-T-S} satisfies the Cayley-Hamilton ${\boldsymbol{\rm CH}}_2$ condition \eqref{sec simple CH: CH2 condition} with
\begin{align}\label{delta-i-phi}
\delta_0 = 0 \,,
\qquad
\delta_1 = Y_\phi \,.
\end{align}
For the traces defined in \eqref{t-Trace-def}, we have
\begin{align}
[t]=Y_\phi \,,
\qquad
[t^2]=Y_\phi^2 \,.
\end{align}
Note that $[t^2] = [t]^2$, indicating that $[t^2]$ is not an independent building block. Therefore, it suffices to consider $C$ and $D$ as functions of $\phi$ and $Y_\phi$ only.

Substituting \eqref{delta-i-phi} in \eqref{sol-inverse-G2}, the inverse metric \eqref{dis-T-inv-G2} for \eqref{dis-T-S} takes the following form
\begin{align}\label{dis-T-inv-S}
g^{\mu\nu} = \frac{1}{C} \left( h^{\mu\nu} - \frac{D}{C+Y_\phi D} u^\mu u^\nu \right) \,;
\qquad
u^\mu\equiv h^{\mu\alpha}u_\alpha \,.
\end{align}

In order to find the Hawking-Ellis type, we need to find the Jordan decomposition of \eqref{t-phi}. Before doing so, let us look at the characteristic polynomial \eqref{sec simple CH: characteristic polynomial} for \eqref{t-phi}:
\begin{align}\label{characteristic polynomial-SF}
&P(\lambda) = (\lambda_1 - \lambda)^3 (\lambda_2 - \lambda) \,;
&&\lambda_1 = 0 \,,
\qquad
\lambda_2 = Y_\phi \,.
\end{align}
There are two eigenvalues, one of which is three times degenerate. Looking at \Cref{sec class: table of CH degrees}, for the Cayley-Hamilton degree ${\boldsymbol{\rm CH}}_2$, the only possibilities are Type I $[1,(111)]$ and Type I $[(1,11)1]$. If $Y_\phi=0$ and $\partial_\mu\phi\neq0$, the gradient is null and the tensor is Type II $[(211)]$. If $\partial_\mu\phi=0$, then $t_{\mu\nu}=0$ and the tensor is Type I. It is easy to confirm this by the explicit Jordan decomposition of \eqref{t-phi} which gives
\begin{align}\label{CH2: Jordan forms-SF}
&\begin{array}{c}
\; \left(
\begin{array}{cccc}
Y_\phi & 0 & 0 & 0 \\
0 & 0 & 0 & 0 \\
0 & 0 & 0 & 0 \\
0 & 0 & 0 & 0 \\
\end{array}
\right) ,
\\[1cm]
\mbox{Type I;}\,\, [1,(111)]\,\,Y_\phi>0
\end{array}
&\begin{array}{c}
\; \left(
\begin{array}{cccc}
0 & 0 & 0 & 0 \\
0 & Y_\phi & 0 & 0 \\
0 & 0 & 0 & 0 \\
0 & 0 & 0 & 0 \\
\end{array}
\right) ,
\\[1cm]
\mbox{Type I;}\,\, [(1,11)1]\,\,Y_\phi<0
\end{array}
&\begin{array}{c}
\; \left(
\begin{array}{cccc}
0 & 1 & 0 & 0 \\
0 & 0 & 0 & 0 \\
0 & 0 & 0 & 0 \\
0 & 0 & 0 & 0 \\
\end{array}
\right) .
\\[1cm]
\mbox{Type II;}\,\, [(211)]\,\,Y_\phi=0,\ \partial_\mu\phi\neq0
\end{array}
\end{align}

The conditions for preserving the Lorentzian signature can be read from \eqref{det-condition} and \eqref{det-normal-form} as follows
\begin{align}\label{conditions-signature-SF}
&C (C + Y_\phi D) > 0 \,;
&& \mbox{if}\,\, Y_\phi \neq 0 \,,
\\
&
C \neq 0 \,;
&& \mbox{if}\,\, Y_\phi = 0 \,.
\end{align}

Let us now derive the mimetic energy-momentum tensor for the disformal transformation given in Eq.~\eqref{dis-T-S}. Since $t_{\mu\nu}$, defined in \eqref{t-phi}, is independent of $h_{\mu\nu}$, all the coefficients $c_i$ in Eq.~\eqref{Dth-contract-CH2} vanish, i.e., $c_0 = c_1 = c_2 = 0$. Using this together with \eqref{delta-i-phi} in Eq. \eqref{eigenvalue-system-CH2-final} yields
\begin{align}\label{eigenvalue-system-CH2-SF}
\begin{split}
aC\, h^{\mu\nu}
+
\big[
b \left(C-Y_\phi C_{Y_\phi}- Y_\phi^2 D_{Y_\phi}\right)-a \left(4 C_{Y_\phi}+Y_\phi D_{Y_\phi}\right)
\big] u^\mu u^\nu = 0 \,,
\end{split}
\end{align}
where we have used the fact that
\begin{align}\nonumber
C_{[t]} = C_{Y_\phi} \,,
\qquad
D_{[t]} = D_{Y_\phi} \,,
\qquad
C_{[t^2]} = 0 \,,
\qquad
D_{[t^2]} = 0 \,.
\end{align}
As $h^{\mu\nu}$ and $u^\mu u^\nu$ are independent and $C\neq0$, Eq. \eqref{eigenvalue-system-CH2-SF} has the following non-trivial solution
\begin{align}\label{sol-SF}
a = 0 \,,
\qquad
C-Y_\phi C_{Y_\phi}- Y_\phi^2 D_{Y_\phi} = 0 \,.
\end{align}
Substituting the above solution together with \eqref{t-phi} in \eqref{dual-tensor-CH2}, we find the dual eigentensor
\begin{align}\label{dual-tensor-SF}
\zeta_{\star}^{\mu\nu} = b\, u^\mu u^\nu \,.
\end{align}
On a local branch with $Y_\phi\neq0$, the last condition in \eqref{sol-SF} is the zero-mode condition \eqref{lambda-star} and integrates to
\begin{align}\label{lambda-0-S}
D = - \frac{C}{Y_\phi} - \frac{1}{\epsilon(\phi)} \,,
\end{align}
where $\epsilon(\phi)\neq0$ is arbitrary. It can be normalized locally to $\epsilon=\pm1$ by a field redefinition. The case $Y_\phi=0$ is not covered by this integration and must be treated separately. After the normalization $\epsilon^2=1$, substituting \eqref{dis-T-S} and \eqref{dual-tensor-SF} into \eqref{EMT-star} gives
\begin{align}\label{EMT-star-SF}
T^\star_{\mu\nu} = b \tilde{\rho}Y_\phi^2 \partial_\mu \phi \partial_\nu \phi \,,
\end{align}
where we used \eqref{lambda-0-S}. On the branch $Y_\phi\neq0$ considered here, both $t_{\mu\nu}$ and $T^\star_{\mu\nu}$ are Type I, including the exceptional case in which the overall coefficient of $T^\star_{\mu\nu}$ vanishes. The nonzero null branch $Y_\phi=0$ is not covered by the integration leading to \eqref{lambda-0-S} and requires a separate zero-mode analysis.

Let us further elaborate on the physical properties of the mimetic energy-momentum tensor \eqref{EMT-star-SF}. Since $T^\star_{\mu\nu}$ should be interpreted in terms of the physical metric $g_{\mu\nu}$, it is useful to define the counterpart of $Y_\phi$ in terms of the physical metric
\begin{align}\label{physical-invariant-SF}
X_\phi \equiv g^{\alpha\beta} \partial_\alpha \phi \partial_\beta \phi = \frac{Y_\phi}{C+DY_\phi} \,,
\end{align}
which obeys $X_\phi=-\epsilon$ on the branch \eqref{lambda-0-S}. All fluid contractions and covariant derivatives below use $g_{\mu\nu}$. The above result is well-known in the context of mimetic dark matter \cite{Golovnev:2013jxa}. Assuming that $\partial_\mu \phi$ is timelike, i.e., $\epsilon = -1$, the mimetic energy-momentum tensor can be rewritten as
\begin{align}\label{EMT-S}
T^\star_{\mu\nu} = {\rho}_{\phi} \, U_\mu U_\nu \,;
\qquad
U_\mu \equiv -\frac{\partial_\mu \phi}{\sqrt{X_\phi}} = - \partial_\mu \phi
\,,
\end{align}
where we defined $\rho_\phi\equiv{b}\tilde{\rho}Y_\phi^2$. Comparing with \eqref{fluid-perfect} and \eqref{fluid-decomposition-definitions}, one has $p=0$, $q_\mu=0$, and $\pi_{\mu\nu}=0$. The four-velocity $U_\mu$ satisfies $U_\mu U^\mu = 1$, with the scalar field acting as a velocity potential. The equation of motion for $\phi$ can be obtained from \eqref{EMT-star-cons} as
\begin{align}\label{EoM-phi}
\nabla_\mu\left( \rho_\phi \nabla^\mu \phi \right) = 0 \,.
\end{align}
Note that the four-velocity is a gradient of constant unit norm and thus satisfies the geodesic equation $U^\nu \nabla_\nu U^\mu = 0$. As shown above, the new scalar degree of freedom induced by the disformal transformation \eqref{dis-T-S} behaves like dust, with the energy-momentum tensor given in \eqref{EMT-S}. In cosmology, dust is often considered a candidate for dark matter, which is why this setup is referred to as mimetic dark matter. However, an important issue arises: the four-velocity of dust follows geodesics, and geodesic flows generically develop caustic singularities~\cite{Gorji:2020ten,Gorji:2025ajb}. While the formation of caustics is not a fundamental problem in the standard cosmological model, where dark matter is modeled as a pressureless perfect fluid, in the mimetic scenario, the scalar field $\phi$ is intended to provide a fundamental description. The emergence of caustics therefore poses a more serious concern. In the next subsection, we consider a gauge field instead of a scalar field; the induced energy-momentum tensor is then not tied to a single geodesic velocity-potential flow.

\subsection{Gauge field}\label{sec-example-gauge}

In this subsection, we consider the case where $t_{\mu\nu}$ is constructed from an antisymmetric rank-$2$ tensor $F_{\mu\nu}=-F_{\nu\mu}$ (2-form) \cite{Gorji:2018okn,Gumrukcuoglu:2019ebp,DeFelice:2019hxb} as
\begin{align}\label{dis-T-G}
g_{\mu\nu} = C(Y_1,Y_2)\, h_{\mu\nu} + D(Y_1,Y_2)\, h^{\alpha\beta} F_{\alpha\mu} F_{\beta\nu} \,,
\end{align}
where
\begin{align}\label{Y1-Y2}
Y_1 \equiv h^{\alpha\rho} h^{\beta\sigma} F_{\rho\sigma} F_{\alpha\beta}\,,
\qquad
Y_2\equiv F_{\mu\nu}\widetilde F_h^{\mu\nu}\,.
\end{align}
The Hodge duals associated with the two metrics are
\begin{align}\label{gauge-dual-def}
 \widetilde F_h^{\mu\nu}&\equiv\frac12\epsilon_h^{\mu\nu\alpha\beta}F_{\alpha\beta}\, ,
 &\epsilon_h^{\mu\nu\alpha\beta}&\equiv-\frac{[\mu\nu\alpha\beta]}{\sqrt{-h}}\, ,\nonumber\\
 \widetilde F_g^{\mu\nu}&\equiv\frac12\epsilon_g^{\mu\nu\alpha\beta}F_{\alpha\beta}\, ,
 &\epsilon_g^{\mu\nu\alpha\beta}&\equiv-\frac{[\mu\nu\alpha\beta]}{\sqrt{-g}}\, .
\end{align}
Here $[\mu\nu\alpha\beta]$ is the totally antisymmetric symbol, with the same orientation for both metrics. Indices on the auxiliary-frame field strengths are raised and lowered with $h_{\mu\nu}$.
	
Comparing \eqref{dis-T-G} with the normal form of the disformal transformation \eqref{DT-simplified}, we find
\begin{align}\label{t-GF}
t_{\mu\nu} = h^{\alpha\beta} F_{\alpha\mu} F_{\beta\nu} \,.
\end{align}
It is straightforward to show that \eqref{t-GF} satisfies the Cayley-Hamilton ${\boldsymbol{\rm CH}}_2$ condition \eqref{sec simple CH: CH2 condition} with
\begin{align}\label{delta-i-GF}
\delta_0 = \frac{1}{16}Y_2^2 \,,
\qquad
\delta_1 = \frac{1}{2} Y_1 \,,
\end{align}
where we have used the following identities
\begin{align}
F^{\lambda\mu} F_{\lambda\nu}
= \frac{1}{2} Y_1 \delta^\mu_\nu +  \widetilde F_h^{\lambda\mu}(\widetilde F_h)_{\lambda\nu} \,,
\qquad
\widetilde F_h^{\lambda\mu}F_{\lambda\nu} = \frac{1}{4} Y_2 \delta^\mu_\nu \,.
\end{align}
For the traces we find
\begin{align}\label{t-poly-G}
[t] = Y_1 \,,
\qquad
[t^2] = \frac{1}{2} Y_1^2 + \frac{1}{4} Y_2^2 \,,
\end{align}
which show that, contrary to the scalar field case, $[t]$ and $[t^2]$ are independent building blocks and we should consider $C$ and $D$ to be functions of both $[t]$ and $[t^2]$ or, on a fixed-sign patch with $Y_2\neq0$, equivalently $Y_1$ and $Y_2$.
	
The inverse contravariant metric for \eqref{dis-T-G} can be found from \eqref{dis-T-inv-G2} as follows
\begin{align}\label{dis-T-inv-G}
g^{\mu\nu} = \frac{(C+\delta _1 D)\, h^{\mu\nu} - D \, h^{\alpha\beta} F_\alpha{}^{\mu} F_{\beta}{}^{\nu}}{C^2+C \delta_1 D-\delta _0 D^2} \,,
\end{align}
where $\delta_{0,1}$ are given by \eqref{delta-i-GF}.
	
As we saw in the scalar field case (see \Cref{sec class: table of CH degrees}), it is sometimes possible to determine the Hawking-Ellis type of a disformal transformation solely from the Cayley-Hamilton degree. For the scalar field, the Jordan decomposition of \eqref{t-phi} is straightforward, whereas for the gauge field case \eqref{t-GF}, the decomposition is more involved. Fortunately, as we will show below, the explicit Jordan decomposition is not required in the generic non-null gauge-field case. This highlights the utility and convenience of the results summarized in \Cref{sec class: table of CH degrees}. The characteristic polynomial \eqref{sec simple CH: characteristic polynomial} for \eqref{t-GF} is given by
\begin{align}\label{characteristic polynomial-GF}
P(\lambda) \propto (\lambda_+ - \lambda)^2 (\lambda_- - \lambda)^2 \,;
\qquad
\lambda_\pm = \frac{1}{4}\left( Y_1\pm\sqrt{Y_1^2+Y_2^2}\right) \,.
\end{align}
There are two eigenvalues, each of which is two times degenerate. For $(Y_1,Y_2)\neq(0,0)$, looking at \Cref{sec class: table of CH degrees}, for the Cayley-Hamilton degree ${\boldsymbol{\rm CH}}_2$, the corresponding possibility is Type I $[(1,1)(11)]$
\begin{align}\label{CH2: Jordan forms-GF}
&\begin{array}{c}
		\left(\begin{array}{cccc}
			\lambda_-&0&0&0\\
			0&\lambda_-&0&0\\
			0&0&\lambda_+&0\\
			0&0&0&\lambda_+
		\end{array}\right)\\[2mm]
		\mbox{Type I;}\ [(1,1)(11)]
\end{array}
\end{align}
One can explicitly check that the results \eqref{t-poly-G} are consistent with \eqref{CH2-t-t2}. In the null case $Y_1=Y_2=0$, however, one has $t^2_{\mu\nu}=0$ while $t_{\mu\nu}\neq0$, so the tensor belongs instead to the Type II subclass $[(211)]$.
	
The conditions for preserving the Lorentzian signature can be read from \eqref{det-condition} and \eqref{det-normal-form} as follows. Since both eigenvalues are twofold degenerate, ${\det g}/{\det h}=\left[\left(C+D\lambda_+\right)\left(C+D\lambda_-\right)\right]^2$ is a perfect square and \eqref{det-condition} only demands that it does not vanish
\begin{align}\label{conditions-signature-GF}
C + D \lambda_{\pm} \neq 0
\qquad \Longleftrightarrow \qquad
C^2 + \frac{1}{2} C D\, Y_1 - \frac{1}{16} D^2 Y_2^2 \neq 0 \,,
\end{align}
which is nothing but the non-vanishing of the denominator of the inverse metric \eqref{dis-T-inv-G}. The null case $Y_1=Y_2=0$ is included and simply gives $C\neq0$.
	
Now, let us find the mimetic energy-momentum tensor for the disformal transformation \eqref{dis-T-G}. For the sake of simplicity, we restrict our setup to conformal transformations with $D=0$ and $C>0$. Contrary to the scalar field case \eqref{t-phi}, \eqref{t-GF} depends on $h_{\mu\nu}$, and the coefficients $c_i$ in \eqref{Dth-contract-CH2} can be found as
\begin{align}
c_0 = - \delta_0 \,,
\qquad
c_1 = -\delta_1 \,,
\qquad
c_2 = -1 \,.
\end{align}
Using the above results together with \eqref{delta-i-GF} in Eq. \eqref{eigenvalue-system-CH2-final} yields
\begin{align}\label{eigenvalue-system-CH2-GF}
\begin{split}
\Big[a C-\frac{1}{2} Y_2 C_{Y_2} \left(4 a+b Y_1\right)\Big] h^{\mu\nu}
+
\Big[
b \left(C-2 Y_1 C_{Y_1}\right)-8 a C_{Y_1}
\Big] h_{\alpha\beta} F^{\alpha\mu} F^{\beta\nu} = 0 \,,
\end{split}
\end{align}
where we have used the fact that
\begin{align}\nonumber
C_{[t]} = C_{Y_1} - 2 \left(\frac{Y_1}{Y_2}\right) C_{Y_2} \,,
\qquad
C_{[t^2]} = \frac{2}{Y_2} C_{Y_2} \,.
\end{align}
Provided $Y_2 C_{Y_2}\neq0$, Eq. \eqref{eigenvalue-system-CH2-GF} has the following non-trivial solution
\begin{align}\label{sol-GF}
b = \frac{4 C_{Y_1}}{Y_2 C_{Y_2}} a \,,
\qquad
C - 2Y_1 C_{Y_1} - 2Y_2 C_{Y_2} = 0 \,.
\end{align}
Substituting the above solution together with \eqref{t-GF} in \eqref{dual-tensor-CH2}, we find the dual eigentensor
\begin{align}\label{dual-tensor-GF}
\zeta_{\star}^{\mu\nu} = a \left( h^{\mu\nu} + \frac{4 C_{Y_1}}{Y_2 C_{Y_2}} \, h_{\alpha\beta} F^{\alpha\mu} F^{\beta\nu} \right) \,.
\end{align}
The last condition in \eqref{sol-GF} is nothing but the condition $\lambda_\star=0$ in Eq. \eqref{lambda-star} which can be integrated to give
\begin{align}\label{lambda-0-GF}
C = \left(\sqrt{\epsilon_1 Y_1} + \sqrt{\epsilon_2 Y_2} \right) f_{\rm G}\left(\frac{Y_2}{Y_1}\right) \,,
\end{align}
where $f_{\rm G}$ is an arbitrary non-vanishing function and the square roots are understood on a local branch where they are real. Substituting \eqref{dual-tensor-GF} in \eqref{EMT-star} we find
\begin{align}\label{EMT-star-GF}
T^\star_{\mu\nu} = \bar{\tilde{\rho}} C \left(
\frac{1}{4} X_2 C_{X_2} g_{\mu\nu}
+ C_{X_1} g^{\alpha\beta} F_{\beta\mu} F_{\alpha\nu} \right) \,;
\qquad
\bar{\tilde{\rho}} \equiv \frac{4a\tilde{\rho}}{X_2C_{X_2}} \,,
\end{align}
where
\begin{align}\label{X1-X2}
X_1 \equiv g^{\alpha\rho} g^{\beta\sigma} F_{\rho\sigma} F_{\alpha\beta} = C^{-2} Y_1 \,,
\qquad
X_2\equiv F_{\mu\nu}\widetilde F_g^{\mu\nu} = C^{-2} Y_2 \,.
\end{align}
The last equalities in \eqref{X1-X2} use the conformal relation $g_{\mu\nu}=Ch_{\mu\nu}$. More generally, \eqref{gauge-dual-def} gives
\begin{align}\label{gauge-volume-invariant}
 \sqrt{-g}\,X_2=\sqrt{-h}\,Y_2 =-\frac12[\mu\nu\alpha\beta]F_{\mu\nu}F_{\alpha\beta}\, .
\end{align}
Thus the density in \eqref{gauge-volume-invariant} is metric independent even for a non-conformal map. Here $C_{X_i}$ denotes the derivative of the same function $C$, evaluated at $(X_1,X_2)$ before imposing the mimetic constraint. Indices in the physical-frame source and conservation equation are raised with $g^{\mu\nu}$. Note that, up to the overall undetermined factor ${\bar{\tilde{\rho}}}C$, the energy-momentum tensor is completely written in terms of $g_{\mu\nu}$. Moreover, in the generic non-null branch it has the same Hawking-Ellis class as the disformal tensor \eqref{t-GF} shown in \eqref{CH2: Jordan forms-GF}.
	
Since the generic non-null branch is Type I, its physical properties can also be described using the fluid decomposition reviewed in appendix~\ref{app:fluid}. In the principal frame, Eqs.~\eqref{fluid-decomposition-general} and \eqref{fluid-principal-degenerate} show that the energy flux vanishes and the two doubly degenerate eigenspaces give an anisotropic fluid with $p_\parallel=-\rho$ and equal transverse pressures.
	
Solution \eqref{lambda-0-GF} together with relations \eqref{X1-X2} implies
\begin{align}\label{mimetic-constraint-GF}
\left(\sqrt{\epsilon_1 X_1} + \sqrt{\epsilon_2 X_2} \right) f_{\rm G}\left(\frac{X_2}{X_1}\right) = 1 \,.
\end{align}
This is because \eqref{lambda-0-GF} makes $C$ homogeneous of degree $1/2$, one then has $C_{X_i}=C\,C_{Y_i}$ and $C(X_1,X_2)=1$ which is nothing but \eqref{mimetic-constraint-GF}. 
	
Note also that \eqref{sol-GF} gives
\begin{align}\label{sol-GF-X}
1 - 2X_1 C_{X_1} - 2X_2 C_{X_2} = 0 \,.
\end{align}
	
The equation of motion for $F_{\mu\nu}$ can be obtained from \eqref{EMT-star-cons} as follows
\begin{align}\label{EoM-G}
\nabla_\mu \left[ \bar{\tilde \rho} C \left( \frac{1}{4} X_2 C_{X_2} g^{\mu\nu} + C_{X_1} F^{\lambda\mu} F_{\lambda}{}^{\nu} \right) \right] = 0 \,.
\end{align}

The family \eqref{lambda-0-GF} was already found in Ref.~\cite{Colleaux:2025vtm}. Related invertibility conditions were studied in Ref.~\cite{Gumrukcuoglu:2019ebp}, while the one-invariant limits appeared in
Refs.~\cite{Gorji:2018okn,Hammer:2020dqp}. Since it contains an arbitrary function $f_{\rm G}$, it describes a broad class of singular gauge-field transformations with potentially different physical properties. A systematic study of this general family is beyond the scope of the present work and will be left for future work. Here, to better understand the mimetic energy-momentum tensor \eqref{EMT-star-GF}, we focus on two simple subsets which isolate the roles of the terms $g^{\alpha\beta}F_{\beta\mu}F_{\alpha\nu}$ and $g_{\mu\nu}$. The branch $C=C(Y_1)$ is singular in the ratio in \eqref{sol-GF} because $C_{Y_2}=0$, while the branch $C=C(Y_2)$ sets $b=0$ and removes the $g^{\alpha\beta}F_{\beta\mu}F_{\alpha\nu}$ contribution. They must therefore be treated separately.	
	
\subsubsection{\texorpdfstring{$C=C(Y_1)$}{C=C(Y1)}}
As seen from \eqref{lambda-0-GF}, the branch $C = C(Y_1)$ corresponds to the special case with $C_{Y_2}=0$. In this case, Eq. \eqref{eigenvalue-system-CH2-GF} gives
\begin{align}
a=0 \,,
\qquad
C - 2Y_1 C_{Y_1}=0 \,,
\nonumber
\end{align}
while $b$ remains arbitrary. The mimetic constraint \eqref{mimetic-constraint-GF} then simplifies to
\begin{align}\label{mimetic-constraint-GF-X1}
\epsilon_1 X_1 = 1 \,.
\end{align}
After absorbing the overall normalization into $\bar{\tilde \rho}$, the mimetic energy-momentum tensor takes the form
\begin{align}\label{EMT-star-GF-X1}
T^\star_{\mu\nu} = \frac{1}{2} \epsilon_1 \bar{\tilde{\rho}} C \, g^{\alpha\beta} F_{\beta\mu} F_{\alpha\nu} \,.
\end{align}
	
This energy-momentum tensor was derived in Refs.~\cite{Gorji:2018okn,Gorji:2019ttx}, where it was shown that generalizing the setup to non-Abelian $SU(2)$ or global $O(3)$ symmetries, allowing for an isotropic background, results in a mimetic energy-momentum tensor component that effectively mimics spatial curvature in a cosmological background.
	
\subsubsection{\texorpdfstring{$C=C(Y_2)$}{C=C(Y2)}}
Now, we focus on the branch $C = C(Y_2)$, for which $C_{Y_1}=0$. Equation \eqref{eigenvalue-system-CH2-GF} then gives
\begin{align}
b=0 \,,
\qquad
C - 2Y_2 C_{Y_2}=0 \,,
\nonumber
\end{align}
while $a$ remains arbitrary. The mimetic constraint \eqref{mimetic-constraint-GF} simplifies to
\begin{align}\label{mimetic-constraint-GF-X2}
\epsilon_2 X_2 = 1 \,.
\end{align}
The conservation law \eqref{EMT-star-cons} implies that the overall coefficient is constant, and therefore the mimetic energy-momentum tensor reduces to
\begin{align}\label{EMT-star-GF-X2}
T^\star_{\mu\nu} = - \Lambda g_{\mu\nu} \,;
\qquad
\Lambda = \mathrm{constant} \,.
\end{align}
Comparing with \eqref{fluid-perfect}, one has $\rho=-\Lambda$ and $p=\Lambda$. This is the result found in Ref. \cite{Hammer:2020dqp}.

\subsection{Scalar and gauge fields}\label{sec-example-scalar-gauge}

We now combine the scalar disformal tensor \eqref{t-phi} with the gauge
invariant $Y_2$ defined in \eqref{Y1-Y2}, allowing $D$ to depend on a quantity not determined by the traces of the scalar tensor. The corresponding conformal
construction was studied in Ref.~\cite{Jirousek:2022mke}. Setting $C=1$
in \eqref{dis-T-S}, we take
\begin{align}\label{dis-T-SG}
 g_{\mu\nu}=h_{\mu\nu}+D(Y_\phi,Y_2)u_\mu u_\nu\, .
\end{align}
Here $u_\mu$ and $Y_\phi$ retain the definitions in \eqref{t-phi} and
\eqref{invariants-SV}, while $F_{\mu\nu}=2\partial_{[\mu}A_{\nu]}$ is
an Abelian field strength. The scalar and gauge fields enter through
the transformation, with no independent scalar or Maxwell term in
the seed action. The inverse follows from \eqref{dis-T-inv-S}. We work
on a local branch with $Y_\phi>0$, $Y_2>0$, and
\begin{align}\label{regularity-SG}
 {\cal A}\equiv1+Y_\phi D>0\, ,\qquad
 \frac{\det g}{\det h}={\cal A}\, .
\end{align}
Thus \eqref{conditions-signature-SF} guarantees that the physical metric
is non-degenerate and Lorentzian.

Using \eqref{J-def} and \eqref{lambda-star}, with the additional
$Y_2$ dependence included, the singularity condition is
\begin{align}\label{singularity-SG}
 Y_\phi^2D_{Y_\phi}+\frac{Y_\phi Y_2}{2}D_{Y_2}=1\, .
\end{align}
For $D_{Y_2}=0$, this reduces to \eqref{sol-SF} with $C=1$.
Its general local solution within \eqref{dis-T-SG} is
\begin{align}\label{D-general-SG}
 D=-\frac1{Y_\phi}+f_{\rm SG}(s)\, ,\qquad
 s\equiv\frac{Y_2}{\sqrt{Y_\phi}}\, ,
\end{align}
where $f_{\rm SG}$ is an arbitrary differentiable function and
${\cal A}=Y_\phi f_{\rm SG}(s)>0$. A dual zero mode is
\begin{align}\label{dual-tensor-SG}
 \zeta_\star^{\mu\nu}
 =\left(\frac1{Y_\phi^2}-\frac{s f'_{\rm SG}(s)}{2Y_\phi}\right)u^\mu u^\nu
 +\frac{s f'_{\rm SG}(s)}2h^{\mu\nu}\, .
\end{align}
This has the form \eqref{zeta-CH2}, with indices raised as in
\eqref{dis-T-inv-S}. Although $D$ depends on the additional invariant $Y_2$, its full metric variation remains a linear combination of $h^{\mu\nu}$ and $t^{\mu\nu}$, as follows from \eqref{metric-variations-SG}. Thus, the polynomial condition \eqref{poly-response-app} is satisfied.
The derivation in appendix~\ref{app:scalar-gauge} also shows that this
mode spans the full Jacobian kernel.

The physical scalar contraction $X_\phi$ of \eqref{physical-invariant-SF}
and gauge invariant $X_2$ of \eqref{X1-X2} now obey
\begin{align}\label{physical-frame-SG}
 X_\phi=\frac{Y_\phi}{{\cal A}}=\frac1{f_{\rm SG}(s)}\, ,\qquad
 X_2=\frac{Y_2}{\sqrt{\cal A}}\, ,\qquad
 U_\mu=-\frac{u_\mu}{\sqrt{X_\phi}}\, .
\end{align}
The second relation follows from \eqref{gauge-volume-invariant}, not
from the conformal scaling in \eqref{X1-X2}. The velocity is normalized
with $g^{\mu\nu}$, as in \eqref{EMT-S}. Substituting
\eqref{dual-tensor-SG} in \eqref{EMT-star} gives the perfect-fluid form
\eqref{fluid-perfect},
\begin{align}\label{EMT-star-SG}
 T^\star_{\mu\nu}&=(\rho+p)U_\mu U_\nu-pg_{\mu\nu}\, ,\nonumber\\
 \rho&=\tilde\rho f_{\rm SG}(s)\, ,\qquad
 p=-\frac{\tilde\rho}{2}s f'_{\rm SG}(s)\, .
\end{align}

The projections \eqref{fluid-decomposition-definitions} give
$q_\mu=0$ and $\pi_{\mu\nu}=0$. In particular, the choice
\begin{align}\label{D-w-SG}
 f_{\rm SG}(s)=b s^{-2w}\, ,\qquad
 D^{(w)}=bY_\phi^wY_2^{-2w}-\frac1{Y_\phi}\, ,\qquad p=w\rho\, ,
\end{align}
with constants $b>0$ and $w$, gives a constant equation of state.
The constant $b$ has the dimensions required by \eqref{D-w-SG}.
The case $w=0$ recovers the dust form \eqref{EMT-S}, while $w=-1$
gives the metric-proportional source \eqref{EMT-star-GF-X2}, with
$\rho=-\Lambda$ constant by \eqref{EMT-star-cons}.

The radiation case $w=1/3$ is also regular. Its constraint and
energy-momentum tensor are
\begin{align}\label{radiation-SG}
	X_2=b^{3/2}X_\phi^2\, ,\qquad
	T^\star_{\mu\nu}=\frac{\rho}{3}
	\left(4U_\mu U_\nu-g_{\mu\nu}\right)\, .
\end{align}
This is an isotropic radiation perfect fluid. Its field equations and
the FLRW dilution law $\rho\propto a^{-4}$ are derived in
appendix~\ref{app:scalar-gauge}. To our knowledge, the explicit scalar-gauge family
\eqref{D-general-SG}, including the regular radiation realization
\eqref{radiation-SG}, has not been presented before.

Since \eqref{EMT-star-SG} has the perfect-fluid form
\eqref{fluid-perfect}, $T^\star_{\mu\nu}$ is Type I on the timelike
branch considered here. The scalar disformal tensor
$t_{\mu\nu}=u_\mu u_\nu$ is also Type I for $Y_\phi>0$, as shown in
\eqref{CH2: Jordan forms-SF}. Thus this example preserves the
Hawking-Ellis type and does not realize the Type II and Type IV
reductions in \Cref{table-EMT}. We turn to two scalar fields to
illustrate these changes explicitly.

\subsection{Two scalar fields}\label{sec-example-two-scalars}

A disformal transformation involving two scalar fields was considered in
Ref.~\cite{Firouzjahi:2018xob}. Here we consider the normal form
\eqref{DT-simplified} with
\begin{align}\label{dis-T-2SF}
g_{\mu\nu}=C h_{\mu\nu}+D t_{\mu\nu} \, , 
\qquad
t_{\mu\nu}=2\partial_{(\mu}\phi\partial_{\nu)}\chi 
=u_\mu v_\nu+v_\mu u_\nu \, ,
\end{align}
where $u_\mu$ and $Y_\phi$ are defined in \eqref{t-phi} and
\eqref{invariants-SV}. For the second scalar and mixed contraction, define
\begin{align}\label{invariants-2SF}
 v_\mu\equiv\partial_\mu\chi\, ,\qquad
 Y_\chi\equiv h^{\mu\nu}v_\mu v_\nu\, ,\qquad
 Y_{\phi\chi}\equiv h^{\mu\nu}u_\mu v_\nu\, .
\end{align}
As in \eqref{dis-T-inv-S}, indices on the gradients and powers of
$t_{\mu\nu}$ are raised with $h^{\mu\nu}$, in particular
$v^\mu=h^{\mu\nu}v_\nu$. Possible explicit dependence
of the coefficient functions on $\phi$ and $\chi$ is left implicit.
Using \eqref{t-HO}, we find
\begin{align}\label{powers-2SF}
t^2_{\mu\nu}&=Y_\chi u_\mu u_\nu+Y_\phi v_\mu v_\nu+Y_{\phi\chi} t_{\mu\nu} \, ,
\nonumber\\
t^3_{\mu\nu}&=2Y_{\phi\chi} t^2_{\mu\nu}+(Y_\phi Y_\chi-Y_{\phi\chi}^2)t_{\mu\nu} \, .
\end{align}
Thus the ${\boldsymbol{\rm CH}}_3$ condition
\eqref{sec simple CH: CH3 condition} holds with
\begin{align}\label{delta-2SF}
\delta_0=0 \, ,\qquad \delta_1=Y_\phi Y_\chi-Y_{\phi\chi}^2 \, ,\qquad \delta_2=2Y_{\phi\chi} \, .
\end{align}
The traces \eqref{t-Trace-def} and characteristic polynomial \eqref{sec simple CH: characteristic polynomial} give
\begin{align}\label{traces-2SF}
I_1\equiv[t]=2Y_{\phi\chi} \, ,\qquad
I_2\equiv[t^2]=2(Y_{\phi\chi}^2+Y_\phi Y_\chi) \, ,
\end{align}
\begin{align}\label{characteristic-2SF}
P(\lambda)=\lambda^2\big[(\lambda-Y_{\phi\chi})^2-Y_\phi Y_\chi\big] \, ,
\qquad \lambda_0=0 \, ,\qquad\lambda_\pm=Y_{\phi\chi}\pm\sqrt{Y_\phi Y_\chi} \, .
\end{align}
Unlike the scalar and gauge field examples, this tensor realizes all four
Hawking-Ellis types. The Jordan structures and simple field configurations
are given in appendix~\ref{app:two-scalar-types} and \Cref{table-2SF}.

Substituting \eqref{delta-2SF} into \eqref{sol-inverse-G3} gives
\begin{align}\label{inverse-2SF}
g^{\mu\nu}=\frac{1}{C}h^{\mu\nu}
-\frac{D(C+2DY_{\phi\chi})}{C{\cal Q}}t^{\mu\nu}
+\frac{D^2}{C{\cal Q}}t^{2\,\mu\nu} \, ,
\end{align}
where
\begin{align}\label{Q-2SF}
{\cal Q}\equiv(C+DY_{\phi\chi})^2-D^2Y_\phi Y_\chi \, .
\end{align}

Equation \eqref{det-normal-form} yields
\begin{align}\label{det-2SF}
\frac{\det g}{\det h}=C^2{\cal Q} \, ,
\end{align}
so the Lorentzian condition \eqref{det-condition} is $C^2{\cal Q}>0$.

Let us now find the mimetic energy-momentum tensor. As in the gauge field
case, we restrict to conformal transformations with $D=0$ and $C>0$.
We first allow $C=C(Y_\phi,Y_\chi,Y_{\phi\chi})$, relaxing the trace-dependence and polynomial restrictions used in the general analysis. Including these scalar arguments in \eqref{J-def} and using \eqref{lambda-star}, we find
\begin{align}\label{singular-conformal-2SF}
C-Y_\phi C_{Y_\phi}-Y_\chi C_{Y_\chi}-Y_{\phi\chi} C_{Y_{\phi\chi}}=0 \, ,
\end{align}
\begin{align}\label{dual-tensor-2SF}
\zeta_\star^{\mu\nu}
=b_\star\left(C_{Y_\phi} u^\mu u^\nu+C_{Y_\chi} v^\mu v^\nu
+\frac{C_{Y_{\phi\chi}}}{2}t^{\mu\nu}\right) \, ,
\end{align}
where $b_\star$ is arbitrary. The homogeneity condition
\eqref{singular-conformal-2SF}, derived in Ref.~\cite{Firouzjahi:2018xob},
integrates locally, for $Y_\phi\neq0$, to
\begin{align}\label{general-C-XYZ-2SF}
C=Y_\phi f_{\rm 2S}\!\left(\frac{Y_\chi}{Y_\phi},\frac{Y_{\phi\chi}}{Y_\phi}\right) \, ,
\end{align}
with $f_{\rm 2S}$ arbitrary, as obtained in Ref.~\cite{Zheng:2022vwm}.
Other nonzero invariants can be used instead of $Y_\phi$ on the remaining
patches. Extending $X_\phi$ from \eqref{physical-invariant-SF} to the second scalar,
\begin{align}\label{physical-invariants-2SF}
 X_\phi&=\frac{Y_\phi}{C}\, ,\qquad
 X_\chi\equiv g^{\mu\nu}v_\mu v_\nu=\frac{Y_\chi}{C}\, ,\nonumber\\
 X_{\phi\chi}&\equiv g^{\mu\nu}u_\mu v_\nu=\frac{Y_{\phi\chi}}{C}\, ,
\end{align}
homogeneity gives the mimetic constraint
\begin{align}\label{mimetic-constraint-2SF}
C(X_\phi,X_\chi,X_{\phi\chi})=X_\phi f_{\rm 2S}\!\left(\frac{X_\chi}{X_\phi},\frac{X_{\phi\chi}}{X_\phi}\right)=1 \, .
\end{align}
Substituting \eqref{dual-tensor-2SF} into \eqref{EMT-star}, and absorbing
the overall factor into ${\cal N}\equiv b_\star\tilde\rho C^2$, gives
\begin{align}\label{EMT-star-2SF}
T^\star_{\mu\nu} 
={\cal N}\left(C_{X_\phi} u_\mu u_\nu+C_{X_\chi} v_\mu v_\nu +\frac{C_{X_{\phi\chi}}}{2}t_{\mu\nu}\right) \, , \qquad T^{\star\mu}{}_{\mu}=g^{\mu\nu}T^\star_{\mu\nu} ={\cal N} \, .
\end{align}
Here $C_{X_\phi}$, $C_{X_\chi}$, and $C_{X_{\phi\chi}}$ are derivatives
of the same function $C$, evaluated at the physical contractions
\eqref{physical-invariants-2SF} before imposing
\eqref{mimetic-constraint-2SF}. This is the general mimetic tensor for
the homogeneous two-field conformal sector \cite{Zheng:2022vwm}. The terms $u_\mu u_\nu$ and $v_\mu v_\nu$ have independent coefficients, whereas \eqref{powers-2SF} fixes their ratio in a polynomial of $t^\mu{}_\nu$. This is why the polynomial conditions in \Cref{app:polynomial} must be checked separately from trace dependence. For this example, we apply \Cref{table-EMT} only to the trace-dependent restriction considered below.

To understand its physical properties, suppose that the two independent
gradients span a Lorentzian plane with respect to the physical metric,
$X_\phi X_\chi-X_{\phi\chi}^2<0$, equivalently $Y_\phi Y_\chi-Y_{\phi\chi}^2<0$, since the two metrics are conformally related. Choose the orthonormal frame of \eqref{fluid-two-plane} on this plane.
Equation \eqref{EMT-star-2SF} then takes the form
\begin{align}\label{2SF-fluid-main}
T^\star_{\mu\nu}=\rho U_\mu U_\nu +2q_\parallel U_{(\mu}n_{\nu)}+p_\parallel n_\mu n_\nu \, ,
\qquad 
\rho-p_\parallel={\cal N} \,,
\end{align}
where $\rho$, $q_\parallel$, and $p_\parallel$ are the projections \eqref{fluid-two-plane-definitions} of $T^\star_{\mu\nu}$. The mean pressure, energy flux, and anisotropic stress follow from \eqref{fluid-two-plane-projections}. Both transverse pressures vanish. The source is therefore generally anisotropic, with a flux along $n_\mu$, rather than a perfect fluid.
The regular single-field limit, and more generally parallel timelike gradients, recover the dust form \eqref{EMT-S}, provided the stress tensor is regular and the physical metric remains non-degenerate. This also explains the dust-like behavior for homogeneous scalar fields found in Refs.~\cite{Firouzjahi:2018xob,Zheng:2022vwm}.

\subsubsection{Linear and monomial conformal factors}

The linear family studied in Ref.~\cite{Firouzjahi:2018xob} is
$C=aY_\phi+bY_\chi+2cY_{\phi\chi}$. In particular, for $c=0$, Eq.~\eqref{EMT-star-2SF} gives
\begin{align}\label{linear-EMT-2SF}
aX_\phi+bX_\chi=1 \, ,
\qquad
T^\star_{\mu\nu}={\cal N}(a u_\mu u_\nu+b v_\mu v_\nu) \, .
\end{align}
For $a,b>0$, $X_\phi,X_\chi>0$, and ${\cal N}>0$, this has the algebraic form of two
pressureless streams. Their relative motion produces a longitudinal pressure in the energy frame. The nonlinear family
$C\propto Y_\phi^{m_1}Y_\chi^{m_2}Y_{\phi\chi}^{1-m_1-m_2}$ of Ref.~\cite{Zheng:2022vwm}, with constant exponents $m_1,m_2$, is also contained in \eqref{general-C-XYZ-2SF}. For example, $C=c_{\rm m}\sqrt{Y_\phi Y_\chi}$ with $c_{\rm m}>0$ on $Y_\phi,Y_\chi>0$ gives
\begin{align}\label{2SF-two-stream-T}
T^\star_{\mu\nu}=\frac{{\cal N}}{2} \left(\frac{u_\mu u_\nu}{X_\phi}+\frac{v_\mu v_\nu}{X_\chi}\right) \, .
\end{align}
These algebraic identifications do not require the two contributions to be separately conserved. 

\subsubsection{Trace-dependent conformal factor}

We now return to trace-dependent coefficients, $C=C(I_1,I_2)$. Since $t_{\mu\nu}$ is metric independent and $C>0$, \eqref{poly-forced-app} ensures that every dual zero mode is polynomial. Equation \eqref{traces-2SF} imposes the additional condition
$Y_\phi C_{Y_\phi}=Y_\chi C_{Y_\chi}$. On a patch with
\begin{align}\label{S-2SF}
{\cal S}\equiv\sqrt{I_1^2-I_2}=\sqrt{2(Y_{\phi\chi}^2-Y_\phi Y_\chi)}>0 \, ,
\end{align}
the general local solution becomes
\begin{align}\label{general-C-2SF}
C={\cal S}f_{\rm tr}\!\left(\frac{I_1}{{\cal S}}\right) \, ,
\end{align}
where $f_{\rm tr}$ is arbitrary. Equation \eqref{EMT-star-2SF} now has precisely the
polynomial structure \eqref{poly-map-Tstar},
\begin{align}\label{poly-conformal}
T^{\star\mu}{}_{\nu}=\frac{{\cal N}}{C}p_C(t)^\mu{}_{\nu} \, , 
\qquad 
p_C(\lambda)=C_{I_1}\lambda+2C_{I_2}\lambda^2 \, ,
\end{align}
where $C_{I_1}$ and $C_{I_2}$ are evaluated at the traces
\eqref{traces-2SF} computed with $h_{\mu\nu}$. To illustrate the type changes, consider
\begin{align}\label{C-alpha-2SF}
C^{(\alpha)}={\cal S}+\alpha I_1>0 \, ,
\end{align}
where $\alpha$ is constant. The corresponding polynomial is
\begin{align}\label{q-alpha-2SF}
p_\alpha(\lambda)=\left(\frac{I_1}{{\cal S}}+\alpha\right)\lambda -\frac{\lambda^2}{{\cal S}} \, .
\end{align}
For the Type II branch $Y_\phi=0$, $Y_\chi\neq0$, $Y_{\phi\chi}\neq0$,
\eqref{matrix-2SF} and \eqref{traces-2SF} give the block eigenvalue
$\lambda_1=Y_{\phi\chi}=I_1/2$. Equation \eqref{q-alpha-2SF} therefore gives
\begin{align}\label{derivative-II-2SF}
p_\alpha'(\lambda_1)=\frac{I_1-2\lambda_1}{{\cal S}}+\alpha=\alpha \, .
\end{align}
For the Type IV branch $Y_\phi Y_\chi<0$,
\eqref{characteristic-2SF} and \eqref{traces-2SF} give
$\lambda_++\lambda_-=I_1$. The divided difference \eqref{divided-difference}
for \eqref{q-alpha-2SF} is then
\begin{align}\label{Delta-IV-2SF}
\Delta_3=\frac{p_\alpha(\lambda_+)-p_\alpha(\lambda_-)}{\lambda_+-\lambda_-}=\alpha \, .
\end{align}
Thus \eqref{poly-map-blocks-2} and \eqref{divided-difference} immediately give
\begin{align}\label{transition-II-2SF}
\mathrm{Type~II}&\ \to
\begin{cases}
\mathrm{Type~II} & \alpha\neq0 \, ,\\
\mathrm{Type~I} & \alpha=0 \, ,
\end{cases}
\\ 
\label{transition-IV-2SF}
\mathrm{Type~IV}&\ \to
\begin{cases}
\mathrm{Type~IV} & \alpha\neq0 \, ,\\
\mathrm{Type~I} & \alpha=0 \, .
\end{cases}
\end{align}
Here ${\cal N}\neq0$ is understood. These are the reductions summarized in
\Cref{table-EMT}. 

The $\alpha=0$ branch also has a simple matter interpretation,
\begin{align}\label{2SF-string-T}
T^\star_{\mu\nu}=\frac{{\cal N}}{2} (U_\mu U_\nu-n_\mu n_\nu) \, .
\end{align}
The projections \eqref{fluid-two-plane-definitions} give
$\rho={\cal N}/2$, $p_\parallel=-\rho$, and $q_\parallel=0$, with
vanishing transverse pressures. Unlike the vacuum-energy source
\eqref{EMT-star-GF-X2}, this tensor has the anisotropic stress in
\eqref{fluid-two-plane-projections}. Its algebraic form coincides with
that of a cloud of geometric strings, for which the longitudinal tension
equals the energy density \cite{Letelier:1979ej}.

For the Type III representative in \Cref{table-2SF}, one has $I_1=I_2=0$ and therefore ${\cal S}=0$. The particular family \eqref{C-alpha-2SF} then gives $C^{(\alpha)}=0$, so the corresponding conformal metric is degenerate. Thus, while the two-scalar tensor itself realizes Type III, the singular conformal family used above does not provide a non-degenerate Type III example. This does not affect the general possibilities summarized in \Cref{table-EMT}.


\section{Summary}\label{summary}

Invertible disformal transformations are very useful tools in the context of modified gravity theories like scalar-tensor and vector-tensor theories. Non-singular disformal transformations can be used to clarify the relation or equivalence between apparently different theories, while singular disformal transformations can yield new theories. In this paper, being agnostic about the field content of a gravitational theory, we have considered a general disformal transformation of the form \eqref{sec inv: general disformal transformation} or \eqref{DT-simplified}. Using the Cayley-Hamilton theorem, we find an explicit expression for the corresponding inverse disformed metric, with the general formula given in \eqref{sol-inverse}. This result makes it possible to systematically find the inverse disformed metric for any consistent disformal transformation. We separately studied cases with simpler Cayley-Hamilton degrees as defined in \eqref{sec simple CH: CH3 condition} and \eqref{sec simple CH: CH2 condition}. The latter includes many disformal transformations that are studied in the literature. We also separated the pointwise determinant test from the stronger two-metric requirement: in four dimensions, $\det g/\det h>0$ is necessary and sufficient for Lorentzian inertia up to overall sign, while the same $(+,-,-,-)$ convention and local causal compatibility hold if and only if $C+D\lambda_i>0$ for every real eigenvalue of $t^{\mu}{}_\nu$; the Type IV complex pair adds no inequality.

Implementing the Hawking-Ellis classification, we classified disformal transformations into four types, Type I, II, III, IV and by looking at the eigenvalues of the system, we further classified each type into the corresponding Segre subclasses. For singular transformations, we found the mimetic energy-momentum tensor $T^\star_{\mu\nu}$ for the polynomial modes of the Jacobian kernel, see Eqs.~\eqref{EMT-CH4},~\eqref{EMT-CH3},~\eqref{EMT-CH2}. In \Cref{app:polynomial}, we derived sufficient conditions under which all dual zero modes have this polynomial form. Within this sector, the Hawking-Ellis type is always preserved for Type I. For Types II and III it can change according to the vanishing of the nilpotent coefficients in the polynomial relating $T^{\star\mu}{}_{\nu}$ to $t^\mu{}_\nu$. Type IV is preserved unless the complex-conjugate pair collapses to a repeated real eigenvalue, in which case it reduces to Type I. We found explicit links between the Cayley-Hamilton degree of the disformal tensor and its Hawking-Ellis type shown in \Cref{sec class: table of CH degrees}. Based on this result, by knowing the Cayley-Hamilton degree, we can determine possible Hawking-Ellis types. The link is very restrictive such that, in some cases, it is possible to determine the Hawking-Ellis type by knowing only the Cayley-Hamilton degree and without performing an explicit Jordan decomposition. This is useful in practice since performing Jordan decomposition is usually more cumbersome than finding the Cayley-Hamilton degree.

We applied our setup first to the two widely studied scalar and gauge field examples. In the gauge-field case, the general singularity condition can be integrated explicitly, giving the family \eqref{lambda-0-GF} with an arbitrary function, previously obtained in Ref.~\cite{Colleaux:2025vtm}. This solution describes a broad class of singular gauge-field transformations and corresponding mimetic energy-momentum tensors. We presented two simple branches, $C=C(Y_1)$ and $C=C(Y_2)$, which isolate the different tensor structures in the energy-momentum tensor, while a systematic study of the general family is left for future work. We then combined a scalar and a gauge field in the pure-disformal singular family \eqref{D-general-SG}. Its explicit dual zero mode \eqref{dual-tensor-SG} gives the perfect-fluid source \eqref{EMT-star-SG}. The power-law subclass \eqref{D-w-SG} realizes $p=w\rho$ for an arbitrary constant $w$, including dust, radiation,
and vacuum energy. On the timelike branch, both the scalar disformal
tensor and the mimetic source are Type I. Finally, the two-scalar tensor \eqref{dis-T-2SF} realizes all four Hawking-Ellis types and gives explicit singular transformations for which Type II and Type IV are either preserved or reduced to Type I, as shown in \eqref{transition-II-2SF} and \eqref{transition-IV-2SF}. Our general setup can be implemented to systematically study kinematical and dynamical properties of many invertible and non-invertible disformal transformations with different field contents.
	
\vspace{0.7cm}
	
{\bf Acknowledgments:} The work of MAG and MY was supported by IBS under the project code, IBS-R018-D3. The work of AV was supported by project 24-13079S of the Czech Science Foundation (GA\v{C}R). PJ acknowledges funding from the South African Research Chairs Initiative of the Department of Science and Technology and the National Research Foundation of South Africa. We would like to thank Keigo Shimada for the initial collaboration on this project.
	
\vspace{0.7cm}


\appendix

\section{Inverse contravariant metric}\label{app-inverse}
In this section, we implement the Cayley-Hamilton theorem to systematically find the inverse contravariant disformed metric $g^{\mu\nu}$. We do this case by case for ${\boldsymbol{\rm CH}}_4$, ${\boldsymbol{\rm CH}}_3$, ${\boldsymbol{\rm CH}}_2$.

\subsection{\texorpdfstring{${\boldsymbol{\rm CH}}_4$}{CH4} transformations}
\subsubsection{Standard form}
Our task is to find $g^{\mu\nu}$ which satisfies \eqref{def-inverse-metric} with $g_{\mu\nu}$ given by \eqref{sec inv: general disformal transformation}. Looking at \eqref{sec inv: general disformal transformation} as a disformal map $h\to{ g}$, we consider another disformal map $h\to{\tilde g}$ such that
\begin{align}\label{gtilde}
{\tilde g}_{\mu\nu} = {\tilde C} \, h_{\mu\nu} + \sum_{I=1}^3 {\tilde D}_{I}\, t^I_{\mu\nu} \,.
\end{align}
Using \eqref{sec inv: general disformal transformation} and \eqref{gtilde}, we have
\begin{align}\nonumber
h^{\alpha\beta} g_{\mu\alpha} {\tilde g}_{\nu\beta}
&=
C \tilde{C}\, h_{\mu \nu }
+
\left(D_{1} \tilde{C}+C \tilde{D}_{1}\right) t_{\mu \nu }
+
\left(D_{2} \tilde{C}+C \tilde{D}_{2}+D_{1}
\tilde{D}_{1}\right) t^2_{\mu\nu}
\\ \nonumber
&+
\left( D_{3} \tilde{C}+C \tilde{D}_{3}+D_{2}
\tilde{D}_{1}+D_{1}
\tilde{D}_{2}\right) t^3_{\mu\nu}
+\left(D_{3} \tilde{D}_{1}+D_{2}
\tilde{D}_{2}+D_{1} \tilde{D}_{3}\right)
t^4_{\mu\nu}
\\
&+ \left(D_{3}\tilde{D}_{2}+D_{2} \tilde{D}_{3}\right)
t^5_{\mu\nu}
+ D_{3} \tilde{D}_{3} \,
t^6_{\mu\nu} \label{matrix-p} \,.
\end{align}
Using the Cayley-Hamilton theorem \eqref{sec inv: CH4 theorem} in \eqref{matrix-p}, we can get rid of $t_{\mu\nu}^{4}, t_{\mu\nu}^{5}, t_{\mu\nu}^{6}$ to get
\begin{align}\label{gbar}
{\bar g}_{\mu\nu}
\equiv h^{\alpha\beta} g_{\mu\alpha} {\tilde g}_{\nu\beta}
= {\bar C} \, h_{\mu\nu} + \sum_{i=1}^3 {\bar D}_{i}\, t^i_{\mu\nu} \,,
\end{align}
where
\begin{align}\nonumber
&{\bar C} \equiv
C \tilde{C}+\delta_0
\left(\tilde{D}_{3} \left(D_{1}+\delta_2
D_{3}\right)+\left(D_{3} \delta_3+D_{2}\right)
\left(\delta_3
\tilde{D}_{3}+\tilde{D}_{2}\right)+D_{3}
\tilde{D}_{1}\right) \,,
\\[8pt] \nonumber
&{\bar D}_{1} \equiv
D_{1} \tilde{C}+C \tilde{D}_{1}+\delta_1 D_{3} \tilde{D}_{1}+\delta_1 D_{2}
\tilde{D}_{2}+\delta_0 \left(\tilde{D}_{3}
\left(D_{3} \delta_3+D_{2}\right)+D_{3}
\tilde{D}_{2}\right)
\\ \nonumber
&
+\delta_1 \left(\delta_2
D_{3} \tilde{D}_{3}+\delta_3 \left(D_{3}
\delta_3 \tilde{D}_{3}+D_{3}
\tilde{D}_{2}+D_{2} \tilde{D}_{3}\right)+D_{1}
\tilde{D}_{3}\right) \,,
\\[8pt] \nonumber
&{\bar D}_{2} \equiv
\tilde{D}_{2}\left(C+\delta_2 D_{2}\right)+D_{2}\tilde{C}+\delta_2^2 D_{3}\tilde{D}_{3}
+\delta_1 D_{3}\tilde{D}_{2}+\delta_1 D_{2}\tilde{D}_{3}+\delta_0 D_{3}\tilde{D}_{3}
\\ \nonumber
&+D_{1}\left(\tilde{D}_{1}+\delta_2\tilde{D}_{3}\right)
+\delta_2\left(D_{3}\delta_3\tilde{D}_{2}+\delta_3\tilde{D}_{3}\left(D_{3}\delta_3+D_{2}\right)+D_{3}\tilde{D}_{1}\right)
+\delta_1 D_{3}\delta_3\tilde{D}_{3}
\,,
\\[8pt] \nonumber
&{\bar D}_{3} \equiv
D_{3} \tilde{C}+C \tilde{D}_{3}+\delta_2 D_{3} \tilde{D}_{2}+\delta_1 D_{3}
\tilde{D}_{3}+D_{3} \delta_3^3
\tilde{D}_{3}
+\delta_3^2 \left(D_{3}
\tilde{D}_{2}+D_{2} \tilde{D}_{3}\right)
\\ \nonumber
&
+\delta_2 \tilde{D}_{3} \left(2 D_{3}
\delta_3+D_{2}\right)+D_{2} \delta_3
\tilde{D}_{2}+D_{1} \delta_3
\tilde{D}_{3}+\tilde{D}_{1} \left(D_{3}
\delta_3+D_{2}\right)+D_{1}
\tilde{D}_{2}
\,.
\end{align}	
Now, ${\bar g}_{\mu\nu}$ has the same (disformal) form as $g_{\mu\nu}$ in \eqref{sec inv: general disformal transformation}. Schematically, this result can be shown as $g\times{\tilde g}\Rightarrow{\bar g}$ where $g$, ${\tilde g}$, and ${\bar g}$ are given by \eqref{sec inv: general disformal transformation}, \eqref{gtilde}, and \eqref{gbar} respectively. The multiplication symbol $\times$, defined by means of $h$ in \eqref{matrix-p}, is ordinary matrix multiplication in the polynomial algebra generated by $t^{\mu}{}_{\nu}$. In order to see this explicitly, we define matrices $\boldsymbol{\tilde {\cal C}}^{(4)}=\big( {\tilde C}, {\tilde D}_{i}\big)$ and $\boldsymbol{\bar {\cal C}}^{(4)}=\left( {\bar C}, {\bar D}_{i} \right)$ in terms of which equations \eqref{gbar} can be rewritten in the following matrix form
\begin{align}\label{G4-multiplication}
\boldsymbol{\tilde {\cal C}}^{(4)}\cdot\boldsymbol{\cal G}^{(4)} = \boldsymbol{\bar {\cal C}}^{(4)} \,,
\end{align}
where the components of matrix $\boldsymbol{{\cal G}}^{(4)}$, ${\cal G}^{(4)}_{IJ}$ with $I,J=0,1,2,3$ are given by
\begin{align}
{\cal G}^{(4)}_{00} &= C \,,
\hspace{1cm}
{\cal G}^{(4)}_{0i} = D_{i} \,,
\hspace{1cm}
{\cal G}^{(4)}_{10} = \delta_0 D_{3} \,,
\hspace{1cm}
{\cal G}^{(4)}_{20} = \delta_0 \left(D_{3}\delta_3+D_{2}\right) \,,
\nonumber \\
{\cal G}^{(4)}_{30} &= \delta_0 \left[\delta_3
\left(D_{3} \delta_3+D_{2}\right)+D_{1} +\delta_2 D_{3} \right] \,,
\hspace{1cm}
{\cal G}^{(4)}_{11} = C+\delta_1 D_{3} \,,
\hspace{1cm}
{\cal G}^{(4)}_{12} = D_{1}+\delta_2 D_{3} \,,
\nonumber \\
{\cal G}^{(4)}_{13} &= D_{3} \delta_3+D_{2} \,,
\hspace{.5cm}
{\cal G}^{(4)}_{21} = \delta_1 \left(D_{3} \delta_3+D_{2}\right)+\delta_0 D_{3} \,,
\hspace{.5cm}
{\cal G}^{(4)}_{22} = C+\delta_1 D_{3}+\delta_2 \left(D_{3}\delta_3+D_{2}\right) \,,
\nonumber  \\
{\cal G}^{(4)}_{23} &= \delta_2 D_{3}+\delta_3 \left(D_{3}
\delta_3+D_{2}\right)+D_{1} \,,
\hspace{.5cm}
{\cal G}^{(4)}_{31} = \delta_1 \left(\delta_2 D_{3}+\delta_3 \left(D_{3}\delta_3+D_{2}\right)+D_{1}\right)+\delta_0 \left(D_{3} \delta_3+D_{2}\right) \,,
\nonumber  \\
{\cal G}^{(4)}_{32} &= \delta_2^2 D_{3}+\delta_0 D_{3}+\delta_2\left(\delta_3 \left(D_{3}\delta_3+D_{2}\right)+D_{1}\right)+\delta_1 \left(D_{3} \delta_3+D_{2}\right) \,,
\nonumber  \\ \label{G4-T}
{\cal G}^{(4)}_{33} &= C+D_{3} \left(\delta_1 + \delta_3^3\right)+\delta_2 \left(2 D_{3}\delta_3+D_{2}\right)+\delta_3 \left(D_{2}\delta_3+D_{1}\right) \,,
\end{align}
in which $\delta_{i}$ are defined in \eqref{sec inv: CH4 deltas}. Thus, at fixed $t_{\mu\nu}$, the coefficient vectors form a finite-dimensional commutative algebra under this multiplication. For $D_1=D_2=D_3=0$, the multiplication operator is $C\boldsymbol I_4$; the identity element is the special case $C=1$. An element is invertible if and only if ${\cal G}^{(4)}\equiv\det\boldsymbol{{\cal G}}^{(4)}\neq0$.  The sign ${\cal G}^{(4)}>0$ is necessary for a continuous path in the full operator space from $\boldsymbol I_4$, but it is not sufficient by itself to characterize the identity component of the restricted family. Only the invertible elements form a group.

The tensor $\tilde g_{\mu\nu}$ in \eqref{gtilde} is an auxiliary covariant tensor. If $g_{\mu\alpha}h^{\alpha\beta}\tilde g_{\beta\nu}=h_{\mu\nu}$, then raising both indices of $\tilde g_{\mu\nu}$ with $h^{\mu\nu}$ gives the actual inverse $g^{\mu\nu}$. Therefore, imposing
\begin{align}\label{eqs-4-inverse}
{\bar C} = 1 \,,
\qquad
{\bar D}_{1} = 0 \,,
\qquad
{\bar D}_{2} = 0 \,,
\qquad
{\bar D}_{3} = 0 \,,
\end{align}
${\tilde g}_{\mu\nu}$ will be the inverse of $g_{\mu\nu}$. Looking at Eq. \eqref{G4-multiplication}, we see that the problem of finding the inverse contravariant metric is equivalent to finding the inverse of matrix $\boldsymbol{{\cal G}}^{(4)}$. Solving Eq. \eqref{G4-multiplication} subject to \eqref{eqs-4-inverse} we find
\begin{align}
\begin{split}
{\tilde C} &= \frac{1}{{\cal G}^{(4)}} \left(
{\cal G}^{(4)}_{21} {\cal G}^{(4)}_{32} {\cal G}^{(4)}_{13}
-{\cal G}^{(4)}_{31} {\cal G}^{(4)}_{22} {\cal G}^{(4)}_{13}
+{\cal G}^{(4)}_{31}{\cal G}^{(4)}_{12} {\cal G}^{(4)}_{23}
-{\cal G}^{(4)}_{11} {\cal G}^{(4)}_{32} {\cal G}^{(4)}_{23}
-{\cal G}^{(4)}_{21} {\cal G}^{(4)}_{12}{\cal G}^{(4)}_{33}
+{\cal G}^{(4)}_{11} {\cal G}^{(4)}_{22} {\cal G}^{(4)}_{33}
\right) ,
\\
{\tilde D}_{1} &= \frac{1}{{\cal G}^{(4)}} \left(
{\cal G}^{(4)}_{31} {\cal G}^{(4)}_{22} {\cal G}^{(4)}_{03}
-{\cal G}^{(4)}_{21} {\cal G}^{(4)}_{32} {\cal G}^{(4)}_{03}
-{\cal G}^{(4)}_{31}{\cal G}^{(4)}_{02} {\cal G}^{(4)}_{23}
+{\cal G}^{(4)}_{01} {\cal G}^{(4)}_{32} {\cal G}^{(4)}_{23}
+{\cal G}^{(4)}_{21} {\cal G}^{(4)}_{02}{\cal G}^{(4)}_{33}
-{\cal G}^{(4)}_{01} {\cal G}^{(4)}_{22} {\cal G}^{(4)}_{33}
\right),
\\
{\tilde D}_{2} &= \frac{1}{{\cal G}^{(4)}} \left(
{\cal G}^{(4)}_{11} {\cal G}^{(4)}_{32} {\cal G}^{(4)}_{03}
-{\cal G}^{(4)}_{31} {\cal G}^{(4)}_{12} {\cal G}^{(4)}_{03}
+{\cal G}^{(4)}_{31}{\cal G}^{(4)}_{02} {\cal G}^{(4)}_{13}
-{\cal G}^{(4)}_{01} {\cal G}^{(4)}_{32} {\cal G}^{(4)}_{13}
-{\cal G}^{(4)}_{11} {\cal G}^{(4)}_{02}{\cal G}^{(4)}_{33}
+{\cal G}^{(4)}_{01} {\cal G}^{(4)}_{12} {\cal G}^{(4)}_{33}
\right),
\\
{\tilde D}_{3} &= \frac{1}{{\cal G}^{(4)}} \left(
{\cal G}^{(4)}_{21} {\cal G}^{(4)}_{12} {\cal G}^{(4)}_{03}
-{\cal G}^{(4)}_{11} {\cal G}^{(4)}_{22} {\cal G}^{(4)}_{03}
-{\cal G}^{(4)}_{21}{\cal G}^{(4)}_{02} {\cal G}^{(4)}_{13}
+{\cal G}^{(4)}_{01} {\cal G}^{(4)}_{22} {\cal G}^{(4)}_{13}
+{\cal G}^{(4)}_{11} {\cal G}^{(4)}_{02}{\cal G}^{(4)}_{23}
-{\cal G}^{(4)}_{01} {\cal G}^{(4)}_{12} {\cal G}^{(4)}_{23}
\right).
\end{split}
\label{sol-inverse}
\end{align}

Therefore, the inverse map for \eqref{sec inv: general disformal transformation} under the matrix multiplication turns out to be
\begin{align}\label{dis-T-inv-2}
g^{\mu\nu} = {\tilde C}\, h^{\mu\nu} + \sum_{I=1}^3 {\tilde D}_{I}\, t^{I\,\mu\nu} \,,
\end{align}
where the explicit forms of ${\tilde C}$ and ${\tilde D}_{I}$ are given by \eqref{sol-inverse}.

\subsubsection{Normal form}

The inverse contravariant metric for the normal form \eqref{DT-simplified}
\begin{align}\label{DT-simplified-app}
g_{\mu\nu} = C h_{\mu\nu} + D t_{\mu\nu} \,,
\end{align}
can be easily obtained from the results of the previous subsection by setting $D_2=0=D_3$ and identifying $D_1=D$.

The inverse contravariant metric of \eqref{DT-simplified-app} takes the same form as \eqref{dis-T-inv-2}
\begin{align}\label{dis-T-inv-sim}
g^{\mu\nu} = {\tilde C}\, h^{\mu\nu} + \sum_{I=1}^3 {\tilde D}_{I}\, t^{I\,\mu\nu} \,,
\end{align}
but with the following simple expressions for the coefficients
\begin{align}
\begin{split}
{\tilde C} &= \frac{C^3 + C^2 D \delta_3 - C D^2 \delta_2 + D^3 \delta_1}{{\cal G}^{(4)}} \,,
\hspace{1cm}
{\tilde D}_{1} = -\frac{D\left(C^2+C D \delta_3 - D^2 \delta_2 \right)}{{\cal G}^{(4)}} \,,
\\
{\tilde D}_{2} &= \frac{D^2\left(C+D \delta_3 \right)}{{\cal G}^{(4)}} \,,
\hspace{4cm}
{\tilde D}_{3} = -\frac{D^3}{{\cal G}^{(4)}} \,,
\end{split}
\label{sol-inverse-sub}
\end{align}
where
\begin{align}\nonumber
{\cal G}^{(4)}=C^4 + D \left( C^3 \delta_3 - C^2 D \delta_2 + C  D^2 \delta_1 - D^3 \delta_0 \right) \,,
\end{align}
and $\delta_{I}$ are defined in \eqref{sec inv: CH4 deltas}. As can be seen in \eqref{dis-T-inv-sim}, even if we only have the linear term in $t_{\mu\nu}$ in the disformal transformation \eqref{DT-simplified-app}, the quadratic and cubic terms, characterized by ${\tilde D}_{2}$ and ${\tilde D}_{3}$ respectively, show up in the inverse contravariant metric due to the ${\boldsymbol{\rm CH}}_4$ nature of the transformation.

\subsection{\texorpdfstring{${\boldsymbol{\rm CH}}_3$}{CH3} transformations}

In this simplified version of the Cayley-Hamilton theorem, we have \eqref{sec simple CH: CH3 condition} which means the highest independent power of $t_{\mu\nu}$ is $t^2_{\mu\nu}$. Therefore, starting with the normal disformal transformation \eqref{DT-simplified-app}, the corresponding contravariant inverse metric takes the form
\begin{align}\label{dis-T-inv-G3}
g^{\mu\nu} = {\tilde C}\, h^{\mu\nu} + \sum_{I=1}^2 {\tilde D}_{I}\, t^{I\,\mu\nu} \,.
\end{align}
Following the same procedure as in the previous subsection, we find
\begin{align}
{\tilde C} = \frac{C^2+C \delta _2 D-\delta _1 D^2}{{\cal G}^{(3)}} \,,
\qquad
{\tilde D}_{1} = -\frac{D \left(C+\delta _2 D\right)}{{\cal G}^{(3)}} \,,
\qquad
{\tilde D}_{2} = \frac{D^2}{{\cal G}^{(3)}} \,,
\label{sol-inverse-G3}
\end{align}
where
\begin{align}
{\cal G}^{(3)} = C^3+C^2 \delta _2 D-C \delta_1 D^2+\delta _0 D^3 \,,
\end{align}
and
\begin{align}
\delta_0 = \lambda_1 \lambda_2 \lambda_3 \,,
\qquad
\delta_1 = - \left( \lambda_1 \lambda_2 + \lambda_1 \lambda_3 + \lambda_2 \lambda_3 \right) \,,
\qquad
\delta_2 = \lambda_1 + \lambda_2 + \lambda_3 \,,
\end{align}
in which $\lambda_i$ are the three roots of the minimal polynomial, repeated when required by a non-trivial Jordan block. In general, we cannot uniquely find $\delta_I$ in terms of the trace of powers of $t^{\mu}_{\ \nu}$.

\subsection{\texorpdfstring{${\boldsymbol{\rm CH}}_2$}{CH2} transformations}\label{subsec inv CH2}

In this case, the Cayley-Hamilton theorem simplifies to \eqref{sec simple CH: CH2 condition} which means that the highest independent power is $t_{\mu\nu}$ itself. Starting with the simplified disformal transformation \eqref{DT-simplified-app},
the corresponding contravariant inverse metric is
\begin{align}\label{dis-T-inv-G2}
g^{\mu\nu} = {\tilde C}\, h^{\mu\nu} + {\tilde D} \, t^{\mu\nu} \,,
\end{align}
where
\begin{align}
{\tilde C} = \frac{C+\delta _1 D}{C^2+C \delta _1 D-\delta _0 D^2} \,,
\qquad
{\tilde D} = -\frac{D}{C^2+C \delta_1 D-\delta _0 D^2} \,,
\label{sol-inverse-G2}
\end{align}
where
\begin{equation}
\delta_{0}=-\lambda_{1}\lambda_{2}\ ,
\qquad 
\delta_{1}=\lambda_{1}+\lambda_{2} \,,
\end{equation}
in which $\lambda_i$ are the two roots of the minimal polynomial, which may coincide for a non-trivial size-two Jordan block.

	
\section{Segre-Pleba\'{n}ski classification}\label{app:SC}
The so-called Segre-Pleba\'{n}ski classification answers the following question: to what extent can we diagonalize a matrix associated to a symmetric rank-$2$ tensor on a Lorentzian manifold? This question has a very simple and well known answer on Euclidean manifolds, where the corresponding matrix is symmetric and, consequently, fully diagonalizable. Furthermore, its eigenvalues are necessarily real valued and its eigenvectors associated to different eigenvalues are necessarily orthogonal to each other. When we consider the Lorentzian signature instead of the Euclidean one we will see that this is no longer the case. Indeed, the matrix form of any given rank $2$ tensor $t_{ab}$ on a Lorentzian manifold\footnote{Note that in this section we will work exclusively in an orthonormal co-basis $e^{a}{}_{\mu}$. The spacetime metric is $h_{\mu\nu}=\eta_{ab}e^{a}{}_{\mu}e^{b}{}_{\nu}$. The orthonormal basis is signified by the use of lowercase Latin indices.} is given as
\begin{equation}
t^{a}_{\ b}=\eta^{ac}t_{cb} \,,
\end{equation}
which is in general not a symmetric matrix ($\eta^{ab}\neq\delta^{ab}$). Nevertheless, the origin of $t^{a}_{\ b}$ in a symmetric tensor has severe consequences on the admissible eigenvalues and their associated eigenvectors. The possible Jordan normal forms of the matrix $t^{a}_{\ b}$ are limited as well. In this section we will demonstrate these limitations and we will explicitly show why $t^{a}_{\ b}$ cannot take certain Jordan normal forms. Finally we explain the notation of the Segre-Pleba\'{n}ski classification, which classifies the remaining admissible forms.

Since the eigenvalues and eigenvectors of a generic real matrix are not necessarily real themselves, it is necessary to first extend the scalar product on the manifold to complex valued vectors. We do this in the standard way as follows
\begin{equation}
V\cdot W = \eta_{ab}\bar{V}^{a}W^{b}\ .\label{app SC: complex inner product}
\end{equation}
Here $V^{a}$ and $W^{b}$ are possibly complex valued vectors and the bar denotes complex conjugation of their components. Note that this scalar product is sesquilinear (linear in  $W$ but conjugate linear in $V$) and, in contrast to the usual linear scalar product, it can no longer be represented as a simple contraction of the vector components with some metric tensor. For example, $V\cdot{W}\neq{W}\cdot{V}$ but instead $V\cdot{W}=\overline{{W}\cdot{V}}$. Nevertheless, this scalar product allows us to define null directions in the usual fashion. We say that $\xi^{a}$ is null whenever
\begin{equation}
\bar{\xi}^{a}\xi^{b}\eta_{ab}=0\ .
\end{equation}
Similarly, we take $V^{a}$ to be timelike when $\bar{V}^{a}V^{b}\eta_{ab}>0$ and spacelike when $\bar{V}^{a}V^{b}\eta_{ab}<0$. Crucially, any two null vectors are orthogonal only if they are proportional to each other as is the case for real valued null vectors. That is, for $\xi^{a}$ and $\zeta^{a}$ both being null vectors
\begin{equation}
\bar{\xi}^{a}\zeta_{a}=0\ \implies\ \zeta^{a}\propto\xi^{a}\ .\label{app SC: null orthogonality}
\end{equation}
Furthermore, any timelike vector has a non-vanishing scalar product with any null vector. Note that the above simple fact is extremely important for the rest of this section. Indeed, all restrictions on the spectrum of $t^{a}_{\ b}$ and on its possible Jordan normal form can and will be traced to contradicting \eqref{app SC: null orthogonality}.

The eigenvectors $V^{a}$ and their associated eigenvalues $\lambda_{V}$ of the matrix $t^{a}_{\ b}$ are defined as the solutions of the equation
\begin{equation}
\left(t^{a}_{\ b}-\lambda_{V}\delta^{a}_{\ b}\right)V^{b}=0\ ,
\end{equation}
or equivalently
\begin{equation}
\left(t_{ab}-\lambda_{V}\eta_{ab}\right)V^{b}=0\ .
\end{equation}
Since the components of the matrix $t_{ab}$ are real it follows that
\begin{equation}
0=\bar{V}^{a}\left(\bar{t}_{ab}-t_{ab}\right)V^{b}=\left(\bar{\lambda}_V-\lambda_V\right)\bar V^{a}V_{a}\ .\label{app SC: reality of eigenvalues}
\end{equation}
Consequently, 
\begin{enumerate}
\item\label{item-timelike-spacelike} {\it The eigenvalues associated to the timelike and spacelike eigenvectors with $\bar V^{a}V_{a}\neq0$, are necessarily real valued}, like in the Euclidean case.

\item\label{item-null} However, eigenvalues corresponding to null eigenvectors $\bar V^{a}V_{a}=0$ can have non-vanishing imaginary part $\bar{\lambda}_V\neq\lambda_V$. Thus, {\it an eigenvector with a non-real eigenvalue is necessarily a null vector}. 
\end{enumerate}
Analogously, for two eigenvectors $V^{a}$ and $W^{a}$ we find
\begin{equation}
0=\bar{W}^{a}\left(\bar{t}_{ab}-t_{ab}\right)V^{b}=(\bar{\lambda}_{W}-\lambda_{V})\bar{W}^{a}V_{a}\ .\label{app SC: orthogonality of eigenvectors}
\end{equation}
This implies that eigenvectors associated to $\lambda_{V}\neq\bar{\lambda}_{W}$ are necessarily orthogonal to each other. This can only be violated if $\lambda_{V}=\bar{\lambda}_{W}$.

This has an immediate consequence for the spectrum of $t^{a}_{\ b}$: the matrix
$t^{a}_{\ b}$ can only have zero or two non-real eigenvalues. In the latter case
the eigenvalues are necessarily complex conjugates. Indeed, for two non-real
eigenvalues that are {\it not} complex conjugates of one another,
$\lambda_{\xi}\neq\bar{\lambda}_{\zeta}$, the associated eigenvectors $\xi$ and
$\zeta$ must be orthogonal to each other due to
\eqref{app SC: orthogonality of eigenvectors}. Furthermore, since their
eigenvalues are non-real, they must both be null vectors (item \ref{item-null} above).
This is, however, in contradiction with \eqref{app SC: null orthogonality}.
Hence, any two non-real eigenvalues can only be complex conjugates of one
another. A second, distinct pair is therefore excluded: since $t^{a}_{\ b}$ is
real, its non-real eigenvalues come in conjugate pairs, so four of them would
form two distinct pairs, and selecting one eigenvalue from each pair would
reproduce a forbidden non-conjugate couple $\lambda_{\xi}\neq\bar{\lambda}_{\zeta}$. Only a single pair can thus occur, i.e. zero or two non-real eigenvalues. Additionally, only one eigenvector can be associated to each of these eigenvalues, since any two eigenvectors $\xi_{1}$ and $\xi_{2}$ sharing the same eigenvalue $\lambda_{\xi}$ would also have to satisfy \eqref{app SC: orthogonality of eigenvectors} while both being null, again contradicting \eqref{app SC: null orthogonality}. In summary, 
\begin{enumerate}[start=3]
\item\label{item-complex-conjugate-eigens} {\it There can be at most two complex eigenvalues, forming a single conjugate pair, with only one null eigenvector associated to each.}
\end{enumerate}

Any square matrix, including $\boldsymbol{t}$, can be brought into its Jordan normal form by means of a non-singular matrix ${\boldsymbol P}$ as follows \cite{Carrell:2017}
\begin{align}\label{Jordan-decomposition-app}
{\boldsymbol P}^{-1}\cdot\boldsymbol{t}\cdot{\boldsymbol P} =
\left(
\begin{array}{cccc}
{\boldsymbol J}_1 & 0 & \cdots & 0 \\
0 & {\boldsymbol J}_2 & \ddots & \vdots \\
\vdots & \ddots & \ddots & 0 \\
0 & \cdots & 0 & {\boldsymbol J}_p \\
\end{array}
\right) \,,
\end{align}
where the diagonal elements ${\boldsymbol J}_k$ are Jordan blocks associated with an eigenvalue $\lambda_{k}$. These matrices are square matrices of size $s_{k}\times s_{k}$ with the form
\begin{align}
{\boldsymbol J}_k =
	\left(
	\begin{array}{cccc}
		\lambda_k & 1 & 0 & \cdots \\
		0 & \lambda_k & \ddots & 0 \\
		\vdots & \ddots & \ddots & 1 \\
		0 & \cdots & 0 & \lambda_k \\
	\end{array}
	\right) \,.
\end{align}
With each Jordan block, there is an associated set of vectors $V_{i}^{a}$, for $1\leq i\leq s_{k}$, called \emph{Jordan chain}, which satisfy
\begin{equation}
(t^{a}_{\ b}-\lambda_{V}\delta^{a}_{\ b})V^{b}_{i}=V^{a}_{i+1},
\qquad
\forall i:\ 1\leq i< s_{k}\ .\label{app SC: jordan chain}
\end{equation}
Here we have denoted $\lambda_{V}=\lambda_{k}$ to stress the association with the vectors $V_{i}^{a}$. These vectors are linearly independent and the last vector, $V^a_{s_k}$, is an ordinary eigenvector. With this terminating-chain convention, the ordered basis $(V_1,\ldots,V_{s_k})$ gives ones on the first subdiagonal. The standard block with ones on the first superdiagonal is obtained in the reversed basis $(V_{s_k},\ldots,V_1)$. Consequently, for any vector $V^a_{i}$ from the chain we have
\begin{equation}\label{app:t-Jordan-form}
\left(t^{a}_{\ b}-\lambda_V \delta^{a}_{\ b}\right)^{m}V^{b}_{i}=0\ ;\qquad m=s_{k}-i+1\ .
\end{equation}
The existence of a non-trivial chain has a direct consequence on the norm of the associated eigenvector. Let us first assume that the eigenvalue $\lambda_{V}$ is real and that the block is non-trivial, $s_k>1$. Then we have
\begin{align}
\begin{split}
|V^{a}_{s_{k}}|^{2}=&\,\bar{V}^{a}_{s_{k}}\eta_{ab}V^{b}_{s_{k}} \\
=&\,\bar{V}^{a}_{s_{k}-1}\left(t_{ab}-\lambda_V \eta_{ab}\right)V^{b}_{s_{k}} \\
=&\,0\ ,
\end{split}
\label{app SC: lightlike eigenvector jordan block}
\end{align}
due to $V^{b}_{s_{k}}$ being an eigenvector. When the imaginary part of $\lambda_V$ is nonzero, the norm is forced to vanish due to \eqref{app SC: reality of eigenvalues}. Consequently, the eigenvector associated with a non-trivial real Jordan block, or with any non-real eigenvalue, is necessarily a null vector.

For a similar reason, the size of a Jordan block of a matrix $t^{a}_{\ b}$ with a real eigenvalue is limited. Indeed, consider a scalar product of two vectors from the chain $V^{a}_{m}$ and $V^{a}_{n}$
\begin{align}
\begin{split}
\bar{V}^{a}_{m}\eta_{ab}V^{b}_{n}=&\,\bar{V}_{1a}\left(t^{a}_{\ b}-\lambda_V\delta^{a}_{\ b}\right)^{m+n-2}V^{b}_{1} \\
=&\,\bar{V}_{1a}\left(t^{a}_{\ b}-\lambda_V\delta^{a}_{\ b}\right)^{m+n-1-s_{k}}V^{b}_{s_{k}} \\
=&\,0\ ,\qquad\mathrm{if}\qquad m+n-1-s_{k}\geq 1\ .
\end{split}
\label{app SC: vanishing scalar product jordan chain}
\end{align}
If $m=n=s_{k}-1$ then the above result implies that for $s_{k}\geq 4$ the vector $V_{s_{k}-1}$ is a null vector and necessarily $V_{s_{k}-1}\cdot V_{s_{k}}=0$. This is however in conflict with \eqref{app SC: null orthogonality}. It follows that
\begin{enumerate}[start=4]
\item\label{app SC: size of Jordan block} {\it For real eigenvalues, the size of a Jordan block should be smaller than the size of $t^a_{\ b}$: $s_{k}< 4$}.
\end{enumerate}
For eigenvalues with non-vanishing imaginary part the situation is even more restrictive. Indeed, if we have any non-trivial Jordan block associated to a non-real eigenvalue $\lambda_{k}$ then the reality of $t^{a}_{\ b}$ implies that there must exist a Jordan block of the same size associated to $\bar{\lambda}_{k}$. It immediately follows that the eigenvectors $V^{a}_{s_{k}}$ and $W^{a}_{s_{k}}$ corresponding in order to the two respective Jordan blocks are orthogonal to each other. Indeed, we get
\begin{align}
\begin{split}
\bar{V}^{a}_{s_{k}}\eta_{ab}W^{b}_{s_{k}}=&\,\bar{V}^{a}_{s_{k}-1}\left(t_{ab}-\lambda_{k}\eta_{ab}\right)^{*}W^{b}_{s_{k}}\ ,\\
=&\,\bar{V}^{a}_{s_{k}-1}\left(t_{ab}-\bar{\lambda}_{k}\eta_{ab}\right)W^{b}_{s_{k}}\ ,\\
=&\,0\ .
\end{split}
\end{align}
Since both eigenvectors are necessarily null, they cannot be simultaneously orthogonal to each other due to \eqref{app SC: null orthogonality}. It follows that 
\begin{enumerate}[start=5]
\item\label{app SC: no Jordan block for complex eigenvalues} {\it There cannot be a non-trivial Jordan block associated with a non-real eigenvalue.}
\end{enumerate}

Finally, we note that there can only be a single non-trivial Jordan block in the decomposition of $t^{a}_{\ b}$. This follows from the fact that all eigenvectors associated to non-trivial Jordan blocks are necessarily null and the associated eigenvalues are real. If the eigenvalues are also distinct, then we find that they must be orthogonal to each other due to \eqref{app SC: orthogonality of eigenvectors}, which is again in conflict with \eqref{app SC: null orthogonality}. If they are equal then we find the same conclusion, however, for a different reason. Indeed since either of the two eigenvectors $V^{a}$ and $W^{a}$ is a part of a Jordan chain we find
\begin{equation}
V^{a}_{s_{k}}\eta_{ab}W^{b}=V^{a}_{s_{k}-1}\left(t_{ab}-\lambda_V\eta_{ab}\right)W^{b}=0\,,
\end{equation}
where we have used the fact that $\lambda_W=\lambda_V$ to conclude the last equality. Hence we get the same contradiction. Therefore,
\begin{enumerate}[start=6]
\item\label{app SC: number of Jordan block} {\it There can only be a single non-trivial Jordan block in the decomposition of $t^{a}_{\ b}$.}
\end{enumerate}

\subsection{Restricting possible Jordan normal forms}
Let us now list all the possible Jordan normal forms a generic $4\times 4$ matrix $t^a_{\ b}$ can take. We will then show that many of these possibilities are in some way in conflict with the results we obtained in the previous part. Consequently, such forms cannot occur for a symmetric and real $t_{ab}$ on the Lorentzian manifold. All candidate Jordan normal forms are
\begin{equation}\label{app SC: all jordan normal forms}
\begin{gathered}
		\begin{array}{ccc}
			\left(\begin{array}{cccc}
				\lambda_1&0&0&0\\0&\lambda_2&0&0\\0&0&\lambda_3&0\\0&0&0&\lambda_4
			\end{array}\right)
			&
			\left(\begin{array}{cccc}
				\lambda_1&1&0&0\\0&\lambda_1&0&0\\0&0&\lambda_2&0\\0&0&0&\lambda_3
			\end{array}\right)
			&
			\left(\begin{array}{cccc}
				\lambda_1&1&0&0\\0&\lambda_1&1&0\\0&0&\lambda_1&0\\0&0&0&\lambda_2
			\end{array}\right)
			\\[2mm]
			(a)\ [1,111]&(b)\ [211]&(c)\ [31]
		\end{array}
		\\[5mm]
		\begin{array}{cc}
			\left(\begin{array}{cccc}
				\lambda_1&1&0&0\\0&\lambda_1&1&0\\0&0&\lambda_1&1\\0&0&0&\lambda_1
			\end{array}\right)
			&
			\left(\begin{array}{cccc}
				\lambda_1&1&0&0\\0&\lambda_1&0&0\\0&0&\lambda_2&1\\0&0&0&\lambda_2
			\end{array}\right)
			\\[2mm]
			(d)\ [4]&(e)\ [22]
		\end{array}
		\\[5mm]
		\begin{array}{ccc}
			\left(\begin{array}{cccc}
				z&0&0&0\\0&\bar z&0&0\\0&0&\lambda_1&0\\0&0&0&\lambda_2
			\end{array}\right)
			&
			\left(\begin{array}{cccc}
				z&0&0&0\\0&\bar z&0&0\\0&0&\lambda&1\\0&0&0&\lambda
			\end{array}\right)
			&
			\left(\begin{array}{cccc}
				z_1&0&0&0\\0&\bar z_1&0&0\\0&0&z_2&0\\0&0&0&\bar z_2
			\end{array}\right)
			\\[2mm]
			(f)\ [z\bar z11]&(g)\ [z\bar z2]&(h)\ [z_1\bar z_1z_2\bar z_2]
		\end{array}
\end{gathered}
\end{equation}
Note that we have distinguished these forms also on the basis of whether they have strictly real eigenvalues, denoted as $\lambda_{i}$, or whether they also have non-real eigenvalues, here denoted by $z_{i}$. Hence, we have already imposed condition \ref{app SC: no Jordan block for complex eigenvalues} and due to the reality of the matrix $t^{a}_{\ b}$ any eigenvalue $z_{i}$ is necessarily accompanied by its conjugate $\bar{z}_{i}$, with the same algebraic multiplicity. 

By applying our results \ref{item-timelike-spacelike}--\ref{app SC: number of Jordan block} from the previous part, we can immediately rule out several of the forms presented in \eqref{app SC: all jordan normal forms}. Form $(d)$ is ruled out because its single Jordan block has size $4$, contradicting result~\ref{app SC: size of Jordan block}. Form $(e)$ contains two non-trivial Jordan blocks and is therefore rejected by result~\ref{app SC: number of Jordan block}. Form $(g)$ is excluded by \eqref{app SC: null orthogonality}: the eigenvector associated with the non-real eigenvalue $z$ is null by result~\ref{item-null}, whereas the eigenvector of the non-trivial real block $\lambda$ is null by \eqref{app SC: lightlike eigenvector jordan block}; being associated with the distinct eigenvalues $z\neq\lambda$, the two must also be orthogonal due to \eqref{app SC: orthogonality of eigenvectors}, which contradicts \eqref{app SC: null orthogonality}. Finally, form $(h)$ possesses four non-real eigenvalues, i.e.\ two distinct conjugate pairs, in direct conflict with result~\ref{item-complex-conjugate-eigens}. Thus, we are left with the forms $(a)$, $(b)$, $(c)$ and $(f)$.

In the Segre-Pleba\'{n}ski classification, each numerical entry in the bracket is the size of one Jordan block. Entries in parentheses denote blocks that share the same eigenvalue. Thus $[211]$ contains one size-two block and two size-one blocks, while $[(21)1]$ states that the size-two block and one size-one block have the same eigenvalue. In the Type I convention, the comma separates the one-dimensional block containing the timelike eigenvector from the three spacelike blocks. Non-real conjugate eigenvalues are denoted by $z\bar z$. With these conventions, the admissible forms $(a)$, $(b)$, $(c)$, and $(f)$ in \eqref{app SC: all jordan normal forms} are denoted by $[1,111]$, $[211]$, $[31]$, and $[z\bar z11]$, respectively.


\section{Hawking-Ellis classification}\label{app:HEC}

The Hawking-Ellis classification states that every real symmetric tensor $t_{ab}$ can be brought by a Lorentz transformation to one of four canonical forms:
\begin{equation}\label{app HE: HE canonical forms}
	\setlength{\arraycolsep}{4pt}
	\begin{gathered}
		\begin{array}{c}
			\left(\begin{array}{cccc}
				\lambda_1&0&0&0\\0&-\lambda_2&0&0\\0&0&-\lambda_3&0\\0&0&0&-\lambda_4
			\end{array}\right)\\[2mm]\mbox{Type I}
		\end{array}
		\quad
		\begin{array}{c}
			\left(\begin{array}{cccc}
				\lambda_1\pm f&\pm f&0&0\\\pm f&-\lambda_1\pm f&0&0\\0&0&-\lambda_2&0\\0&0&0&-\lambda_3
			\end{array}\right)\\[2mm]\mbox{Type II}
		\end{array}
		\quad
		\begin{array}{c}
			\left(\begin{array}{cccc}
				\lambda_1&f&0&0\\f&-\lambda_1&f&0\\0&f&-\lambda_1&0\\0&0&0&-\lambda_2
			\end{array}\right)\\[2mm]\mbox{Type III}
		\end{array}
		\\[5mm]
		\begin{array}{c}
			\left(\begin{array}{cccc}
				\mathrm{Re}[z]&-\mathrm{Im}[z]&0&0\\-\mathrm{Im}[z]&-\mathrm{Re}[z]&0&0\\0&0&-\lambda_1&0\\0&0&0&-\lambda_2
			\end{array}\right)\\[2mm]\mbox{Type IV}
		\end{array}
	\end{gathered}
\end{equation}
where the factor $f$ represents a residual Lorentz freedom and $\lambda_{i}$, $z$ and $\bar{z}$ are the eigenvalues of the associated mixed tensor $t^{a}{}_{b}$. As we will demonstrate momentarily, these four forms are directly related to the four different Jordan normal forms $(a), (b), (c), (f)$ shown in \eqref{app SC: all jordan normal forms}, which we have found in the Segre-Pleba\'{n}ski classification in the previous section.

In order to show this relation we first revisit the Jordan chains \eqref{app SC: jordan chain}. As we have stated, the vectors from the chain are linearly independent and can be used as a basis of the subspace associated with the Jordan block. In fact, the Jordan normal form of a matrix is just the matrix written in this particular basis. However, it is important to note that the Jordan chains are not uniquely given. Indeed, given any chain $V_{i}^{a}$ we can form new chains of the same length by simply defining a new lead vector as a linear combination of the vectors from the chain. That is
\begin{equation}
W^{a}_{1}=\sum_{i=1}^{s}\alpha_{i}V^{a}_{i}\ ,\label{app HE: new Jordan chain}
\end{equation}
where $s$ is the length of the original chain and $\alpha_{i}$ are arbitrary real constants with $\alpha_{1}\neq 0$. We generate the rest of the chain by simply acting on $W^{a}_{1}$ with $\left(t^{a}_{\ b}-\lambda\delta^{a}_{\ b}\right)$ as in \eqref{app SC: jordan chain}. This results in a novel Jordan chain, which, as long as $\alpha_{1}\neq 0$, has the same length as the original. The constants $\alpha_{i}$ can be chosen so that the basis $W^{a}_{i}$ is null orthonormal. Since the chain size in the present case is limited to at most $s=3$, we only need to consider two possibilities: $s=2$ and $s=3$. Let us start with $s=2$. In this case we need to enforce
\begin{align}
\begin{split}
W_{1}\cdot W_{1}&=0\ ,\\
W_{1}\cdot W_{2}&=\pm 1\ .
\end{split}
\nonumber
\end{align}
Note that the sign of the last relation is fixed by the sign of $V_{1}\cdot V_{2}$ and it cannot be changed by the redefinition \eqref{app HE: new Jordan chain}. Hence there are two subcases of this form depending on this sign. To solve the above we find
\begin{align}
\begin{split}
		\alpha_{1}&=\frac{1}{\sqrt{\pm V_{1}\cdot V_{2}}}\ ,\\
		\frac{\alpha_{2}}{\alpha_{1}}&=-\frac{1}{2}\frac{V_{1}\cdot V_{1}}{V_{1}\cdot V_{2}}\ .
\end{split}
\nonumber
\end{align}
In the $s=3$ case we automatically have $W_{2}\cdot W_{3}=W_{3}\cdot W_{3}=0$ due to \eqref{app SC: vanishing scalar product jordan chain}. Furthermore, the same reasoning used in \eqref{app SC: vanishing scalar product jordan chain} can be easily used to show that $W_{1}\cdot W_{3}=W_{2}\cdot W_{2}$ always holds. Hence, we only need to set up $\alpha_{i}$ so that
\begin{align}
\begin{split}
W_{1}\cdot W_{1}&=0\ ,\\
W_{1}\cdot W_{2}&=0\ ,\\
W_{1}\cdot W_{3}&=-1\ .
\end{split}
\nonumber
\end{align}
Note that unlike in the $s=2$ case the scalar product must be negative since the opposite sign would indicate that $W_{2}$ is a timelike vector. Since $W_{2}$ is simultaneously orthogonal to a null vector $W_{3}$, such a possibility is inconsistent. The solution of the above conditions for $\alpha_{i}$ is
\begin{align}
\begin{split}
\alpha_{1}&=\frac{1}{\sqrt{-V_{1}\cdot V_{3}}}\ ,\\
\frac{\alpha_{2}}{\alpha_{1}}&=-\frac{1}{2}\frac{V_{1}\cdot V_{2}}{V_{1}\cdot V_{3}}\ ,\\
\frac{\alpha_{3}}{\alpha_{1}}&=\frac{1}{2}\left[\frac{3}{4}\frac{(V_{1}\cdot V_{2})^{2}}{(V_{1}\cdot V_{3})^{2}}-\frac{V_{1}^{2}}{V_{1}\cdot V_{3}}\right]\ .
\end{split}
\nonumber
\end{align}
Hence we can always form\footnote{The situation can be slightly more complicated when there is an additional eigenvector sharing the eigenvalue with the Jordan block. However, even in such cases an appropriate null orthonormal basis can always be found.} a null orthonormal basis out of the ordinary and generalized eigenvectors of the matrix $t^{a}_{\ b}$.

Now we can show the equivalence between the Jordan forms $(a), (b), (c), (f)$ that we have found in \eqref{app SC: all jordan normal forms} in the previous section and the Hawking-Ellis canonical forms \eqref{app HE: HE canonical forms}. In order to do so, we note that in the form $(a)$ we can always take one of the eigenvectors to be timelike, while the rest is spacelike. In the cases $(b)$ and $(c)$ we have one null eigenvector with the rest being spacelike. The null eigenvector associated with the Jordan block is often referred to as double- or triple-null respectively. In case $(f)$ we have two complex null-eigenvectors and two spacelike.\footnote{In some works this case is said to have no null eigenvectors since they are necessarily complex valued and thus do not correspond to physical directions.}

For the form $(a)$ we can always find a single timelike eigenvector $U_{a}$ and three spacelike eigenvectors $n^{a}_{i}$, $i=1,2,3$, with a unit length, which form an orthonormal basis. The matrix $t^{a}_{\ b}$ can therefore be written as
\begin{equation}
t^{a}_{\ b}=\lambda_{1}U^{a}U_{b}-\lambda_{2}n^{a}_{1}n_{1b}-\lambda_{3}n^{a}_{2}n_{2b}-\lambda_{4}n^{a}_{3}n_{3b}\ .\label{app HE: jordan chain tensor form a}
\end{equation}
By simply lowering the upper index, we immediately get the canonical form of Type I Hawking-Ellis class
\begin{equation}
t_{ab}=\lambda_{1}U_{a}U_{b}-\lambda_{2}n_{1a}n_{1b}-\lambda_{3}n_{2a}n_{2b}-\lambda_{4}n_{3a}n_{3b}\,,
\qquad
\mbox{Type I} \,.
\label{app HE: Type I HE class}
\end{equation}
Note that all orthonormal bases are connected via Lorentz transformations, hence, we can arrive at this form by Lorentz transformations only.

For the form $(b)$ we have two spacelike eigenvectors $n^{a}_{2}$ and $n^{a}_{3}$ and a Jordan chain $V^{a}_{1}$, $V^{a}_{2}$ associated with the Jordan block. In this case we can take $V_{1}$ and $V_{2}$ to be null and to satisfy $V_{1}\cdot V_{2}=\pm1$. The spacelike eigenvectors can be normalized to have unit length. Let us focus in detail only on the case with the plus sign, as the remaining minus-sign case is completely analogous. In the above basis the matrix $t^{a}_{\ b}$ can be written as
\begin{equation}
t^{a}_{\ b}=+\lambda_{1}\left(V^{a}_{1}V_{2b}+V^{a}_{2}V_{1b}\right)+V^{a}_{2}V_{2b}-\lambda_{2}n^{a}_{2}n_{2b}-\lambda_{3}n^{a}_{3}n_{3b}\ .\label{app HE: jordan chain tensor form b plus}
\end{equation}
Now we form a unit timelike vector $U^{a}$ and a unit spacelike vector $n^{a}_{1}$ as
\begin{align}
\begin{split}
U^{a}=\sqrt{f}\left(V^{a}_{1}+\frac{1}{2f}V^{a}_{2}\right)\ ,\\
n^{a}_{1}=\sqrt{f}\left(V^{a}_{1}-\frac{1}{2f}V^{a}_{2}\right)\ ,
\end{split}
\label{app HE:V1-V2-def}
\end{align}
which along with $n^{a}_{2}$ and $n^{a}_{3}$ form an orthonormal basis. Note that the value of $f$ is strictly positive. By lowering the indices of \eqref{app HE: jordan chain tensor form b plus} and using the above relations we find
\begin{equation}
t_{ab}=(\lambda_{1}+f)U_{a}U_{b}-(\lambda_{1}-f)n_{1a}n_{1b}- f\left(U_{a}n_{1b}+n_{1a}U_{b}\right)-\lambda_{2}n_{2a}n_{2b}-\lambda_{3}n_{3a}n_{3b}\,,
\qquad
\mbox{Type II} \,.
\label{app HE: Type II HE class plus}
\end{equation}
This is exactly the Hawking-Ellis canonical form of Type II. For the second case characterized by $V_{1}\cdot V_{2}=-1$, we find the form of the matrix to be
\begin{equation}
t^{a}_{\ b}=-\lambda_{1}\left(V^{a}_{1}V_{2b}+V^{a}_{2}V_{1b}\right)-V^{a}_{2}V_{2b}-\lambda_{2}n^{a}_{2}n_{2b}-\lambda_{3}n^{a}_{3}n_{3b}\ .\label{app HE: jordan chain tensor form b minus}
\end{equation}
A completely analogous construction gives the form
\begin{equation}
t_{ab}=(\lambda_{1}-f)U_{a}U_{b}-(\lambda_{1}+f)n_{1a}n_{1b}+ f\left(U_{a}n_{1b}+n_{1a}U_{b}\right)-\lambda_{2}n_{2a}n_{2b}-\lambda_{3}n_{3a}n_{3b}\,,
\qquad
\mbox{Type II} \,.
\label{app HE: Type II HE class minus}
\end{equation}
Notice that in this case the sign of the $f$ term in the diagonal terms is flipped in comparison with \eqref{app HE: Type II HE class plus}. Since $f$ is strictly positive one cannot connect the two cases using a Lorentz transformation even when we consider parity and time reversal transformations.

For the form $(c)$ we have a single unit spacelike eigenvector $n^{a}_{3}$ and a Jordan chain $V^{a}_{1}$, $V^{a}_{2}$, $V^{a}_{3}$ associated with the Jordan block. The vectors $V_{1}$ and $V_{3}$ can both be taken to be null and to satisfy $V_{1}\cdot V_{3}=V_{2}\cdot V_{2}=-1$. In such a basis the matrix $t^{a}_{\ b}$ can be written as
\begin{equation}
t^{a}_{\ b}=-\lambda_{1}\left(V^{a}_{1}V_{3b}+V^{a}_{3}V_{1b}+V^{a}_{2}V_{2b}\right)-V^{a}_{2}V_{3b}-V^{a}_{3}V_{2b}-\lambda_{2}n^{a}_{3}n_{3b}\ .\label{app HE: jordan chain tensor form c}
\end{equation}
Similarly to the previous case, we define an orthonormal basis
\begin{align}
\begin{split}
U^{a}&=f\left(V^{a}_{1}-\frac{1}{2f^{2}}V^{a}_{3}\right)\ ,\\
n^{a}_{2}&=f\left(V^{a}_{1}+\frac{1}{2f^{2}}V^{a}_{3}\right)\ , \\
n_{1}^{a}&=-V^{a}_{2} .
\end{split}
\nonumber
\end{align}
By lowering the indices of \eqref{app HE: jordan chain tensor form c} and using the above relations we find
\begin{equation}
t_{ab}=\lambda_{1}U_{a}U_{b}-\lambda_{1}n_{1a}n_{1b}-\lambda_{1}n_{2a}n_{2b}-f\left(U_{a}n_{1b}+n_{1a}U_{b}\right)+f\left(n_{2a}n_{1b}+n_{1a}n_{2b}\right)-\lambda_{2}n_{3a}n_{3b}\,,
\hspace{.2cm}
\mbox{Type III} \,.
\label{app HE: Type III HE class}
\end{equation}
This is the Hawking-Ellis canonical Type III. We would like to stress that the factor $f$ in Type II and Type III represents a residual Lorentz freedom, for which there is no canonical choice like in Type I. Apart from the sign of $f$ in the diagonal terms in Type II, $f$ does not represent any information about the tensor $t_{ab}$ itself. In \cite{Hawking:1973uf}, the authors fix this freedom to $f=1$ in both Type II and Type III.

Finally, form $(f)$ has no non-trivial Jordan chain. It has complex-conjugate eigenvectors $\xi^a$ and $\bar\xi^a$, with eigenvalues $z$ and $\bar z$, and two real spacelike eigenvectors $n_2^a,n_3^a$. The complex eigenvector is null in the Hermitian sense, $\xi\cdot\xi=\eta_{ab}\bar\xi^a\xi^b=0$. The bilinear contraction $\eta_{ab}\xi^a\xi^b=\bar\xi\cdot\xi$ is nonzero and can be normalized to one by a complex rescaling of $\xi^a$.
With these expressions, it is straightforward to see that the matrix $t^{a}_{\ b}$ can be written as
\begin{equation}
t^{a}_{\ b}=z\xi^{a}\xi_{b}+\bar{z}\bar{\xi}^{a}\bar{\xi}_{b}-\lambda_{1}n^{a}_{2}n_{2b}-\lambda_{2}n^{a}_{3}n_{3b} \, ,
\end{equation}
and the corresponding symmetric tensor is
\begin{equation}
t_{ab}=z\;\xi_{a}\xi_{b}+\bar{z}\;\bar{\xi}_{a}\bar{\xi}_{b}-\lambda_{1}n_{2a}n_{2b}-\lambda_{2}n_{3a}n_{3b}\ .
\end{equation}
We can form a real unit timelike vector $U^{a}$ and a real unit spacelike vector $n^{a}_{1}$ as
\begin{align}
\begin{split}
U^{a}=\frac{1}{\sqrt{2}}\left(\xi^{a}+\bar{\xi}^{a}\right)\ ,\\
n^{a}_{1}=\frac{i}{\sqrt{2}} \left(\xi^{a}-\bar{\xi}^{a}\right)\ ,
\end{split}
\nonumber
\end{align}
which, along with $n_{2}$ and $n_{3}$, form an orthonormal basis. Writing the tensor $t_{ab}$ in this basis yields
\begin{equation}
t_{ab}=\mathrm{Re}[z]\left (U_{a}U_{b}-n_{1a}n_{1b}\right )+\mathrm{Im}[z]\left(U_{a}n_{1b}+U_{b}n_{1a}\right)-\lambda_{1}n_{2a}n_{2b}-\lambda_{2}n_{3a}n_{3b}\,,
\qquad
\mbox{Type IV} \,.
\label{app HE: Type IV HE class}
\end{equation}
This is the Hawking-Ellis canonical Type IV. Note that this is not the only form that is found in the literature. For example, we can utilize the Lorentz transformation between $U^{a}$ and $n^{a}_{1}$ in order to eliminate one of the diagonal entries. This brings the form \eqref{app HE: Type IV HE class} to the form presented in \cite{Hawking:1973uf}.


\section{Causally compatible frames}\label{app:causal}
In subsection~\ref{sec:signature} we found that the determinant condition \eqref{det-condition} is necessary and sufficient for the disformal map to preserve the Lorentzian signature, and that this is all that the internal consistency of the map requires. For completeness, in this appendix we impose the stronger, frame-dependent requirement that the two metrics $h_{\mu\nu}$ and $g_{\mu\nu}$ be causally compatible, in the sense that they share a common time direction and a common family of spacelike hypersurfaces and therefore admit compatible local $3+1$ decompositions. This situation arises whenever the two metrics related by the invertible disformal map are both regarded as physical at the same time, with matter minimally coupled to one of them and hence non-minimally to the other. We do not attach a specific physical application to it; our aim is simply to record the conditions it imposes and to show that, as for the signature, they are organized by the Hawking-Ellis type of $t_{\mu\nu}$.

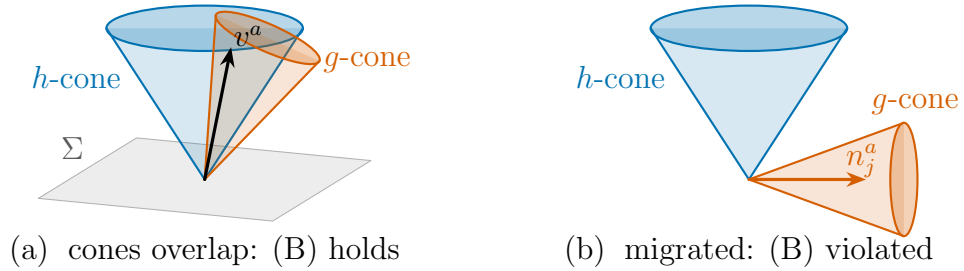
\begin{figure}[!htb]
\centering
\begin{tikzpicture}[>={Stealth[length=2.6mm]},line cap=round,line join=round]			
		\begin{scope}[shift={(0,0)}]
			\fill[gray!14] (-2.2,-0.25) -- (-0.9,0.55) -- (2.2,0.25) -- (0.9,-0.55) -- cycle;
			\draw[gray!60]  (-2.2,-0.25) -- (-0.9,0.55) -- (2.2,0.25) -- (0.9,-0.55) -- cycle;
			\node[gray!70!black] at (-1.75,0.42) {$\Sigma$};
			
			\futurecone{hblue}{1.3}{2}{0.30}
			\node[hblue] at (-1.70,1.35) {$h$-cone};
			
			\begin{scope}[rotate=-24]
				\futurecone{gorange}{0.75}{2.05}{0.18}
			\end{scope}
			\node[gorange] at (2.15,1.55) {$g$-cone};
			
			\draw[->,very thick,black] (0,0) -- (0.35,1.75);
			\node[black] at (0.58,1.90) {$v^a$};
			
			\node at (0,-0.95) {(a)\; cones overlap: (B) holds};
		\end{scope}
		
		\begin{scope}[shift={(7.2,0)}]
			\futurecone{hblue}{1.3}{2}{0.30}
			\node[hblue] at (-1.70,1.35) {$h$-cone};
			
			\begin{scope}[rotate=-90]
				\futurecone{gorange}{0.75}{2.05}{0.18}
			\end{scope}
			\node[gorange] at (2.20,1.00) {$g$-cone};
			
			\draw[->,very thick,gorange] (0,0) -- (1.55,0.0);
			\node[gorange] at (1.5,0.25) {$n_j^a$};
			
			\node at (0,-0.95) {(b)\; migrated: (B) violated};
		\end{scope}
	\end{tikzpicture}
	\caption{Schematic illustration of requirement (B). (a) The cones of $h_{\mu\nu}$ (blue) and $g_{\mu\nu}$ (orange) overlap: a vector $v^a$ timelike with respect to both metrics and a hypersurface $\Sigma$ spacelike with respect to both exist, so requirement (B) holds. (b) A migrated Type I configuration, in which the cone of $g_{\mu\nu}$ opens along a spatial leg $n_j$ of $h_{\mu\nu}$: the interiors are disjoint and (B) is violated.}
	\label{fig:cones-AB}
\end{figure}

Working in the fixed convention $(+,-,-,-)$, two requirements beyond \eqref{det-condition} then arise:
\begin{itemize}
\item[(A)] $g_{\mu\nu}$ must have the same signature $(+,-,-,-)$ as $h_{\mu\nu}$. On the reversed branch, a field with a standard-form kinetic term minimally coupled to $g_{\mu\nu}$ carries kinetic energy of the opposite sign relative to the fields associated with $h_{\mu\nu}$.
	
\item[(B)] There must exist at least one hypersurface spacelike with respect to both metrics and at least one vector timelike with respect to both. These conditions play different roles. A hypersurface with conormal $\partial_\mu S$ can carry initial data for the coupled system if and only if $h^{\mu\nu}\partial_\mu S\,\partial_\nu S>0$ and $g^{\mu\nu}\partial_\mu S\,\partial_\nu S>0$. Thus, the well-posedness of the joint Cauchy problem is determined solely by the cones of the contravariant metrics. The common-timelike-vector condition instead guarantees the existence of subsonic observers, for whom the Hamiltonian of the coupled fluctuations is bounded, and fixes the relative time orientation of the two cones. If it fails, no observer measures bounded energies in both sectors, and two inequivalent relative time orientations remain \cite{Sawicki:2024ryt}.\footnote{In bimetric theories, where two independent metrics interact through a square-root potential, the same geometric condition is called causal coupling and is part of the definition of the theory \cite{HassanKocic:2017}.}
\end{itemize}
We emphasize that (B) does not require every $h$-timelike vector to remain $g$-timelike as such a requirement would impose containment of the timelike cones. Under a disformal transformation the cone of $g_{\mu\nu}$ tilts, widens or narrows relative to that of $h_{\mu\nu}$, and vectors close to the cones change their causal character; condition (B) only demands that the two cone systems overlap. The two situations are illustrated schematically in \Cref{fig:cones-AB}.

In the case-by-case analysis below we implement (A) and (B) for each Hawking-Ellis type.
\begin{itemize}
\item {\bf Type I:}
In the orthonormal eigenbasis $(U^a,n_1^a,n_2^a,n_3^a)$ of \eqref{app HE: Type I HE class},
\begin{equation}
g_{ab} = C\eta_{ab} + D \left(
\lambda_{1}U_{a}U_{b}-\lambda_{2}n_{1a}n_{1b}-\lambda_{3}n_{2a}n_{2b}-\lambda_{4}n_{3a}n_{3b}
\right) \,,
\nonumber
\end{equation}
which takes the diagonal matrix form
\begin{equation}
g_{ab}=
\begin{pmatrix}
C+D\lambda_1 & 0 & 0 & 0\\
0 & -(C+D\lambda_2) & 0 & 0\\
0 & 0 & -(C+D\lambda_3) & 0\\
0 & 0 & 0 & -(C+D\lambda_4)
\end{pmatrix} \,.
\nonumber
\end{equation}
Requirement (A) is satisfied either when all four factors $C+D\lambda_i$ are positive, or when $C+D\lambda_1<0$ and exactly one of the three spatial factors is negative. In the second case the unique timelike direction of $g_{\mu\nu}$ lies along an $h$-spacelike eigenvector. We call this a migrated configuration. It cannot satisfy both parts of (B). To see this, normalize the relevant two-dimensional block to $\mathrm{diag}(-{\mathrm a},{\mathrm b})$, with ${\mathrm a},{\mathrm b}>0$, in an $h$-orthonormal $(U,n_j)$ plane. A common timelike vector exists if and only if ${\mathrm b}>{\mathrm a}$, whereas a common timelike covector exists if and only if ${\mathrm b}< {\mathrm a}$. At ${\mathrm a}={\mathrm b}$ neither strict condition holds.
	
Thus (A) and (B) together require that the original timelike and spacelike eigenvectors retain their causal character,
\begin{align}\label{conditions-signature}
g_{ab}U^aU^b&>0 \, ,
&g_{ab}n_1^an_1^b&<0 \, ,
&g_{ab}n_2^an_2^b&<0 \, ,
&g_{ab}n_3^an_3^b&<0 \, ,
\end{align}
which is equivalent to
\begin{align}\label{conditions-signature-TypeI}
C+D\lambda_1&>0 \, ,
&C+D\lambda_2&>0 \, ,
&C+D\lambda_3&>0 \, ,
&C+D\lambda_4&>0 \, .
\end{align}
	
\item {\bf Type II:}
In this case, using Eq.~\eqref{app HE: Type II HE class plus}, we find
\begin{equation}
g_{ab} = C\eta_{ab} + D \left[ (\lambda_{1}+f)U_{a}U_{b}-(\lambda_{1}-f)n_{1a}n_{1b}- f\left(U_{a}n_{1b}+n_{1a}U_{b}\right)-\lambda_{2}n_{2a}n_{2b}-\lambda_{3}n_{3a}n_{3b}\right] \,,
\nonumber
\end{equation}
or, in matrix form,
\begin{equation}
g_{ab}=
\begin{pmatrix}
C+D(\lambda_1+f) & Df & 0 & 0\\
Df & -C-D(\lambda_1-f) & 0 & 0\\
0 & 0 & -(C+D\lambda_2) & 0\\
0 & 0 & 0 & -(C+D\lambda_3)
\end{pmatrix} \,.
\nonumber
\end{equation}
The non-trivial $2\times 2$ block in the $(U,n_1)$ sector has determinant
\begin{equation}
\det g\big|_{(U,n_1)} = -\left(C+D\lambda_1\right)^2 \,.
\nonumber
\end{equation}
The inertia for the $2\times2$ block is $(+,-)$ or $(-,+)$ as long as $C+D\lambda_1\neq0$. Requirement (A) therefore holds if and only if $C+D\lambda_2>0$ and $C+D\lambda_3>0$, irrespective of the sign of $C+D\lambda_1$. Note that, once (A) holds, the case $C+D\lambda_1<0$ requires $D\neq0$ and hence $Df\neq0$, since for $D=0$ the transformation is conformal and (A) already enforces $C>0$.
	
To check (B), note that the null eigenvector of the Jordan block is $k^a\equiv U^a-n_1^a$: it is $\eta$-null, $\eta_{ab}k^ak^b=0$, and satisfies $t^{a}{}_{b}\,k^b=\lambda_1\,k^a$. In terms of the null vector, we have
\begin{align}\label{TypeII-decomposition}
g_{ab}=\left(C+D\lambda_1\right)\eta_{ab}+Df\,k_ak_b
+D(\lambda_1-\lambda_2)\,n_{2a}n_{2b}+D(\lambda_1-\lambda_3)\,n_{3a}n_{3b}\,.
\end{align}
Contracting \eqref{TypeII-decomposition} with $k^b$ immediately gives
\begin{align}\label{tangency-null}
g_{ab}\,k^b=\left(C+D\lambda_1\right)\eta_{ab}\,k^b \,,
\end{align}
so $k^a$ is null for both $h_{ab}=\eta_{ab}$ and $g_{\mu\nu}$, and the two cones are tangent along $k^a$. The sign selection follows from two exact identities implied by \eqref{TypeII-decomposition}. Decomposing an arbitrary vector as $v^a=v^0U^a+v^1n_1^a+v^2n_2^a+v^3n_3^a$, and denoting the frame components of an arbitrary covector by $(\omega_0,\omega_1,\omega_2,\omega_3)$, we find
\begin{align}
\begin{split}
g_{ab}v^av^b&=\left(C+D\lambda_1\right)\left[(v^0)^2-(v^1)^2\right]+Df\,(v^0+v^1)^2
-\left(C+D\lambda_2\right)(v^2)^2-\left(C+D\lambda_3\right)(v^3)^2\,,
\\
g^{ab}\omega_a\omega_b&=\frac{\omega_0^2-\omega_1^2}{C+D\lambda_1}-\frac{Df\,(\omega_0-\omega_1)^2}{\left(C+D\lambda_1\right)^2}
-\frac{\omega_2^2}{C+D\lambda_2}-\frac{\omega_3^2}{C+D\lambda_3}\,.
\end{split}
\nonumber
\end{align}
If $C+D\lambda_1<0$ and $Df<0$, every $h$-timelike vector is $g$-spacelike, as illustrated in \Cref{fig:typeII-cones}(b). If $C+D\lambda_1<0$ and $Df>0$, every $h$-timelike covector is $g$-spacelike, and no common spacelike hypersurface exists. The case $Df=0$ is incompatible with (A) and $C+D\lambda_1<0$. Hence (B) fails whenever $C+D\lambda_1<0$. Conversely, if $C+D\lambda_1>0$, vectors sufficiently close to $k^a$ on the common timelike side and covectors sufficiently close to the common null covector are timelike for both metrics. Therefore
\begin{align}\label{conditions-signature-TypeII}
C+D\lambda_1&>0 \, ,
&C+D\lambda_2&>0 \, ,
&C+D\lambda_3&>0 \, .
\end{align}	
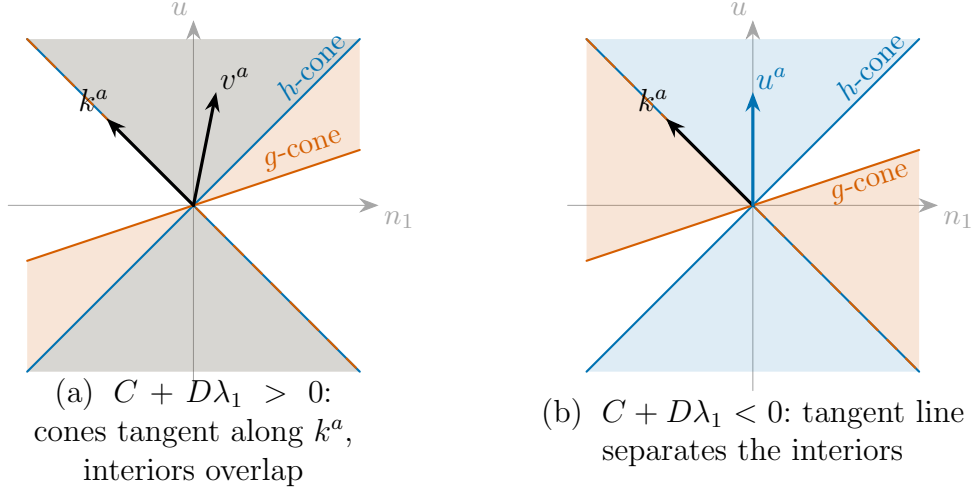
\begin{figure}[!htb]
\centering
\begin{tikzpicture}[>={Stealth[length=2.6mm]},line cap=round,line join=round]
			
			\begin{scope}[shift={(0,0)}]
				\axes
				\fill[gorange,fill opacity=0.16] (0,0) -- (-2.2,2.2) -- (2.2,2.2) -- (2.2,0.7333) -- cycle;
				\fill[gorange,fill opacity=0.16] (0,0) -- (2.2,-2.2) -- (-2.2,-2.2) -- (-2.2,-0.7333) -- cycle;
				\hcone
				\draw[gorange,thick] (-2.2,-0.7333) -- (2.2,0.7333);
				\tangentline
				\draw[->,very thick,black] (0,0) -- (-1.15,1.15);
				\node[black] at (-1.32,1.40) {$k^a$};
				\draw[->,very thick,black] (0,0) -- (0.3,1.5);
				\node[black] at (0.55,1.68) {$v^a$};
				\node[hblue,rotate=45]      at (1.55,1.85) {\small $h$-cone};
				\node[gorange,rotate=18.43] at (1.45,0.72) {\small $g$-cone};
				\node[text width=5.8cm,align=center] at (0,-3.0)
				{(a)\; $C+D\lambda_1>0$:\ cones tangent along $k^a$,\\ interiors overlap};
			\end{scope}
			
			\begin{scope}[shift={(7.4,0)}]
				\axes
				\fill[gorange,fill opacity=0.16] (0,0) -- (2.2,0.7333) -- (2.2,-2.2) -- cycle;
				\fill[gorange,fill opacity=0.16] (0,0) -- (-2.2,-0.7333) -- (-2.2,2.2) -- cycle;
				\hcone
				\draw[gorange,thick] (-2.2,-0.7333) -- (2.2,0.7333);
				\tangentline
				\draw[->,very thick,black] (0,0) -- (-1.15,1.15);
				\node[black] at (-1.32,1.40) {$k^a$};
				\draw[->,very thick,hblue] (0,0) -- (0,1.5);
				\node[hblue] at (0.25,1.68) {$U^a$};
				\node[hblue,rotate=45]      at (1.55,1.85) {\small $h$-cone};
				\node[gorange,rotate=18.43] at (1.55,0.28) {\small $g$-cone}; 
				\node[text width=5.8cm,align=center] at (0,-3.0)
				{(b)\; $C+D\lambda_1<0$:\ tangent line\\ separates the interiors};
			\end{scope}
\end{tikzpicture}
\caption{Sections of the light cones of $h_{\mu\nu}$ (blue) and $g_{\mu\nu}$ (orange) in the $(U,n_1)$ plane for a Type II disformal transformation; shaded wedges indicate the timelike directions of each metric and $s=v^1/v^0$ denotes the slope of $v^a=v^0U^a+v^1n_1^a$. The panels are drawn for $(C+D\lambda_1,\,Df)=(1,\tfrac{1}{2})$ and $(-1,-\tfrac{1}{2})$ respectively, for which the second $g$-null direction lies at $s=3$ in both cases. In accordance with \eqref{tangency-null}, the two cones are tangent along the null eigenvector $k^a=U^a-n_1^a$ of the Jordan block and share the tangent line (dashed). (a) For $C+D\lambda_1>0$ the interiors overlap and a common timelike vector $v^a$ exists. (b) For $C+D\lambda_1<0$ with $Df<0$, the common tangent line separates the interiors: every $h$-timelike vector, e.g.\ $U^a$, is $g$-spacelike; for $Df>0$ requirement (B) fails instead in the covector cones.}\label{fig:typeII-cones}
\end{figure}
	
\item {\bf Type III:}
In this case, the disformal tensor is given by \eqref{app HE: Type III HE class} which yields
\begin{align*}
 g_{ab}&=C\eta_{ab}+D\lambda_1
 \left(U_aU_b-n_{1a}n_{1b}-n_{2a}n_{2b}\right)-D\lambda_2n_{3a}n_{3b}
 \\
 &\quad+Df\left(n_{2a}n_{1b}+n_{1a}n_{2b}-U_an_{1b}-n_{1a}U_b\right)\, .
\end{align*}
or, in matrix form,
\begin{equation}
g_{ab}=
\begin{pmatrix}
C+D\lambda_1 & Df & 0 & 0\\
Df & -(C+D\lambda_1) & Df & 0\\
0 & Df & -(C+D\lambda_1) & 0\\
0 & 0 & 0 & -(C+D\lambda_2)
\end{pmatrix} \,.
\nonumber
\end{equation}
The non-trivial $3\times 3$ block in the $(U,n_1,n_2)$ sector has determinant
\begin{equation}
\det g\big|_{(U,n_1,n_2)} = \left(C+D\lambda_1\right)^3 \,.
\nonumber
\end{equation}
Replacing $Df$ continuously by $\tau Df$, $0\leq\tau\leq1$, leaves this determinant equal to $\left(C+D\lambda_1\right)^3$. If $C+D\lambda_1\neq0$, no eigenvalue can cross zero, so the block has the same inertia as $\operatorname{diag}\bigl( C+D\lambda_1,\allowbreak - (C+D\lambda_1),\allowbreak -(C+D\lambda_1)\bigr)$. Requirement (A) is therefore equivalent to $C+D\lambda_1>0$ and $C+D\lambda_2>0$.
	
The null eigenvector is $k^a=U^a-n_2^a$, and it obeys \eqref{tangency-null}. The vector $U^a$ is timelike for both metrics when $C+D\lambda_1>0$. For covectors, take $\omega_a=k_a+\varepsilon n_{2a}$, with components $(1,0,1-\varepsilon,0)$. Then
\begin{equation*}
h^{ab}\omega_a\omega_b=\varepsilon(2-\varepsilon),
\quad
g^{ab}\omega_a\omega_b=
\frac{\varepsilon(2-\varepsilon)}{C+D\lambda_1}
-\frac{D^2f^2\varepsilon^2}{(C+D\lambda_1)^3} \, .
\end{equation*}
Both expressions are positive for sufficiently small $\varepsilon>0$. Thus (B) follows from (A), and the complete conditions are
\begin{align}\label{conditions-signature-TypeIII}
C+D\lambda_1&>0 \, ,
&C+D\lambda_2&>0 \, .
\end{align}

\item {\bf Type IV:}
Finally, for Type IV, the disformal tensor is given by \eqref{app HE: Type IV HE class} and we find
\begin{equation}
g_{ab} = C\eta_{ab} + D \left[\mathrm{Re}[z]\left (U_{a}U_{b}-n_{1a}n_{1b}\right )+\mathrm{Im}[z]\left(U_{a}n_{1b}+U_{b}n_{1a}\right)-\lambda_{1}n_{2a}n_{2b}-\lambda_{2}n_{3a}n_{3b} \right] \,,
\nonumber
\end{equation}
or, in matrix form,
\begin{equation}
g_{ab}=
\begin{pmatrix}
{\mathrm a}& {\mathrm b} & 0 & 0\\
{\mathrm b} & -{\mathrm a} & 0 & 0\\
0 & 0 & -(C+D\lambda_1) & 0\\
0 & 0 & 0 & -(C+D\lambda_2)
\end{pmatrix} \,,
\nonumber
\end{equation}
where
\begin{align}
{\mathrm a}\equiv C+D\,\mathrm{Re}[z] \,,
\qquad 
{\mathrm b}\equiv -D\,\mathrm{Im}[z] \,.
\end{align}
The non-trivial $2\times2$ block in the $(U,n_1)$ sector is $g\big|_{(U,n_1)}=\left(\begin{smallmatrix} {\mathrm a} & {\mathrm b}\\ {\mathrm b} & -{\mathrm a}\end{smallmatrix}\right)$, with
\begin{equation}
\det g\big|_{(U,n_1)} = -\left({\mathrm a}^2+{\mathrm b}^2\right)<0 \,.
\nonumber
\end{equation}
The block has real nonzero eigenvalues $\pm\sqrt{{\mathrm a}^2+{\mathrm b}^2}$. The $(U,n_1)$ plane is thus automatically Lorentzian, and requirement (A) reduces to the two remaining diagonal directions, giving $C+D\lambda_1>0$ and $C+D\lambda_2>0$.
	
Requirement (B) is automatic as well. For $D=0$, requirement (A) gives $C>0$, so the two causal cones coincide. We therefore take $D\neq0$ in the following. Let $v^a=v^0U^a+v^1 n_1^a$ be an arbitrary vector in the $(U,n_1)$ plane and $s\equiv v^1/v^0$ its slope. Because $h_{ab}=\eta_{ab}$ in this basis, the $h$-cone is the standard one: $h$-null directions have $|s|=1$ (a $45^{\circ}$ cone), $h$-timelike ones $|s|<1$, and $h$-spacelike ones $|s|>1$. The $g$-null directions follow from $g_{ab}v^av^b={\mathrm a}\,(v^0)^2+2{\mathrm b}\,v^0v^1-{\mathrm a}\,(v^1)^2=0$, i.e. ${\mathrm a}+2{\mathrm b}s-{\mathrm a}s^2=0$, whose roots
\begin{equation}
s_\pm=\frac{{\mathrm b}\pm\sqrt{{\mathrm a}^2+{\mathrm b}^2}}{{\mathrm a}}\,,
\qquad
\ell_\pm^a=U^a+s_\pm\,n_1^a\,,
\nonumber
\end{equation}
are real and distinct, and satisfy
\begin{equation}
s_+s_-=-1
\qquad(\text{for }{\mathrm a}=0,\ \ s_\pm=0,\infty)\,.
\nonumber
\end{equation}
Being negative, this product forces $s_+$ and $s_-$ to have opposite signs; being unit in magnitude, $|s_+|\,|s_-|=1$, it forces the two magnitudes to be mutual reciprocals. Hence one root has $|s|<1$ and the other $|s|>1$ (they could coincide at $|s|=1$ only for ${\mathrm b}=0$, the conformal case excluded above). Since the $h$-cone boundary is precisely $|s|=1$, exactly one $g$-null line lies inside the $h$-cone and the other outside, for every ${\mathrm a},{\mathrm b}$, as shown in \Cref{fig:typeIV-cones}. Displacing the interior null line (the one with $|s|<1$) infinitesimally toward its $g$-timelike side keeps it inside the $h$-cone, producing a vector timelike with respect to both metrics. The same conclusion holds for covectors, since the inverse block $g^{ab}\big|_{(U,n_1)}=\tfrac{1}{{\mathrm a}^2+{\mathrm b}^2}\left(\begin{smallmatrix} {\mathrm a} & {\mathrm b}\\ {\mathrm b} & -{\mathrm a}\end{smallmatrix}\right)$ is proportional to the direct block and hence has the identical null structure. Both parts of (B) are therefore satisfied with no restriction on ${\mathrm a},{\mathrm b}$, and the full condition is
\begin{align}\label{conditions-signature-TypeIV}
C + D \lambda_1 > 0 \,,
\qquad
C + D \lambda_2 > 0 \,.
\end{align}
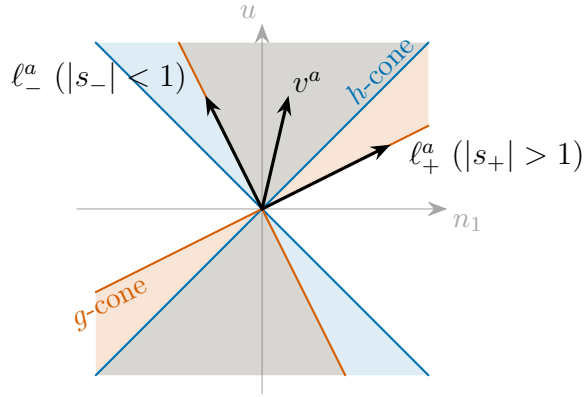
\begin{figure}[!htb]
\centering
\begin{tikzpicture}[>={Stealth[length=2.6mm]},line cap=round,line join=round]			
			\draw[->,gray!70,thin] (-2.45,0) -- (2.45,0);
			\node[gray!70,below right=-2pt] at (2.45,0) {\small $n_1$};
			\draw[->,gray!70,thin] (0,-2.45) -- (0,2.45);
			\node[gray!70,above left=-2pt] at (0,2.45) {\small $U$};
			
			\fill[gorange,fill opacity=0.16] (0,0) -- (-1.1,2.2) -- (2.2,2.2) -- (2.2,1.1) -- cycle;
			\fill[gorange,fill opacity=0.16] (0,0) -- (1.1,-2.2) -- (-2.2,-2.2) -- (-2.2,-1.1) -- cycle;
			
			\fill[hblue,fill opacity=0.13] (0,0) -- (-2.2,2.2) -- (2.2,2.2) -- cycle;
			\fill[hblue,fill opacity=0.13] (0,0) -- (2.2,-2.2) -- (-2.2,-2.2) -- cycle;
			\draw[hblue,thick] (-2.2,-2.2) -- (2.2,2.2);
			\draw[hblue,thick] (-2.2,2.2)  -- (2.2,-2.2);
			
			\draw[gorange,thick] (1.1,-2.2) -- (-1.1,2.2);
			\draw[gorange,thick] (-2.2,-1.1) -- (2.2,1.1);
			
			\draw[->,very thick,black] (0,0) -- (-0.75,1.5);
			\node[black,anchor=east] at (-0.85,1.72) {$\ell_-^a\ (|s_-|<1)$};
			\draw[->,very thick,black] (0,0) -- (1.7,0.85);
			\node[black,anchor=west] at (1.80,0.68) {$\ell_+^a\ (|s_+|>1)$};
			
			\draw[->,very thick,black] (0,0) -- (0.35,1.5);
			\node[black] at (0.60,1.68) {$v^a$};
			
			\node[hblue,rotate=45]      at (1.55,1.85) {\small $h$-cone};
			\node[gorange,rotate=26.57] at (-2.0,-1.28) {\small $g$-cone};
\end{tikzpicture}
\caption{Section of the light cones of $h_{\mu\nu}$ (blue) and $g_{\mu\nu}$ (orange) in the $(U,n_1)$ plane for a Type IV disformal transformation; shaded wedges indicate the timelike directions of each metric. The figure is drawn for $({\mathrm a},{\mathrm b})=(1,\tfrac{3}{4})$, for which the $g$-null directions $\ell^a_\pm=U^a+s_\pm n_1^a$ have slopes $s_-=-\tfrac{1}{2}$ and $s_+=2$, in accordance with $s_+s_-=-1$. Exactly one $g$-null direction lies inside the $h$-cone and the other outside, so the two interiors always overlap and a common timelike vector $v^a$ exists: requirement (B) holds without any restriction on ${\mathrm a}$ and ${\mathrm b}$.}\label{fig:typeIV-cones}
\end{figure}
	
\end{itemize}

The four cases give the compact necessary-and-sufficient condition
\begin{align}\label{conditions-signature-unified}
C+D\lambda_i>0
\qquad
\text{for every real eigenvalue $\lambda_i$ of $t^{\mu}{}_{\nu}$} \, .
\end{align}
For a fixed tensor $t_{\mu\nu}$ and a fixed Hawking-Ellis sector, \eqref{conditions-signature-unified} is necessary and sufficient for $g_{\mu\nu}$ to have the convention $(+,-,-,-)$ and to be locally causally compatible with $h_{\mu\nu}$. The complex pair of Type IV imposes no additional inequality. The region defined by \eqref{conditions-signature-unified} is an intersection of open half-planes in $(C,D)$ and is therefore the component containing the identity $(C,D)=(1,0)$. Reaching another component at fixed $t_{\mu\nu}$ requires at least one real factor to vanish, at which point the metric is singular.

Condition \eqref{det-condition} is weaker. It is the complete pointwise criterion only when one asks whether $g_{\mu\nu}$ is Lorentzian up to overall sign. It also admits the branch $g_{\mu\nu}\mapsto-g_{\mu\nu}$ and, for Type I, the migrated branch. Neither branch can be rejected solely from the determinant. Their physical acceptability depends on the action, the kinetic normalization of the fields, and whether a common local time direction and spacelike hypersurface are required.

Finally, on a branch where the functional inverse exists, $C\neq0$, and the trace map discussed in subsection~\ref{sec:inverse-map} is locally invertible, the inverse coefficients satisfy $C'=1/C$ and $D'=-D/C$. The eigenvalues evaluated with the metric $g_{\mu\nu}$ are $\lambda_i'=\lambda_i/(C+D\lambda_i)$, and $C'+D'\lambda_i'=\frac{1}{C+D\lambda_i}$. Thus the strong inequalities are stable under such an inverse. This last statement is conditional on functional invertibility; pointwise non-degeneracy of $g_{\mu\nu}$ alone is not enough.


\section{Conditions for polynomial zero modes}\label[appendix]{app:polynomial}

This appendix gives sufficient conditions for all dual zero modes to
have the polynomial forms \eqref{zeta-CH4}-\eqref{zeta-CH2}. We use the normal form \eqref{DT-simplified}, with no metric
derivatives, and vary the metric while holding all other fields fixed.
Both metrics are non-degenerate, and we work where the Cayley-Hamilton
degree $n$ does not change. Indices on $t_{\mu\nu}$ and its powers
are raised with $h^{\mu\nu}$ as in \eqref{t-HO}.

There are two questions: whether $C,D$ depend on the metric only
through traces, and whether $t_{\mu\nu}$ itself depends on the metric.
\Cref{table-poly-tests} lists the four cases. Passing all checks for
a given case guarantees that every dual zero mode is polynomial.
Otherwise, polynomiality remains possible and must be checked directly
using \eqref{poly-full-kernel-app}.

\begin{table}[H]
\centering
\renewcommand{\arraystretch}{1.35}
\begin{tabular}{|p{0.24\textwidth}|p{0.23\textwidth}|p{0.40\textwidth}|}
\hline
Metric dependence of $C,D$ & Metric dependence of $t_{\mu\nu}$
& Sufficient checks for all dual zero modes to be polynomial
\\ \hline
Only through traces $[t^I]$ & None & $C\neq0$. No other check is needed.
\\ \hline
Only through traces $[t^I]$ & Present & Closure (T) and invertibility (B). Condition (P) then follows.
\\ \hline
Additional invariants & None & $C\neq0$ and polynomial coefficient derivatives (P). Conditions (T) and (B) follow.
\\ \hline
Additional invariants & Present & Polynomial coefficient derivatives (P), closure (T), and invertibility (B).
\\ \hline
\end{tabular}
\caption{Sufficient conditions for the normal-form map \eqref{DT-simplified}. Checks (P), (T), and (B) are defined in \eqref{poly-response-app}, \eqref{poly-closure-app}, and \eqref{poly-B-invertible-app}. Explicit dependence on fields held fixed in the metric variation is allowed in all four cases.}\label{table-poly-tests}
\end{table}

Conformal transformations with $D$ identically zero admit the simpler
criterion \eqref{poly-conformal-check-app}: only the full metric
derivative of $C$ needs to be checked. Condition (T) is unnecessary,
and (B) follows from non-degeneracy of the conformal metric.
Setting $D=0$ at a single configuration is not enough, since
$\partial D/\partial h_{\mu\nu}$ may still be nonzero.

\subsection{The three checks}\label{app:poly-guide}

For convenience, we introduce the following quantities:
\begin{align}
\label{poly-metric-gradients-app}
R_C^{\rho\sigma}&\equiv-\frac{\partial C}{\partial h_{\rho\sigma}}\, ,
\qquad
R_D^{\rho\sigma}\equiv-\frac{\partial D}{\partial h_{\rho\sigma}}\, ,
\\
\label{poly-B-app}
{\cal B}^{\rho\sigma}_{\alpha\beta}
&\equiv C\delta_\alpha^{(\rho}\delta_\beta^{\sigma)}
+D\frac{\partial t_{\alpha\beta}}{\partial h_{\rho\sigma}}\, ,
\\
\label{poly-contractions-app}
r_h&\equiv h_{\alpha\beta}\zeta_\star^{\alpha\beta}\, ,
\qquad
r_t\equiv t_{\alpha\beta}\zeta_\star^{\alpha\beta}\, .
\end{align}
Here the metric derivatives are taken at fixed underlying fields and
include all metric dependence of $C$ and $D$. With these definitions,
Eq.~\eqref{eigenvalue-system} takes the form
\begin{align}\label{poly-full-kernel-app}
{\cal B}^{\rho\sigma}_{\alpha\beta}\zeta_\star^{\alpha\beta} =r_hR_C^{\rho\sigma}+r_tR_D^{\rho\sigma}\, .
\end{align}

To study polynomial zero modes, we introduce the basis
\begin{align}\label{poly-basis-app}
E_0^{\mu\nu}\equiv h^{\mu\nu}\, ,\qquad
E_j^{\mu\nu}\equiv t^{j\,\mu\nu}\, ,
\qquad j=1,\ldots,n-1\, .
\end{align}
Since $n$ is the Cayley-Hamilton degree, these $n$ tensors are
linearly independent and form a basis for all polynomials in
$t^\mu{}_\nu$.

The three checks listed in \Cref{table-poly-tests} are as follows.

\begin{enumerate}
	
\item[(P)] Polynomial coefficient derivatives. Check whether
\begin{align}\label{poly-response-app}
R_C^{\rho\sigma}=\sum_{j=0}^{n-1}r_{Cj}E_j^{\rho\sigma}\, ,\qquad
R_D^{\rho\sigma}=\sum_{j=0}^{n-1}r_{Dj}E_j^{\rho\sigma}\, .
\end{align}
All metric-dependent arguments must be included in the derivatives. A vanishing derivative is already polynomial.
	
\item[(T)] Closure of the tensor's metric response. Check whether
\begin{align}\label{poly-closure-app}
E_i^{\alpha\beta} \frac{\partial t_{\alpha\beta}}{\partial h_{\rho\sigma}} =\sum_{j=0}^{n-1}\kappa_{ji}E_j^{\rho\sigma}\, ,
\qquad 
i=0,\ldots,n-1\, .
\end{align}
The scalar coefficients $\kappa_{ji}$ are determined by the given $t_{\mu\nu}$. This condition does not follow from the Cayley-Hamilton relation. For metric-independent $t_{\mu\nu}$, it holds with $\kappa_{ji}=0$.
	
\item[(B)] Invertibility of the tensor map. Check whether
\begin{align}\label{poly-B-invertible-app}
\det {\cal B}\neq0\, .
\end{align}
This is the determinant of the $10\times10$ map on symmetric tensors, not $\det g$ or the determinant of the full Jacobian $J$. If $D=0$ or $t_{\mu\nu}$ is metric independent, then ${\cal B}=C\,\mathrm{I}$, and (B) reduces to $C\neq0$.
	
\end{enumerate}

\subsection{Coefficients depending only on traces}\label{app:poly-traces}

Here the metric dependence of $C,D$ enters only through the traces
\eqref{t-Trace-def}. Explicit dependence on the fields held fixed
in the metric variation is allowed.

\subsubsection{Metric-independent tensor}\label{app:poly-independent}

For metric-independent $t_{\mu\nu}$, only $C\neq0$ needs to be checked.
The coefficient $D$ need not vanish.

Varying the inverse metrics in \eqref{t-Trace-def} gives
\begin{align}\label{poly-trace-independent-app}
\frac{\partial[t^I]}{\partial h_{\rho\sigma}}
=-I t^{I\,\rho\sigma}\, ,\qquad I=1,\ldots,4\, .
\end{align}
The $I$ inverse metrics give identical contributions by cyclicity
of the trace. The chain rule and \eqref{poly-full-kernel-app} then give
\begin{align}\label{poly-forced-app}
\zeta_\star^{\rho\sigma}
=\frac1C\sum_{I=1}^{4}I
\left(r_hC_{[t^I]}+r_tD_{[t^I]}\right)t^{I\,\rho\sigma}\, .
\end{align}
Every zero mode is therefore polynomial without making an ansatz.
The Cayley-Hamilton relations reduce its degree to at most $n-1$. The scalar example \eqref{t-phi} and the trace-dependent two-scalar
example \eqref{poly-conformal} belong to this case.

\subsubsection{Metric-dependent tensor}\label{app:poly-dependent}

For metric-dependent $t_{\mu\nu}$, conditions (T) and (B) are
sufficient. Neither follows from trace dependence alone.

The trace variation now reads
\begin{align}\label{poly-trace-general-app}
\frac{\partial[t^I]}{\partial h_{\rho\sigma}}
=-I t^{I\,\rho\sigma}
+I t^{I-1\,\alpha\beta}
\frac{\partial t_{\alpha\beta}}{\partial h_{\rho\sigma}}\, ,
\qquad 
t^{0\,\alpha\beta}\equiv h^{\alpha\beta}\, .
\end{align}
For $I=1,2$, this reproduces \eqref{Dth-CH2}. By (T) and the Cayley-Hamilton relations, both terms are polynomial. Thus, for $C$ and $D$ depending only on the traces $[t^I]$, the closure condition \eqref{poly-closure-app} and the invertibility condition \eqref{poly-B-invertible-app} guarantee that every dual zero mode is polynomial.

\subsection{Coefficients with additional metric-dependent invariants}\label{app:poly-extra}

Here $C,D$ may depend on metric-dependent invariants other than the traces. Their full derivatives \eqref{poly-metric-gradients-app} must therefore be used to check (P).

\subsubsection{Metric-independent tensor}\label{app:poly-extra-independent}

For metric-independent $t_{\mu\nu}$, the closure condition \eqref{poly-closure-app} holds automatically, while the invertibility
condition \eqref{poly-B-invertible-app} holds for $C\neq0$. Therefore, we only need to check that the full metric derivatives
of $C$ and $D$ satisfy \eqref{poly-response-app}.
In this case, \eqref{poly-full-kernel-app} reduces to
\begin{align}\label{poly-independent-full-app}
\zeta_\star^{\rho\sigma}
=\frac{r_hR_C^{\rho\sigma}+r_tR_D^{\rho\sigma}}{C}\, .
\end{align}
By (P), the right-hand side is polynomial, so every dual zero mode
has this form.

For the scalar-gauge example, \eqref{metric-variations-SG} gives
\begin{align}\label{poly-SG-response-app}
R_C^{\mu\nu}=0\, ,\qquad
R_D^{\mu\nu}
=D_{Y_\phi}t^{\mu\nu}
+\frac{Y_2}{2}D_{Y_2}h^{\mu\nu}\, .
\end{align}
Thus (P) holds even though $Y_2$ is not a trace of the scalar tensor. Since $C=1$ and $t_{\mu\nu}$ is metric independent,
\eqref{poly-independent-full-app} gives
$\zeta_\star^{\mu\nu}=r_tR_D^{\mu\nu}$.
Thus \eqref{poly-SG-response-app} shows that every dual zero mode is polynomial.

\subsubsection{Metric-dependent tensor}\label{app:poly-extra-dependent}

When $t_{\mu\nu}$ also depends on the metric, (P), (T), and (B)
are sufficient. The following proof also covers the trace-dependent
case above.

By (T), ${\cal B}$ maps polynomial tensors to polynomial tensors:
\begin{align}\label{poly-restricted-B-app}
{\cal B}^{\rho\sigma}_{\alpha\beta}E_i^{\alpha\beta}
=
\sum_{j=0}^{n-1}M_{ji}E_j^{\rho\sigma}\, ,\qquad
M_{ji}\equiv C\delta_{ji}+D\kappa_{ji}\, .
\end{align}
By (B), the $n\times n$ matrix $M_{ji}$ is invertible. If $M$ were singular, ${\cal B}$ would also have a nonzero
polynomial zero mode, contradicting (B).

For a given dual zero mode, let $r_h$ and $r_t$ be defined by
\eqref{poly-contractions-app}. We seek a polynomial tensor
$\widehat\zeta^{\mu\nu}=\sum_{i=0}^{n-1}b_iE_i^{\mu\nu}$,
with $b_0=a$, satisfying \eqref{poly-full-kernel-app}.
Using \eqref{poly-restricted-B-app} and \eqref{poly-response-app},
we obtain
\begin{align}\label{poly-coefficient-system-app}
\sum_{i=0}^{n-1}M_{ji}b_i
=r_hr_{Cj}+r_tr_{Dj}\, ,
\qquad j=0,\ldots,n-1\, .
\end{align}
Since $M_{ji}$ is invertible, these equations determine $b_i$.
Using \eqref{poly-restricted-B-app},
\eqref{poly-coefficient-system-app}, and \eqref{poly-response-app},
we find
\begin{align}
{\cal B}^{\rho\sigma}_{\alpha\beta}
\widehat\zeta^{\alpha\beta}
=
r_hR_C^{\rho\sigma}+r_tR_D^{\rho\sigma}\, .\nonumber
\end{align}
Comparing with \eqref{poly-full-kernel-app}, the invertibility
condition \eqref{poly-B-invertible-app} implies
$\zeta_\star^{\mu\nu}=\widehat\zeta^{\mu\nu}$.
Thus every dual zero mode is polynomial.

\subsection{Conformal shortcut and the remaining examples}\label{app:poly-examples}

When $D$ vanishes identically, non-degeneracy gives $C\neq0$.
Since $R_D^{\mu\nu}=0$ and ${\cal B}=C\,\mathrm{I}$,
\eqref{poly-full-kernel-app} becomes
\begin{align}\label{poly-conformal-check-app}
\zeta_\star^{\mu\nu}
=\frac{r_h}{C}R_C^{\mu\nu}\, ,\qquad
C=h_{\mu\nu}R_C^{\mu\nu}
\quad\text{for a nonzero dual zero mode}\, .
\end{align}
A nonzero mode requires $r_h\neq0$, so contraction of the first
relation gives the second. Conversely, the second relation makes
$R_C^{\mu\nu}$ a nonzero dual zero mode. The kernel is then
one-dimensional and is polynomial if and only if $R_C^{\mu\nu}$
is a linear combination of the tensors \eqref{poly-basis-app}.
This criterion is exact and requires neither trace dependence nor (T).

For the conformal gauge example, \eqref{eigenvalue-system-CH2-GF} gives
\begin{align}\label{poly-gauge-response-app}
R_C^{\mu\nu}
=2C_{Y_1}t^{\mu\nu}
+\frac{Y_2}{2}C_{Y_2}h^{\mu\nu}\, ,\qquad
R_D^{\mu\nu}=0\, .
\end{align}
Thus the dual zero mode on the singular branch is polynomial,
without checking (T). 

For the general two-scalar conformal factor, $R_C^{\mu\nu}$ is the expression in parentheses in \eqref{dual-tensor-2SF}. The independent coefficients of $u^\mu u^\nu$ and $v^\mu v^\nu$ mean that polynomiality is not automatic. Restricting to $C=C(I_1,I_2)$, \eqref{traces-2SF} gives
\begin{align}\label{poly-2SF-response-app}
R_C^{\mu\nu}
=C_{I_1}t^{\mu\nu}+2C_{I_2}t^{2\,\mu\nu}\, ,\qquad
R_D^{\mu\nu}=0\, .
\end{align}
This is the polynomial branch used in \eqref{poly-conformal},
\eqref{transition-II-2SF}, and \eqref{transition-IV-2SF}.

Finally, for any normal-form transformation, a polynomial dual mode
gives
\begin{align}\label{poly-EMT-product-app}
T^{\star\mu}{}_\nu
&=\tilde\rho\,\zeta_\star^{\mu\alpha}g_{\alpha\nu}=\tilde\rho
\left(a\delta^\mu{}_\alpha
+\sum_{i=1}^{n-1}b_i t^{i\,\mu}{}_\alpha\right)
\left(C\delta^\alpha{}_\nu+D t^\alpha{}_\nu\right)\, ,
\end{align}
by \eqref{EMT-star}, with the index on $T^{\star\mu}{}_\nu$ raised
using $g^{\mu\nu}$. The Cayley-Hamilton relations reduce this product
to \eqref{EMT-CH4}-\eqref{EMT-CH2}. Sums of such sources
have the same form, so the classification in
\Cref{table-EMT} applies to the resulting mimetic source.


\section{Fluid interpretation and example calculations}\label{app:fluid-examples}

\subsection{Fluid quantities}\label{app:fluid}

All index operations in this subsection use the physical metric $g_{\mu\nu}$.
For any symmetric energy-momentum tensor and a unit timelike vector
$U^\mu$, with $U_\mu U^\mu=1$, write
\begin{align}\label{fluid-decomposition-general}
T_{\mu\nu}=\rho U_\mu U_\nu+2U_{(\mu}q_{\nu)} - p\Delta_{\mu\nu}+\pi_{\mu\nu}\, ,
\qquad
\Delta_{\mu\nu}\equiv g_{\mu\nu}-U_\mu U_\nu\, ,
\end{align}
where
\begin{align}\label{fluid-decomposition-definitions}
\rho&=T_{\mu\nu}U^\mu U^\nu\, ,\qquad
q_\mu=\Delta_\mu{}^\alpha T_{\alpha\beta}U^\beta\, ,\nonumber\\
p&=-\frac13\Delta^{\mu\nu}T_{\mu\nu}\, ,\qquad
\pi_{\mu\nu}=\Delta_\mu{}^\alpha\Delta_\nu{}^\beta T_{\alpha\beta} +p\Delta_{\mu\nu}\, .
\end{align}
Here $\rho$ is the energy density, $q_\mu$ the energy flux, $p$ the mean
pressure, and $\pi_{\mu\nu}$ the anisotropic stress measured in the
$U^\mu$ frame. They satisfy $q_\mu U^\mu=0$,
$\pi_{\mu\nu}U^\nu=0$, and $g^{\mu\nu}\pi_{\mu\nu}=0$.
A perfect fluid has $q_\mu=0$ and $\pi_{\mu\nu}=0$, so
\begin{align}\label{fluid-perfect}
 T_{\mu\nu}=(\rho+p)U_\mu U_\nu-pg_{\mu\nu}\, .
\end{align}
Dust corresponds to $p=0$, as in \eqref{EMT-S}, radiation to $p=\rho/3$,
as in \eqref{radiation-SG}, and vacuum energy to $p=-\rho$, as in
\eqref{EMT-star-GF-X2}. These refer to the local perfect-fluid source,
not to a choice of microscopic matter fields.

For a tensor supported on a Lorentzian two-plane, choose a unit spacelike
vector $n^\mu$ with $n_\mu n^\mu=-1$ and $U_\mu n^\mu=0$. Then
\begin{align}\label{fluid-two-plane}
T_{\mu\nu}=\rho U_\mu U_\nu+2q_\parallel U_{(\mu}n_{\nu)}
+p_\parallel n_\mu n_\nu\, .
\end{align}
The longitudinal pressure $p_\parallel$ and flux coefficient $q_\parallel$ are
\begin{align}\label{fluid-two-plane-definitions}
\rho=T_{\mu\nu}U^\mu U^\nu\, ,\qquad
q_\parallel=-T_{\mu\nu}U^\mu n^\nu\, ,\qquad
p_\parallel=T_{\mu\nu}n^\mu n^\nu\, .
\end{align}
Both transverse pressures vanish, and
\eqref{fluid-decomposition-definitions} gives
\begin{align}\label{fluid-two-plane-projections}
p=\frac{p_\parallel}{3}\, ,\qquad q_\mu=q_\parallel n_\mu\, ,\qquad
\pi_{\mu\nu}=p_\parallel\left(n_\mu n_\nu+\frac13\Delta_{\mu\nu}\right)\, .
\end{align}
Thus $p_\parallel$ is not the mean pressure $p$. For the string-like
source \eqref{2SF-string-T}, one has $p_\parallel=-\rho$ and
$q_\parallel=0$, but $\pi_{\mu\nu}\neq0$ for $\rho\neq0$, unlike vacuum energy.

For a Type I energy-momentum tensor, one may choose $U^\mu$ to be its
timelike eigenvector, in which case $q_\mu=0$. If, in addition, the
mixed tensor has one doubly degenerate eigenspace containing the
timelike direction and another doubly degenerate spacelike eigenspace,
an orthonormal principal frame $(U^\mu,n^\mu,e_2^\mu,e_3^\mu)$ gives
\begin{align}\label{fluid-principal-degenerate}
T_{\mu\nu}=\rho U_\mu U_\nu-\rho n_\mu n_\nu
+p_\perp\left(e_{2\mu}e_{2\nu}+e_{3\mu}e_{3\nu}\right)\, .
\end{align}
Here $p_\parallel=-\rho$ and
$p_\perp=T_{\mu\nu}e_2^\mu e_2^\nu=T_{\mu\nu}e_3^\mu e_3^\nu$, so
$p=(p_\parallel+2p_\perp)/3$ by \eqref{fluid-decomposition-definitions}.
This is the form used for the non-null gauge-field source
\eqref{EMT-star-GF}.

\subsection{Scalar-gauge singular transformation}\label{app:scalar-gauge}

For \eqref{dis-T-SG}, metric variation at fixed $\phi$ and $A_\mu$ gives
\begin{align}\label{metric-variations-SG}
\frac{\partial Y_\phi}{\partial h_{\rho\sigma}}
=-u^\rho u^\sigma\, ,\qquad
\frac{\partial Y_2}{\partial h_{\rho\sigma}}
=-\frac{Y_2}{2}h^{\rho\sigma}\, .
\end{align}
The first identity follows from \eqref{invariants-SV}, with $u^\mu$
raised as in \eqref{dis-T-inv-S}. The second follows from the
metric-independent density \eqref{gauge-volume-invariant}.
Thus \eqref{J-def} becomes
\begin{align}\label{Jacobian-SG}
J^{\rho\sigma}_{\mu\nu}
&=\delta_\mu^{(\rho}\delta_\nu^{\sigma)}-u_\mu u_\nu V^{\rho\sigma}\, ,
\qquad
V^{\rho\sigma}\equiv D_{Y_\phi}u^\rho u^\sigma
+\frac{Y_2}{2}D_{Y_2}h^{\rho\sigma}\, ,
\\
\label{Jacobian-det-SG}
\det J&=1-Y_\phi^2D_{Y_\phi}-\frac{Y_\phi Y_2}{2}D_{Y_2}\, .
\end{align}
On the singular branch \eqref{singularity-SG},
$u_\rho u_\sigma V^{\rho\sigma}=1$, so $J^2=J$.
The full ten-component Jacobian has nine unit eigenvalues and one
zero eigenvalue. Its right zero mode is $u_\mu u_\nu$, and its dual
zero mode is $\zeta_\star^{\mu\nu}=V^{\mu\nu}$.

Using $s=Y_2/\sqrt{Y_\phi}$ from \eqref{D-general-SG}, the singularity
condition \eqref{singularity-SG} at fixed $s$ becomes
\begin{align}\label{integration-SG}
\left.\frac{\partial D}{\partial Y_\phi}\right|_s
=D_{Y_\phi}+\frac{Y_2}{2Y_\phi}D_{Y_2}
=\frac1{Y_\phi^2}\, .
\end{align}
Integration gives \eqref{D-general-SG}, and differentiation yields
\eqref{dual-tensor-SG}. There is also a finite local symmetry. Under
$h_{\mu\nu}\to h_{\mu\nu}+\varepsilon u_\mu u_\nu$,
\begin{align}\label{finite-symmetry-SG}
Y_\phi\to\frac{Y_\phi}{1+\varepsilon Y_\phi}\, ,\qquad
Y_2\to\frac{Y_2}{\sqrt{1+\varepsilon Y_\phi}}\, ,\qquad
s\to s\, ,\qquad D\to D-\varepsilon\, .
\end{align}
Hence $g_{\mu\nu}$ is unchanged, provided the shifted auxiliary metric
stays on the same regular branch. This is a pure-disformal, rather than
Weyl, redundancy.

To obtain the source, use
$g_{\mu\alpha}u^\alpha={\cal A}u_\mu$ and
$g_{\mu\alpha}g_{\nu\beta}h^{\alpha\beta}
=g_{\mu\nu}+D{\cal A}u_\mu u_\nu$ in \eqref{EMT-star}.
Together with ${\cal A}=Y_\phi f_{\rm SG}(s)$ from \eqref{D-general-SG},
these identities give \eqref{EMT-star-SG}.
The physical-frame relations \eqref{physical-frame-SG} imply
\begin{align}\label{physical-invariants-SG}
 s=\frac{X_2}{\sqrt{X_\phi}}\, ,\qquad
 X_\phi f_{\rm SG}\!\left(\frac{X_2}{\sqrt{X_\phi}}\right)=1\, .
\end{align}
For \eqref{D-w-SG}, this constraint is
$bX_\phi^{1+w}X_2^{-2w}=1$, including the dust limit $bX_\phi=1$.
For $w=1/3$ it reduces to \eqref{radiation-SG}.

The radiation constraint \eqref{radiation-SG} can be imposed in the
physical frame by
\begin{align}\label{action-radiation-SG}
 S_\star=\frac12\int\D^4x\sqrt{-g}\,\lambda
 \left(X_2-b^{3/2}X_\phi^2\right)\, .
\end{align}
On the constraint, metric variation gives
\begin{align}\label{radiation-source-SG}
 T^\star_{\mu\nu}
 &=\frac{\lambda b^{3/2}}2
 \left(4X_\phi u_\mu u_\nu-X_\phi^2g_{\mu\nu}\right)\, ,\nonumber\\
 \rho&=\frac32\lambda b^{3/2}X_\phi^2\, ,\qquad
 p=\frac12\lambda b^{3/2}X_\phi^2\, .
\end{align}
The fluid projections are those of \eqref{fluid-decomposition-definitions}.
This agrees with \eqref{radiation-SG} after a rescaling of the multiplier.
The gauge and scalar equations are
\begin{align}\label{radiation-eom-SG}
 \widetilde F_g^{\mu\nu}\partial_\mu\lambda=0\, ,\qquad
 \nabla_\mu\left(\lambda X_\phi\nabla^\mu\phi\right)=0\, ,
\end{align}
where the dual is defined in \eqref{gauge-dual-def}, and the covariant
derivative uses $g_{\mu\nu}$. Since $X_2\neq0$, the gauge equation
implies that $\lambda$ is constant, as in the Pontryagin construction
of Ref.~\cite{Hammer:2020dqp}. For nonzero $\lambda$ and homogeneous
$\phi(t)$ in an FLRW metric with scale factor $a(t)$,
\eqref{radiation-eom-SG} gives $\partial_t(a^3\dot\phi^3)=0$.
Hence $\dot\phi\propto a^{-1}$ and $\rho\propto a^{-4}$.
This is an irrotational radiation-fluid realization, not a microscopic
description of photons. These results do not establish perturbative
stability of the full transformed theory.

\subsection{Hawking-Ellis types of the two-scalar disformal tensor}\label{app:two-scalar-types}

The tensor \eqref{dis-T-2SF} can also be written as
\begin{align}\label{field-rotation-2SF}
t_{\mu\nu}=\partial_\mu\varphi\partial_\nu\varphi
-\partial_\mu\psi\partial_\nu\psi \, ,\qquad
\varphi\equiv\frac{\phi+\chi}{\sqrt{2}} \, ,\qquad
\psi\equiv\frac{\phi-\chi}{\sqrt{2}} \, .
\end{align}
This is \eqref{t-SV} with its scalar replaced by $\varphi$,
$A_\mu=\partial_\mu\psi$, $d_A=-1$, and $d_{\phi A}=0$. The relative minus sign
concerns the definition of the disformal tensor, not a kinetic term in
the action, and does not by itself imply a ghost.

For independent gradients, let ${\cal V}=\mathrm{span}\{u^\mu,v^\mu\}$,
with indices raised using $h^{\mu\nu}$, as in \eqref{invariants-2SF}.
Direct application of \eqref{dis-T-2SF} gives
\begin{align}\label{matrix-2SF}
t^\mu{}_\nu\big|_{\cal V}
=\begin{pmatrix}Y_{\phi\chi}&Y_\chi\\Y_\phi&Y_{\phi\chi}\end{pmatrix} \, ,
\qquad
G=\begin{pmatrix}Y_\phi&Y_{\phi\chi}\\Y_{\phi\chi}&Y_\chi\end{pmatrix} \, ,
\end{align}
where $G$ is the Gram matrix of the gradients. If $Y_\phi Y_\chi-Y_{\phi\chi}^2\neq0$, the
orthogonal complement of ${\cal V}$ supplies two independent zero
eigenvectors. The tensor is Type I for $Y_\phi Y_\chi>0$, Type IV for $Y_\phi Y_\chi<0$, and
Type II for $Y_\phi Y_\chi=0$ with $(Y_\phi,Y_\chi)\neq(0,0)$. The exceptional case
$Y_\phi=Y_\chi=0$, $Y_{\phi\chi}\neq0$ is Type I, since the first matrix in
\eqref{matrix-2SF} is proportional to the identity.

For $Y_\phi Y_\chi=Y_{\phi\chi}^2>0$, the gradient plane is degenerate and the tensor has rank
two, while zero has algebraic multiplicity three. Its Jordan structure
is therefore $[(21)1]$, of Type II. For $Y_\phi=Y_{\phi\chi}=0$, $Y_\chi<0$, or the case
obtained by interchanging the gradients, Eq.~\eqref{powers-2SF} instead
gives $t^3=0$ and $t^2\neq0$. This is Type III $[(31)]$.
Dependent gradients give a rank-one tensor, which is Type I for a
non-null direction and Type II for a nonzero null direction. The zero
tensor is Type I.

In a Minkowski chart, \Cref{table-2SF} gives one representative of each
type. The coordinates are $(t,x^1,x^2,x^3)$ and
$h_{\mu\nu}=\mathrm{diag}(1,-1,-1,-1)$.
\begin{table}[H]
\centering
\renewcommand{\arraystretch}{1.3}
\begin{tabular}{|c|c|c|c|c|}
\hline
Hawking-Ellis type & $\phi$ & $\chi$ & $(Y_\phi,Y_\chi,Y_{\phi\chi})$ & Segre bracket
\\ \hline
Type I & $x^1$ & $x^2$ & $(-1,-1,0)$ & $[(1,1)11]$
\\ \hline
Type II & $t+x^1$ & $t$ & $(0,1,1)$ & $[2(11)]$
\\ \hline
Type III & $t+x^1$ & $x^2$ & $(0,-1,0)$ & $[(31)]$
\\ \hline
Type IV & $t$ & $x^1$ & $(1,-1,0)$ & $[z\bar z(11)]$
\\ \hline
\end{tabular}
\caption{Local representatives of the four Hawking-Ellis types of,the two-scalar disformal tensor \eqref{dis-T-2SF}.} \label{table-2SF}
\end{table}

\bibliography{refs}
\bibliographystyle{JHEPmod}
	
\end{document}